\pdfoutput=1
\documentclass[smallextended,natbib]{svjour3}       
\smartqed  

\usepackage{graphicx}
\newcommand{\solphys}{Sol.~Phys. }
\newcommand{\aap}{A\&A }
\newcommand{\aaps}{A\&AS }
\newcommand{\aapr}{A\&A~Rev. }
\newcommand{\apj}{ApJ }
\newcommand{\apjl}{ApJ }
\newcommand{\nat}{Nature }
\newcommand{\physrep}{Phys.~Rep. }

\newcommand{\mnras}{MNRAS }

\newcommand{\la}{\mathrel{\mathchoice {\vcenter{\offinterlineskip\halign{\hfil
$\displaystyle##$\hfil\cr<\cr\sim\cr}}}
{\vcenter{\offinterlineskip\halign{\hfil$\textstyle##$\hfil\cr<\cr\sim\cr}}}
{\vcenter{\offinterlineskip\halign{\hfil$\scriptstyle##$\hfil\cr<\cr\sim\cr}}}
{\vcenter{\offinterlineskip\halign{\hfil$\scriptscriptstyle##$\hfil\cr<\cr\sim\cr}}}}}

\journalname{The Astronomy and Astrophysics Review}

\begin{document}

\title{Solar magnetic fields as revealed by Stokes polarimetry}

\author{J.O. Stenflo}

\institute{J.O. Stenflo \at Institute of Astronomy, ETH Zurich,
  CH-8093 Zurich, and\\
Istituto Ricerche Solari Locarno, Via Patocchi, CH-6605 Locarno Monti,
Switzerland \\ \email{stenflo@astro.phys.ethz.ch} }

\date{Received: 10 July 2013 / Accepted: 4 September 2013}

\maketitle

\begin{abstract}\ 
Observational astrophysics started when spectroscopy could be applied
to astronomy. Similarly, observational work on stellar magnetic fields
became possible with the application of spectro-polarimetry. In recent
decades there have been dramatic advances in the observational tools
for spectro-polarimetry. The four Stokes parameters that provide a
complete representation of partially polarized light can now be
simultaneously imaged with megapixel array detectors with high
polarimetric precision ($10^{-5}$ in the degree of polarization). This
has led to new insights about the nature and properties of the
magnetic field, and has helped pave the way for the use of the Hanle
effect as a diagnostic tool beside the Zeeman effect. The magnetic
structuring continues on scales orders of magnitudes smaller than the
resolved ones, but various types of spectro-polarimetric signatures
can be identified, which let us determine the field strengths and
angular distributions of the field vectors in the spatially unresolved
domain. Here we review the observational properties of the magnetic field,
from the global patterns to the smallest scales at the magnetic
diffusion limit, and relate them to the global and local dynamos. 
\end{abstract}

\section{Historical background}\label{sec:historical}
The discovery and classification by Joseph Fraunhofer of the
absorption lines in the Sun's spectrum and the demonstration by Bunsen
and Kirchhoff that such lines represent ``fingerprints'' of chemical
elements marked the birth of modern astrophysics. Spectral analysis
became the standard tool of astronomers to determine the temperatures,
densities, velocities, and chemical compositions of stellar
atmospheres. 

From the electromagnetic wave theory of Maxwell and Hertz
it follows that the spectral radiation is not only characterized by its
intensity, but also by its polarization. G.G. Stokes showed how the
complete intensity and polarization information of a light beam could
be described in a unified way in the form of a 4-vector, the Stokes
vector. The first vector component represents the ordinary
intensity, the second and third components describe linear
polarization along two directions in the transverse plane, while the
fourth component relates to the circular polarization. 

This representation is very elegant and powerful, since it can
describe any partially polarized light beam, and all the four vector
components are given in the same units (intensity). Going from
spectroscopy to spectro-polarimetry thus means that we increase the
dimensionality of information space from 1-D to 4-D. 

The presence of nonzero polarization implies a breaking of the
symmetry in the source region. In reflection and scattering it is
the relative directions of the incident and the reflected
or scattered beams that break the symmetry. However, of greater
diagnostic importance in astrophysics is the breaking of the spatial
symmetry by a magnetic field. The Lorentz force induces a precession
of the oscillating atomic dipole moment around the magnetic field
vector (Larmor precession), with the result that the transition
frequencies between the different atomic levels get split into
different components, which are polarized in ways that depend on the
strength and orientation of the field. This effect, discovered by
Pieter Zeeman in 1896, led to the discovery of magnetic fields in
sunspots by George Ellery \citet{stenflo-hale08}. 

In the usual theory of the Zeeman effect the split components
superpose incoherently, i.e., they behave like independent
lines. However, in scattering processes there are phase relations
between the different magnetic $m$ substates of the excited state, a
coherent superposition of quantum amplitudes that can be thought of as
a ``Schr\"odinger cat state'', provided that the relative phases do
not get scrambled by collisions during the life time of the excited
state. In 1923 Wilhelm Hanle demonstrated experimentally how an
imposed magnetic field that breaks the $m$ state degeneracy causes
partial decoherence that increases to become complete with increasing
strength of the field \citep{stenflo-hanle24}. His results 
provided guidance to the 
early development of quantum theory, since they represented a direct
demonstration of the concept of linear superposition of quantum
states, which is a cornerstone 
of quantum mechanics. 

The effect of the magnetic field on the $m$ state interference and
decoherence with the accompanying, observable polarization effects in
the scattered radiation, is what is called the Hanle effect. The
resulting polarization phenomena in the Sun's spectrum can be used to
diagnose solar magnetic fields in parameter domains that are not
accessible to the ordinary (incoherent) Zeeman effect. The Hanle and
Zeeman effects are therefore highly complementary. A theoretical
  foundation for the application of the Hanle effect in astrophysics
  was established through the work of
  \citet{stenflo-house70a,stenflo-house70b,stenflo-house71},
  \citet{stenflo-omont73}, \citet{stenflo-bommiersb78}, and
  \citet{stenflo-s78hanle}. The 
  observational programs initially aimed at the diagnostics of 
  magnetic fields in solar prominences \citep[e.g.][]{stenflo-leroy77}.

Recent compilations of articles on various aspects of solar magnetism
have been provided by \citet{stenflo-issi09} and
\citet{stenflo-hasanrutten10}. When the present review journal ({\it
  Astron. Astrophys. Rev.}) was founded, its first article was on the
topic ``Small-scale magnetic structures on the Sun''
\citep{stenflo-s89aar}. The diagnostic use of polarized radiation to
explore solar magnetism has been the topic of a few monographs
\citep{stenflo-book94,stenflo-dti03,stenflo-lanlan04} and has been the focus of a series of
International {\it Solar Polarization Workshops}, which have taken
place every three years since the first one in St. Petersburg, Russia,
in 1995. The proceedings of these Workshops (6 of them so far) provide
a good account of the state of the art in the field
\citep{stenflo-spw1,stenflo-spw2,stenflo-spw3,stenflo-spw4,stenflo-spw5,stenflo-spw6}. 

The present review tries to give a broad account of what we have
learnt about solar magnetic fields from spectro-polarimetry by
addressing a number of key topics, like the global and local dynamos,
intermittency, and turbulence. By selectively focusing on these more
general aspects, we have left out some other important topics, like the magnetic properties of
sunspots, flares, or other active phenomena, which have received much
attention in the solar physics literature. 
The selection of topics is biased by representing areas that have been
of particular interest in the work of the present author. This choice has
made it easier to present a coherent picture of the subject as opposed
to trying to report on and give adequate credit to the diverse and
often diverging world-wide contributions to the field. 

The observed field properties that we are reviewing 
mainly reflect the conditions in the Sun's photosphere, since
most diagnostics with Stokes polarimetry depends on the use of
photospheric spectral lines. In future, attention will shift towards
the chromosphere, transition region, and corona, but so far Stokes
polarimetry of these regions have added relatively little to our
understanding of the fundamental properties of the Sun's magnetic
field. 

Much of contemporary solar physics deals with the properties of solar
magnetic fields, since they are responsible for all of solar and stellar
activity and variability on intermediate time scales and govern space
weather processes. Because the literature on all these various topics
has grown immensely, the present review does not aim for any
completeness in giving credit to all the important contributions in
the field. In the selection of references we generally try to include
references to the older discovery papers of the respective topics in
addition to referring to more recent work, to avoid the impression
that most of what we know comes from recent discoveries.

\section{Overview of Stokes polarimetry}\label{sec:overview}
Electromagnetic radiation consists of an ensemble of wave packets
(photons) with an electric vector that oscillates in the transverse
plane (perpendicular to the propagation direction). Each such wave
packet is 100\,\%\ elliptically polarized, i.e., the trajectory of the
tip of the electric vector describes an ellipse defined by three
parameters: its size, the orientation angle of its major axis, and the
ellipticity with sign (which determines the sense of rotation, left or
right). These parameters are inconvenient to work with, since they
have different dimensions, and the polarization ellipse is incapable
of describing partial polarization. The degree of polarization would
have to be added as a fourth parameter to obtain a full description of
polarized light. 

Except for laser light the radiation that we encounter in nature
represents an incoherent superposition of the 100\,\%\ elliptically
polarized wave packets of the ensemble. Since their phases are not
correlated with each other, the superposition causes the
degree of polarization of the ensemble to usually be much smaller
than 100\,\%. 

\subsection{Stokes parameters and Mueller calculus }\label{sec:stokes}

\begin{figure*}
  \vspace{-6.8cm}
\centering
\includegraphics[width=0.6\textwidth]{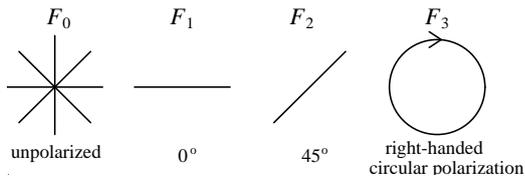}
  \vspace{1mm}
\caption{Symbolic representation of four idealized filters used for
  the definition of the four Stokes parameters. Filter $F_0$ is empty,
  $F_1$ and $F_2$ transmit linear polarization oriented at $0^\circ$ and
  $45^\circ$, while $F_4$ transmits right-handed circularly polarized
  light. 
  From \citep{stenflo-book94}.}\label{fig:stokesym} 
\end{figure*}

There are different formalisms for dealing with polarization, but the
most convenient and powerful description that also relates directly to
the measurement process is in terms of the four Stokes parameters,
which can be combined into a 4-vector, the Stokes vector 
\begin{equation}\label{eq:stokes}
\vec{I}=\left(\matrix{S_0\cr S_1\cr S_2\cr S_3\cr}\right)
\equiv\left(\matrix{I\cr Q\cr U\cr V\cr}\right)\,.
\end{equation}
There are different ways to define these parameters, but the
operational definition in terms of the four ideal filters
$F_k,\,\,k=0,1,2,3$ illustrated in Fig.~\ref{fig:stokesym} is directly
related to the measurement process. If $I_k$ is the intensity that is
measured by a detector placed behind filter $F_k$, then Stokes
parameter $S_k$ is defined as 
\begin{equation}\label{eq:sk}
S_k=2I_k-I_0\,.
\end{equation}
The inversion of this equation is 
\begin{equation}\label{eq:ik}
I_k={1\over 2}(I_0+S_k)\,.
\end{equation}
If we would replace the three polarizing filters with their orthogonal
versions (vertical polarization, $-45^\circ$ polarization, left-handed
circular polarization), then the signs of the corresponding $S_k$
would change. From this it follows that Stokes $Q$ represents the
difference between horizontal and vertical linear polarization, Stokes $U$
 the difference between + and $-45^\circ$ linear polarization, Stokes
 $V$ the difference between right- and left-handed circular
 polarization. 

Note that all four Stokes parameters have the same dimension,
intensity, although $Q$, $U$, and $V$ represent intensity {\it
  differences}. The intensity $I$ (proportional to the number of
photons) can always be considered as the sum of two contributions: 
\begin{equation}\label{eq:iup}
I=I_u+I_p\,,
\end{equation}
where 
\begin{equation}\label{eq:ip}
I_p=\sqrt{Q^2+U^2+V^2}
\end{equation}
represents 100\,\%\ elliptically polarized light, while $I_u$ is
completely unpolarized (with $Q=U=V=0$). The {\it degree of polarization} is then defined as 
\begin{equation}\label{eq:degpol}
p={I_p\over I_u+I_p}={\sqrt{Q^2+U^2+V^2}\over I}\,.
\end{equation}
$Q/I$ and $U/I$ represent the fractional linear polarizations in the
horizontal and $+45^\circ$ directions, $V/I$ the fractional circular
polarization. 

The effect of a medium on the Stokes vector $\vec{I}$ can be described by the $4\times 4$ {\it Mueller matrix} $\vec{M}$: 
\begin{equation}\label{eq:mueller}
\vec{I}^\prime=\vec{M}\vec{I}\,.
\end{equation}
The medium may be of any kind, like a telescope system or a stellar
atmosphere. If the medium is described as a sequence of consecutive
components $i=1,2,\ldots ,n$ (like a sequence of retarders,
polarizers, or modulators in a telescope system, or a sequence of
differential layers in a stellar atmosphere), each with its own Mueller matrix $\vec{M}_i$, then 
\begin{equation}\label{eq:train}
\vec{M}=\vec{M}_n\vec{M}_{n-1}\ldots\vec{M}_2\vec{M}_1\,.
\end{equation}
Index $i$ increases in the propagation direction (the order is
essential). 

Polarized radiative transfer can be formulated in a way similar to
that of unpolarized radiative transfer. The main difference is the
dimensionality of the problem. Instead of a scalar problem we have to
deal with the transfer of a 4-vector, which is operated on by $4\times
4$ matrices, which do not commute. This greatly enhances not only the
complexity but also the richness of the problem and its solutions. In
terms of diagnostics, we go from 1-D to 4-D information space. It is
important to note that the information provided by the new dimensions
($Q$, $U$, and $V$) cannot be derived from Stokes $I$, but each
dimension provides 
a different diagnostic window to the universe. We will see many
examples on this.

\subsection{Spectro-polarimeters}\label{sec:polarimeters}
As we saw in Sect.~\ref{sec:overview}, images of the $Q$, $U$, and $V$
parameters are obtained by forming {\it differences} between images in
orthogonal polarization states. There are two principal ways of
forming such differences: (1) Recording the orthogonal polarization states
simultaneously with a polarizing beam splitter. Advantage: Identical
seeing effects in the two images, which subtract out when forming the
difference. Disadvantages: The images fall on different detector areas
with different gain tables, and it gets complicated when trying to record all Stokes
parameters simultaneously ($Q$, $U$, and $V$ require 6 images). (2)
Doing sequential switching of the polarization state (modulation of the
beam). Advantages: The same detector area and gain table is used for
all the polarization states, and one can modulate all the Stokes
parameters with a single beam. Disadvantage: The images are not
simultaneous and may therefore be subject to different seeing
distortions, which generate spurious polarization signatures in the
difference images. 

If one can modulate faster than the seeing frequencies, then the
modulation approach is superior to the beam splitter approach. A
breakthrough in the mapping of solar magnetic fields came with the
introduction by \citet{stenflo-babcock53} of the solar magnetograph,
which used electro-optical modulation of the circular
polarization. The detector was a photomultiplier in the spectrograph
focal plane, receiving the light selected by exit slits in the wings
of a spectral line sensitive to the Zeeman effect, and the
demodulation was performed with a lock-in amplifier. This technique
was extended by 
\citet{stenflo-stepseverny62} to record all Stokes parameters and 
thereby produce maps of the vector magnetic field. Since the detectors
used at that time were basically 1-pixel devices (photomultipliers),
images had to be built up by mechanical scanning. Line profiles were
not recorded, only selected wavelength bands usually positioned in the
wing of a spectral line. 

Stokes polarimeters that could simultaneously record the full line profiles of the
four Stokes parameters were developed in the 1970s with various types
of detectors (photographic plates or linear arrays), beam splitters,
and modulators
\citep{stenflo-harveyetal72,stenflo-wittmann73,stenflo-houseetal75,stenflo-bauretal80}. This 
development transformed spectroscopy from 1-D (intensity) to 4-D (the
four Stokes parameters). 

The next technological advance came in the 1980s with the introduction
of 2-D CCD type detectors in astronomy. While they allow for
typically a million simultaneous image elements (pixels) in contrast
to the 1-pixel photomultiplier detectors, the readout is slow and
therefore seemed to be incompatible with fast modulation. This is of no particular
concern for magnetographs and polarimeters in space, like SOHO/MDI
\citep{stenflo-scherreretal95}, 
or the various polarimetric instruments on the Hinode spacecraft like 
SOT/SP \citep{stenflo-kosugi07,stenflo-suematsu08,stenflo-tsuneta08},
because of the absence of seeing outside the terrestrial
atmosphere. In the case of ground-based observations, however, seeing
noise becomes a limiting factor when the modulation is not fast
enough. 

The apparent incompatibility problem between the slow readout of
large-area CCD detectors and fast polarization modulation found a
solution with the ZIMPOL (Zurich Imaging
Polarimeter) technology
\citep{stenflo-povel95,stenflo-povel01,stenflo-gandetal04}. Instead of
trying to do readout at the modulation rate, the photo charges are
shifted between the exposed areas and fast hidden buffers in synchrony
with the modulation. Since CCD detectors do not have storage areas
below the pixel layer, the buffer storage is created by depositing in
the manufacturing process a mask on the sensor such that for each
group of four pixel rows, one is open, while the other three are
hidden behind the mask for use as buffers. The photo
charges can be laterally shifted at fast rates and cycled through the
four rows in synchrony with the modulation at kHz rates. This implies
that we create four simultaneous image planes within the same
CCD sensor, each of which represents a different
state of polarization. 

To avoid light losses on the masked detector area an array of
cylindrical microlenses is mounted on top of the sensor. Each
microlens has a width of four pixel rows and ensures that all light
falling on it is directed to the unmasked pixel row. After temporal integration
over a large number (e.g. $10^4$) modulation cycles, all four image
planes are read out. Through linear combinations between the image
planes and use of the polarization calibration information,
simultaneous images of the four Stokes parameters are obtained. The
fractional polarization images $Q/I$, $U/I$, and $V/I$ are free from
gain-table effects (flat-field errors), since the identical pixels
have been used to expose the four image planes. Therefore the
individual pixel gain factors divide out completely when
forming the fractional polarization. This is a main reason why we
for ZIMPOL observations do not represent the observed polarization
images in terms of $Q$, $U$, 
and $V$, but instead in terms of these quantities divided by Stokes
$I$. Flat-fielding is only needed for the Stokes $I$ image. 

With this technology the two main noise sources get eliminated,
seeing noise (because we modulate much faster 
than the seeing frequencies) and gain-table noise (which divides out
in the fractional polarizations). In practice the polarimetric sensitivity is 
only limited by the Poisson statistics of the number of collected photo
charges, which depends on the photon collecting area of the telescope,
the optical transmission and detector efficiencies, 
the effective integration time, and the spatial and spectral
resolutions. Optimizing these parameters ZIMPOL routinely 
achieves a polarimetric precision of $10^{-5}$ in combination with
high spectral resolution. 

\begin{figure*}
  \vspace{-9mm}
\centering
\includegraphics[width=0.7\textwidth]{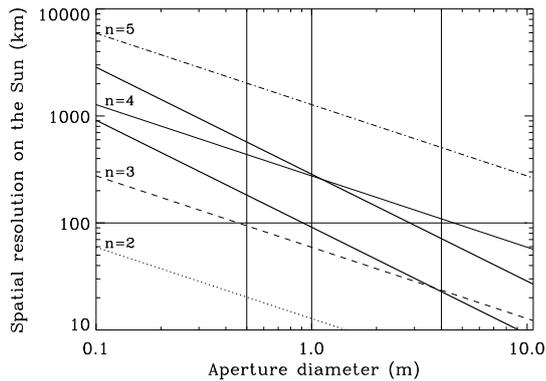}
\vspace{-4.55cm}
\caption{Trade-offs between spatial resolution and polarimetric
  precision as a function of telescope aperture size, assuming a
  spectral resolution of 300,000 and a maximum integration time that
  is limited by the evolutionary time scale for the given spatial
  scale. The thin slanted lines labeled by parameter $n$ show where
  the polarimetric precision is $10^{-n}$. The thick slanted lines give
  the telescope diffraction limit for wavelengths 1.56\,$\mu$m (upper
  line) and 5000\,\AA\ (lower line). The vertical lines mark aperture
  sizes 0.5, 1, and 4\,m. From \citet{stenflo-s99spw2}.}\label{fig:tradeoffs} 
\end{figure*}

\subsection{Trade-offs}\label{sec:tradeoffs}
With a perfect polarimeter the precision is limited by the number of 
collected photons. It is a common misunderstanding that we always receive
enough photons from the Sun because it is so close to us. The number
of photons from a resolved stellar disk per diffraction-limited angular
resolution element is independent of stellar distance and only depends
on the effective temperature of the stellar surface. If we want to do
polarimetry with an angular resolution at the telescope diffraction
limit, then the solar observations are severely photon starved. It is
never possible to independently optimize the four basic observing
parameters spectral, spatial, and temporal resolution and polarimetric
accuracy, major trade-offs are always necessary. It is common that
solar astronomers give priority to the highest possible
spatial resolution to probe the smallest solar structures, but this
can only be done at the expense of polarimetric precision and/or
spectral resolution. Temporal resolution  cannot be chosen
independently of angular resolution, since smaller structures evolve
faster, and the effective integration time must be shorter than the
evolutionary time scale. 

Figure \ref{fig:tradeoffs} gives an overview of the trade-offs in the
mentioned 4-D observational parameter space. We see that with a 4-m
solar telescope like the future ATST it is theoretically possible to
reach a polarimetric precision of $10^{-4}$ in combination with a
spatial resolution of 100\,km, but for a precision of $10^{-5}$, which
is possible with a polarimeter like ZIMPOL and needed for explorations
of the Second Solar Spectrum and the Hanle effect, the angular
resolution cannot be better than about 1\,arcsec even with such a
large telescope. 

\begin{figure*}
\centering
\includegraphics[width=0.75\textwidth]{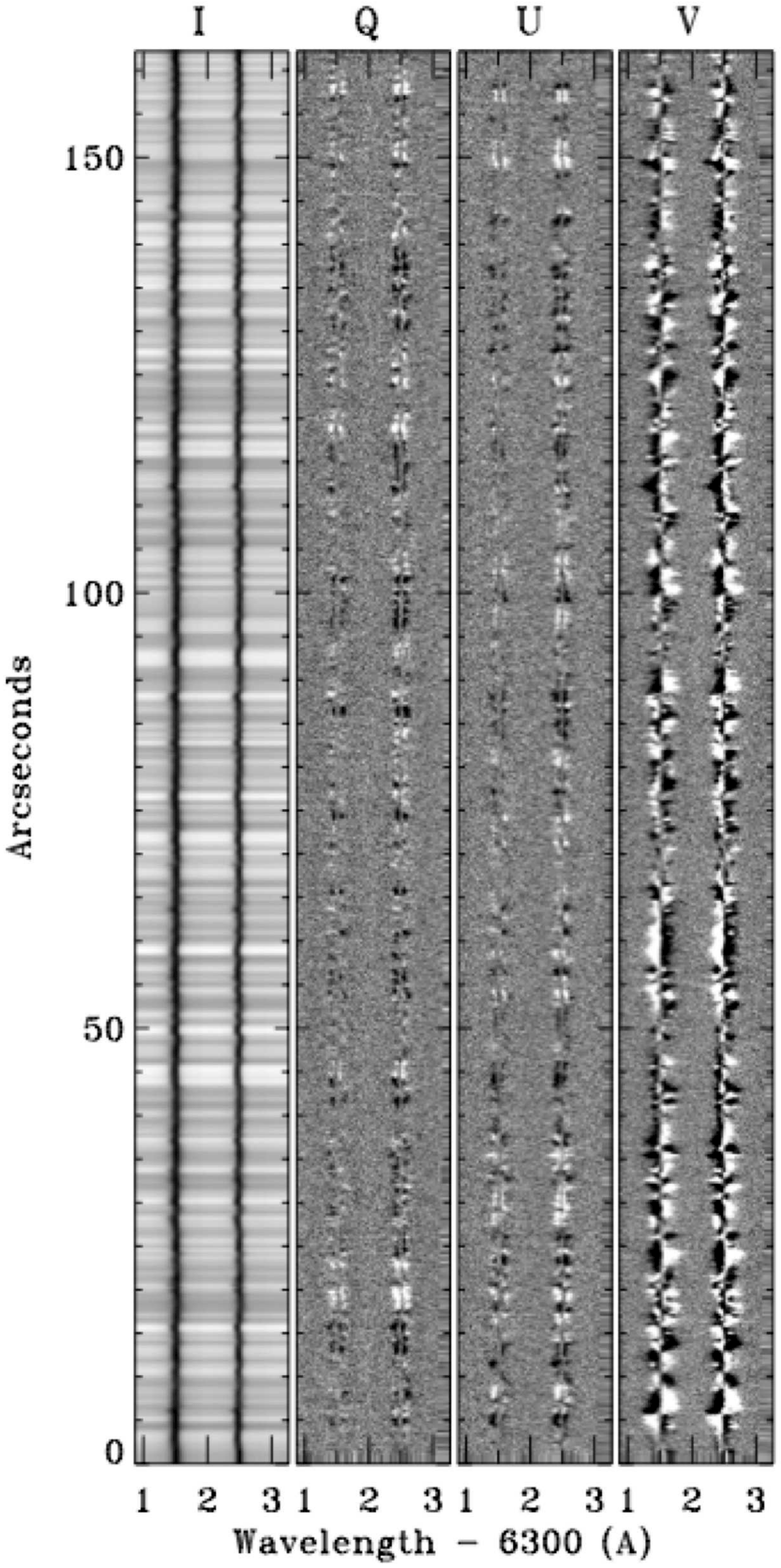}\caption{Example
  of Stokes spectra of the quiet-sun disk center for the spectral
  window with the two Fe\,{\sc i } 6301.5 and 6302.5\,\AA\ lines. The
  recording was made with the Hinode SOT/SP instrument in deep mode (9.6\,s
  integration time) on February 27, 2007. The spatial resolution is
  about 230\,km on the Sun.  From \citet{stenflo-litesetal08}.}\label{fig:stokeshinode} 
\end{figure*}

\section{Zeeman effect}\label{sec:zeeman}
Since its first application in astrophysics little more than a century
ago \citep{stenflo-hale08}, the Zeeman effect has been the prime tool
to gain information about cosmic magnetic fields. With modern imaging
polarimeters we can produce simultaneous images of the four Stokes
parameters, either with narrow-band filters to obtain 2-D maps of the
solar surface in a selected wavelength band, or with spectrographs to
obtain detailed Stokes line profiles for each resolution element along the
spectrograph slit. An example of such simultaneous spectral images of
the four Stokes parameters obtained with high spatial resolution
(0.3\,arcsec) with the SOT/SP instrument on the Hinode spacecraft
is shown in Fig.~\ref{fig:stokeshinode}. The spectral structures in Stokes $V$
are due to the longitudinal Zeeman effect and generally have
anti-symmetric profile shapes, while the structures in $Q$ and $U$ are
caused by the transverse Zeeman effect and have symmetric profile shapes. The
polarized spectra reveal the existence of highly structured magnetic fields everywhere
on the quiet Sun. The intensity variations and line wiggles seen in
Stokes $I$ are due to the solar granulation. 

An intuitive understanding of such Zeeman-effect spectra can be
developed in terms of the basic symmetry properties of the Stokes profiles and
their different sensitivities to the magnetic field. The essential
properties can be sufficiently exposed 
by considering the two idealized cases of a homogeneous
magnetic field oriented (a) along the line of sight, and (b)
perpendicular to the line of sight. No assumptions or explicit
representations for the model atmosphere are needed \citep[cf.][]{stenflo-book94}. 

Through a suitable choice of basis polarization vectors, the four
coupled transfer equations for the four Stokes parameters decouple
(the problem gets diagonalized). For case (a) we choose right-
and left-handed polarized unit vectors, corresponding to the
polarization of the two $\sigma$ components of the longitudinal Zeeman
pattern. For
case (b) we choose two unit vectors linearly polarized along and
perpendicular to the direction of the transverse magnetic field,
corresponding to the polarization of the $\pi$ and $\sigma$ components
of the transverse Zeeman pattern. The decoupling of the transfer
equations implies that the atmosphere decouples into two mutually
independent, noninteracting atmospheres. 

Let us first consider case (a). If $I_0(\Delta\lambda)$ represents the
emergent spectrum from each atmosphere in the limit of vanishing
magnetic field, then in the presence of a homogeneous longitudinal
field with Zeeman splitting $\Delta\lambda_H$ the emergent spectra
$I_\pm$ from the two atmospheres can simply be expressed as  
\begin{equation}\label{eq:longcase}
I_\pm(\Delta\lambda)=I_0(\Delta\lambda\mp\Delta\lambda_H)\,.
\end{equation}
This is an exact result for a normal Zeeman triplet, valid for all field
strengths, line strengths, and static model atmospheres. 

Transforming back to the standard Stokes system (which uses Cartesian basis vectors) we get 
\begin{eqnarray}\label{eq:ivtaylor}
I&=&\textstyle{1\over 2}(I_+ +I_-)\approx I_0 +{\textstyle{1\over 2}}(\Delta\lambda_H)^2\,\partial^2 I_0 /\partial \lambda^2+\ldots\,,\nonumber\\ V&=&\textstyle{1\over 2}(I_+ -I_-)\approx -\Delta\lambda_H\bigl[\,\partial I_0 /\partial \lambda+\,{\textstyle{1\over 6}}(\Delta\lambda_H)^2\,\partial^3 I_0 /\partial \lambda^3+\ldots\,\bigr]\,.
\end{eqnarray}
 For weak magnetic fields (when the Zeeman splitting $\Delta\lambda\ll$ the spectral line width) 
\begin{eqnarray}\label{eq:vweak}
I&\approx & I_0\,,\nonumber\\ V&\approx &-\Delta\lambda_H\,\partial I_0/\partial \lambda\,,
\end{eqnarray}
 where 
\begin{equation}\label{eq:dlh}
\Delta\lambda_H =zg\lambda^2 B
\end{equation}
and $g$ is the Land\'e factor. $z=4.67\times 10^{-13}\,$
\AA$^{-1}\,$G$^{-1}$. This implies that  
\begin{equation}\label{eq:vigrad}
V\sim B\,{\partial I\over\partial \lambda}\,.
\end{equation}
Stokes $V$ therefore has a profile shape that mimics
the anti-symmetric gradient of the Stokes $I$ spectrum, and the $V$
amplitude is proportional to the field strength. Beyond the weak-field
regime, when the Zeeman splitting is no longer small in comparison with
the line width, the relation between Stokes $V$, field-strength $B$,
and intensity profile $I$ becomes nonlinear (Zeeman saturation). This
nonlinearity can be used to infer the existence and strengths of
intermittent flux concentrations at scales far smaller than the
telescope resolution, as we will see in Sect.~\ref{sec:kg}. 

For photospheric lines there are generally fluctuating and systematic
deviations from the proportionality described by
Eq.~(\ref{eq:vigrad}), for two reasons:  (1) The photosphere is
dynamic with correlated magnetic and velocity-field gradients {\it
  along} the line of sight, which leads to Stokes $V$ asymmetries that
do not obey Eq.~(\ref{eq:vigrad})
\citep[cf.][]{stenflo-illing75,stenflo-auer78,stenflo-setal84}, and
(2) much of the photospheric flux is in collapsed, strong-field form,
while photospheric lines in general have small line widths, with the
consequence that lines with large Land\'e factors are affected by
Zeeman saturation (deviation from linearity of the field-strength
dependence). 

\begin{figure*}
\centering
\includegraphics[width=0.75\textwidth]{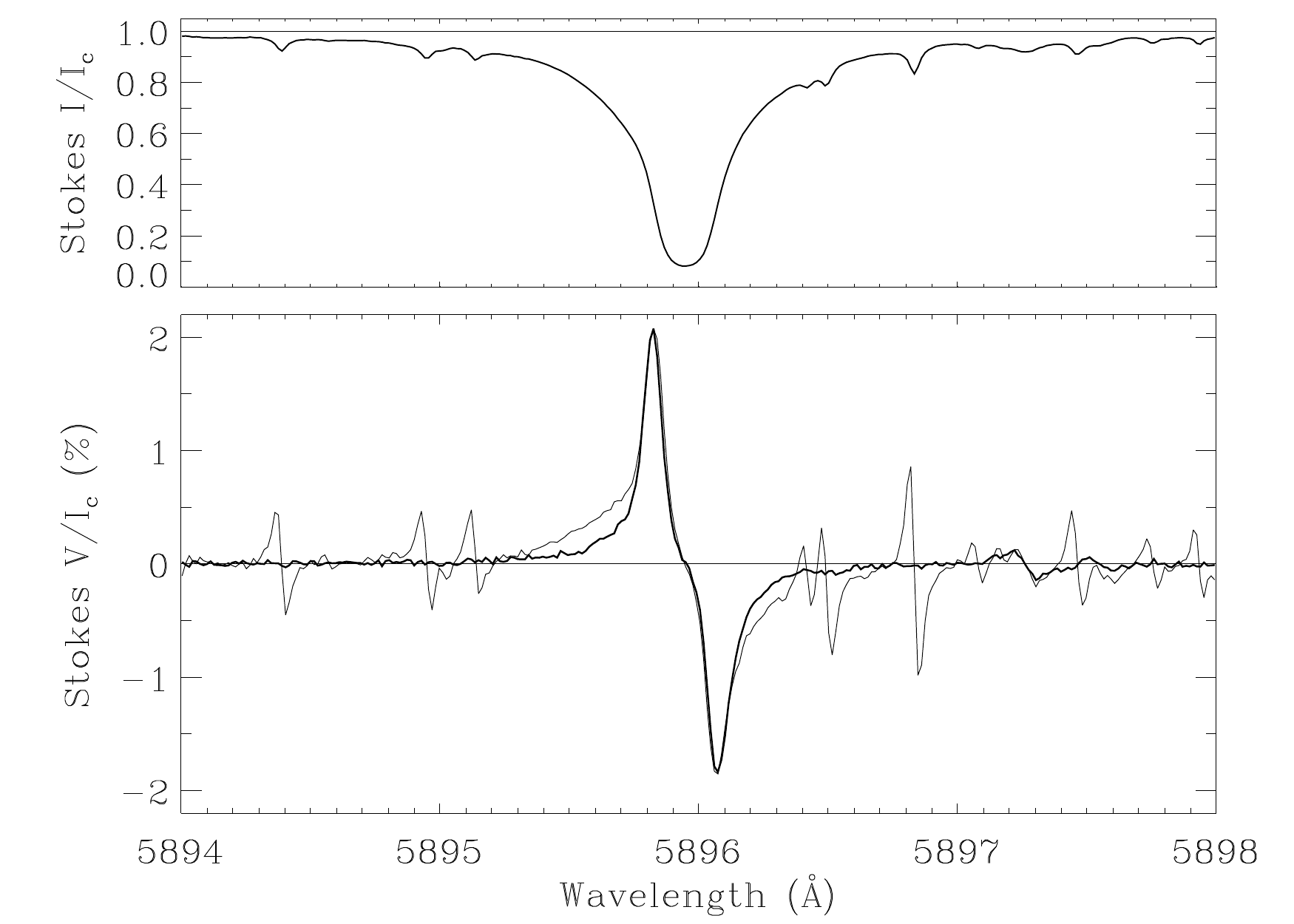}\caption{Example
  of  a small spectral section around the chromospheric Na\,{\sc i} D$_1$ 5896\,\AA\
  line from the Stokes $V$ atlas of a strong plage at disk center recorded
  on April 30, 1979, with the FTS (Fourier transform spectrometer) of
  the NSO McMath-Pierce facility at Kitt Peak. In the bottom panel the
  Stokes $V$ spectrum is shown (thick 
  solid line) with the intensity gradient $-\partial I/\partial\lambda$
  superposed (thin solid line) after it has been normalized to the $V$
  amplitude. Note the numerous spectral features in $-\partial
  I/\partial\lambda$ from water vapor in the terrestrial
  atmosphere. These telluric features are completely absent in the $V$
  spectrum. Adapted
from \citet{stenflo-setal84}.}\label{fig:nad1fts} 
\end{figure*}

For the broad chromospheric lines, however, the $V$ profile shape is
nearly identical to the shape of $-\partial I/\partial\lambda$, as
illustrated in Fig.~\ref{fig:nad1fts} for the well-known Na\,{\sc 
  i} D$_1$ 5896\,\AA\ line. The 
numerous telluric water vapor lines that surround the sodium line are conspicuous in the
$-\partial I/\partial\lambda$ spectrum, but as they are unpolarized
they are entirely absent in the Stokes $V$ spectrum. 

Let us now turn to case (b), a transverse magnetic field
in a static atmosphere. As before, the atmosphere can be treated as
composed of two mutually noninteracting atmospheres, one relating to
the transfer of the $\pi$ component, the other related to the two $\sigma$
components. However, in contrast to the longitudinal case the 
two $\sigma$ components now have the same polarization state. They
therefore 
interact in the radiative-transfer process if the medium is optically
thick. To avoid this complication for the present conceptual
discussion, let us assume that we are dealing with weak (optically
thin) spectral lines. If again $I_0(\Delta\lambda)$ represents the
emergent spectrum from each atmosphere in the limit of vanishing
magnetic field, then in the presence of a homogeneous transverse 
field the emergent spectra $I_\pi$ and $I_\sigma$ from the two
atmospheres become 
\begin{eqnarray}\label{eq:ipisigma}
I_\pi&=&I_0(\Delta\lambda)\,,\nonumber\\ I_\sigma &=&\textstyle{1\over 2}[\,I_0(\Delta\lambda-\Delta\lambda_H)\,+I_0(\Delta\lambda+\Delta\lambda_H)\,]\,.
\end{eqnarray}
 In the standard Stokes system with the positive Stokes $Q$ direction
 defined along the magnetic field, we get 
\begin{eqnarray}\label{eq:iqtaylor}
I&=&\textstyle{1\over 2}(I_\pi +I_\sigma)\approx I_0 +{\textstyle{1\over 4}}(\Delta\lambda_H)^2\,\partial^2 I_0 /\partial \lambda^2+\ldots\,,\nonumber\\ Q&=&\textstyle{1\over 2}(I_\pi -I_\sigma)\approx -{\textstyle{1\over 4}}(\Delta\lambda_H)^2\,\partial^2 I_0 /\partial \lambda^2+\ldots\,.
\end{eqnarray}
 For weak fields 
\begin{equation}\label{eq:b22grad}
Q\sim B^2\,{\partial^2 I\over\partial \lambda^2}\,.
\end{equation}
These expressions show that the linear polarization profiles are
symmetric and in the weak-field and weak-line limit mimic the second
derivative of the Stokes $I$ spectrum. Note in particular that for
weak fields the linear polarization scales with $B^2$, in contrast to
the circular polarization, which is linear in $B$.  

For a field that is oriented with azimuth angle $\chi$ with respect to the
defined $Q$ direction, 
\begin{eqnarray}\label{eq:b22azim}
Q&\sim &B^2\,\cos 2\chi\,\,{\partial^2 I\over\partial \lambda^2}\,,\nonumber\\ U&\sim &B^2\,\sin 2\chi\,\,{\partial^2 I\over\partial \lambda^2}\,.
\end{eqnarray}

If we now go to the general case when the magnetic field vector has an
arbitrary inclination $\gamma$ with respect to the line of sight, then
all the previous equations remain valid if we replace $B$ in the expressions for Stokes
$V$ with $B\cos\gamma$ (the line-of-sight component) and (in the
optically thin limit) replace $B^2$ in the
expressions for $Q$ and $U$ by $B^2\sin^2\gamma$ (the square of the
transverse field component). Although this $\gamma$ dependence is not exact in
the case of strong lines (due to transfer effects in optically thick media and 
magneto-optical effects), it remains a good
approximation. However, the
$Q$ and $U$ profile shapes will differ substantially from the
$\partial^2 I/\partial \lambda^2$ shape when the lines are no
longer optically thin. 

While the circular polarization is related linearly to the line-of-sight component $B_\parallel$
of the field according to Eq.~(\ref{eq:vigrad}), the linear
polarization scales with the square of the transverse component, in
other words with the magnetic energy density of the transverse
component, according to Eq.~(\ref{eq:b22azim}). Thus with the
weak-field approximation, which is valid for practically all lines
when $B\la 0.5$\,kG, 
\begin{eqnarray}\label{eq:bparperp}
B_\parallel\,& \sim &\,V\,,\nonumber\\
B_\perp\,& \sim &\,[\,Q^2 +U^2\,]^{1/4}\,.
\end{eqnarray}

\begin{figure*}
\centering
\includegraphics[width=0.7\textwidth]{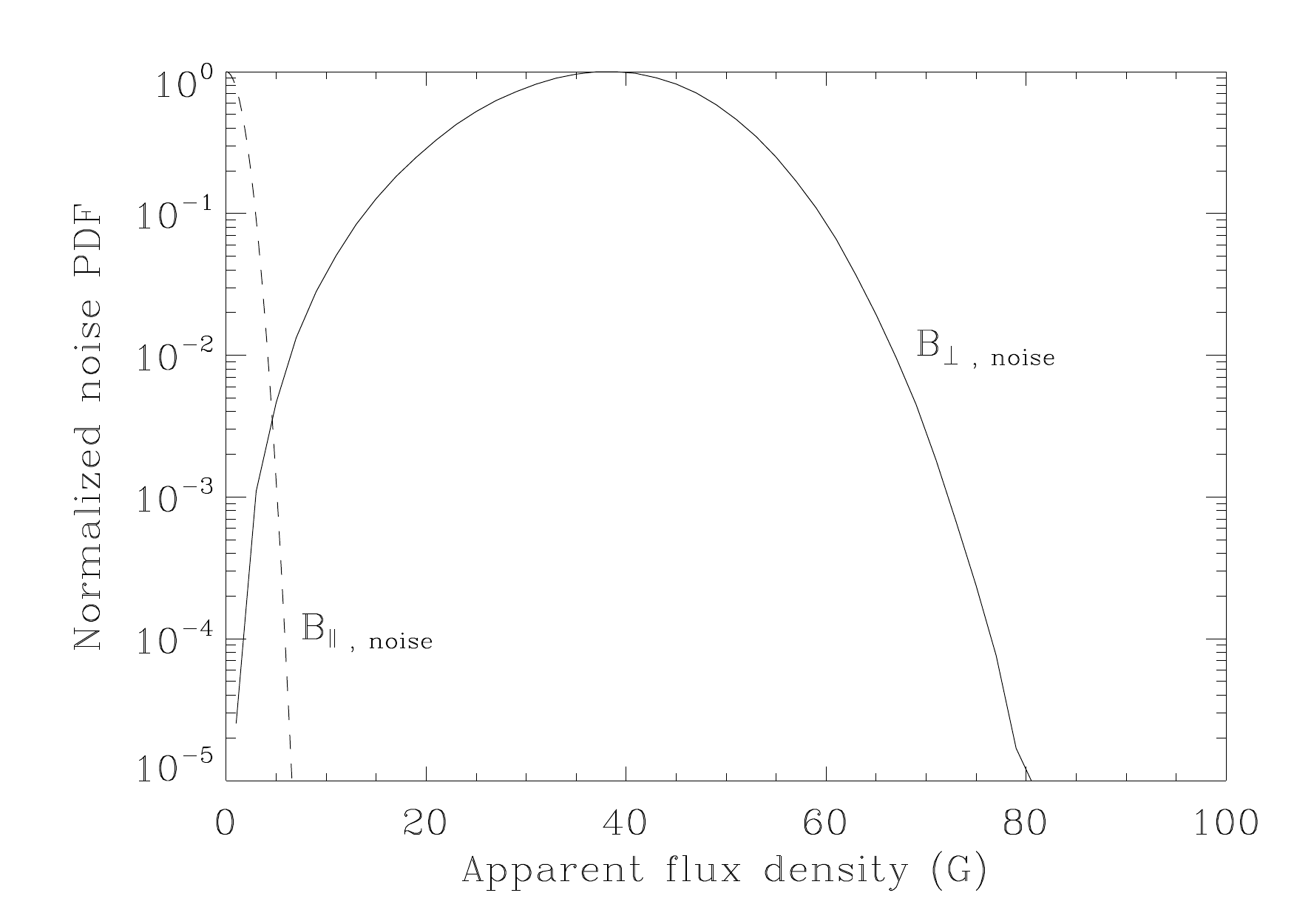}\caption{Histograms
  of the noise in the deep-mode Hinode SOT/SP observations, converted from polarization to field-strength units
  using the weak-field approximation. The measured polarization noise
  is Gaussian with standard deviations 0.035\,\%\ for Stokes $Q$ and
  $U$, 0.047\,\%\ for $V$. Although the noise in the linear
  polarization is smaller, it translates to much larger apparent field
  strengths $B_{\perp ,\,{\rm noise}}$ than the apparent field
  strengths $B_{\parallel ,\,{\rm noise}}$ of the circular
  polarization. Adapted from
\citep[][]{stenflo-s11aa}}\label{fig:pdfnoise} 
\end{figure*}

In addition to its nonlinear field dependence, the linear
polarization has much less field sensitivity than the circular
polarization, which makes the transverse field highly susceptible to
noise. As an example we consider the state-of-the-art Hinode SOT/SP
recording that was illustrated in Fig.~\ref{fig:stokeshinode}. With
radiative-transfer modeling of the Fe\,{\sc i} 6301.5\,\AA\ line the
proportionality constants in Eq.~(\ref{eq:bparperp}) can be estimated
to be 29.4 for the longitudinal and 184 for the transverse equation,
provided that the Stokes parameters are expressed in \%\ of the
continuum intensity and the field strength in G
\citep{stenflo-s11aa}. The measured 1-$\sigma$ noise in Stokes $V$ is
0.047\,\%, while it is somewhat smaller, 0.035\,\%, in each of $Q$ and
$U$. When however the Gaussian noise distributions are translated into
noise distributions for $B_\parallel$ and $B_\perp$, we get the
distributions shown in Fig.~\ref{fig:pdfnoise}. In spite of the
smaller noise in the linear polarization, the noise in the transverse
field is larger by approximately a factor of 25. What is much worse,
however, is the profoundly non-Gaussian nature of the noise
distribution due to the nonlinear field dependence of the linear
polarization. This makes it very difficult to obtain any reliable
determinations of vector magnetic fields in quiet solar regions, where
the magnetic fluxes are weak. If for instance in disk center
observations only a tiny fraction of
the noise in $B_\perp$ would be mistaken for a solar signal, then the
field would appear to be more horizontal than it really is. 

Another serious consequence of the nonlinear response of the
linear polarization to the magnetic field is a model dependence of
the interpretation of spatially averaged quantities. We know that the
magnetic structuring continues far beyond the resolution limit of
current telescopes, where we expect to have both collapsed kG type field concentrations
and turbulent fields that are tangled on small scales. In nearly all cases
(except for large structures like sunspots) the pixels of the detector are
much larger than the solar magnetic structures. Averaging over a pixel
is not a serious problem when the polarimetric response to the magnetic field
is linear, as it approximately is for the circular
polarization. Regardless of the nature of the subpixel structuring,
the pixel average of Stokes $V$ is proportional to the
line-of-sight component of the total
magnetic flux through the pixel area, or, equivalently, to the average
strength of the line-of-sight component. 

A similar type of interpretation is not possible for the transverse
magnetic field. The pixel average of $Q$ or $U$ is in a first
approximation proportional to the average magnetic energy density of
the transverse component, but a translation of average energy
density to average field strength is not possible without an assumed
model for the subpixel structuring, including the subpixel
distributions of field strengths, azimuths, and inclinations. 

The determination of vector
magnetic fields is therefore not a technical problem like how to
develop efficient algorithms for Stokes inversion, but it is
more fundamental, namely how to combine one component (along the
line of sight) that represents flux with another component (the
transverse one) for which the flux is not an observable. The
problem is aggravated by the circumstance that the noise in the
line-of-sight flux is Gaussian and very small in comparison with the
non-Gaussian transverse field noise. 

Although attempts to map vector magnetic fields on the Sun have been
made for more than half a century \citep[starting
with][]{stenflo-stepseverny62}, it is not surprising that most
magnetic-field work has focused on mapping and exploring the
line-of-sight component through recordings of the circular
polarization. With the advent of sensitive imaging polarimeters that
can record the full Stokes line profiles with good spatial resolution,
as illustrated in Fig.~\ref{fig:stokeshinode}, there has been a
revived interest in the determination of vector magnetic fields, but
there has been a tendency to downplay the fundamental limitations just
mentioned when doing formal Stokes inversions to generate maps of the vector
magnetic field.

\section{Hanle effect}\label{sec:hanle}
Polarization is related to some symmetry breaking process. In the
case of the Zeeman effect it is the magnetic field that breaks the
spatial symmetry. In the absence
of magnetic fields the symmetry can be broken in a scattering process,
depending on the directional 
relations between the incident and scattered radiation. The
oscillating electric vector of the incident light excites an
oscillating dipole moment in the scattering atomic system. The emitted
dipole radiation depends on how the dipole was
excited. 

\subsection{Scattering polarization}\label{sec:scattering}
The blue sky is linearly polarized by molecular Rayleigh
scattering. Similarly the solar spectrum is linearly polarized by
scattering processes in the Sun's atmosphere. In classical dipole
scattering the polarization reaches 100\,\%\ when the scattering angle
is $90^\circ$, but on the Sun the polarization is much smaller because
we average over a wide range of scattering angles. The angular
distribution of the incident radiation must be anisotropic in order to
generate any net polarization. For a
spherically symmetric Sun (disregarding local inhomogeneities) the
anisotropy expresses itself as limb darkening if the distribution
has a net preference for the vertical direction. For symmetry reasons
the scattering polarization vanishes at disk center and increases
monotonically as we approach the limb. The emitted radiation is
polarized with the electric vector oriented perpendicular to the
radius vector (parallel to the nearest solar limb). 

The idea of looking for scattering polarization on the Sun is old
\citep[e.g.][]{stenflo-ohman29}. The first survey of the scattering
polarization throughout the solar spectrum
\citep{stenflo-setal83a,stenflo-setal83b}, from the deep UV (near the
atmospheric cut-off around 3165\,\AA) to the near infrared (up to about
9950\,\AA) revealed a 
wealth of unfamiliar polarized spectral structures, for instance the 
feature that extends over more than 200\,\AA\ around the Ca\,{\sc ii}
K and H lines at 3933 and 3968\,\AA\ and which is caused by quantum
interference between the $J=3/2$ and 1/2 upper states of the K and H
transitions \citep{stenflo-s80}. It was like uncovering an
entirely new and unfamiliar spectrum and one had to start over again to
identify the various structures and their underlying physical
mechanisms. This prompted V.V. Ivanov of St. Petersburg to introduce
the now established name {\it Second Solar Spectrum} for the linearly polarized
spectrum that is exclusively caused by coherent scattering processes \citep{stenflo-ivanov91}. 

The polarimetric noise in this initial survey was of order
$10^{-3}$. Only the most prominent polarizing lines protruded well
above this level, but it was clear that they represented only the ``tips of the
icebergs''. With the implementation of the ZIMPOL
technology in the 1990s (see Sect.~\ref{sec:polarimeters}), with
which the noise level could be reduced to $10^{-5}$, the full
wealth of structures in the Second Solar Spectrum became accessible to
observation \citep{stenflo-sk96,stenflo-sk97}. 

Scattering is coherent and a source of linear polarization if there is
a phase relation between the excitation and emission process, in other
words, if the emitting particle retains a ``memory'' about how it was
excited. This memory can be erased by collisions during the excited
state, which scramble the phase and therefore depolarize. There is a
competition between the radiative decay rate and the collisional
rate. The fraction of emission processes that are undisturbed by
collisions represents coherent, polarizing scattering.

\subsection{Principles of the Hanle effect}\label{sec:principleshanle}
\begin{figure*}
\vspace{-5mm}
\centering
\includegraphics[width=0.85\textwidth]{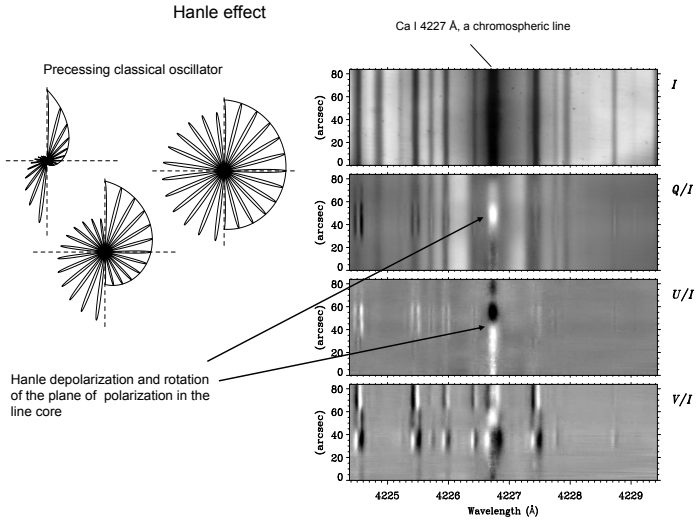}
\vspace{-3mm}
\caption{Illustration
of the Hanle effect, from \citet{stenflo-evershed10}. The three
rosette patterns to the left show the classical vector trajectories of
a damped dipole oscillator that is precessing around a magnetic field
that is oriented along the line of sight, for three values of the
field strength, increasing from left to right. When converted to the
Fourier domain the net effect of the magnetic field on the scattering
polarization in the Doppler core of the considered spectral line is to
depolarize and rotate the plane of polarization, as indicated by the
spectral signatures seen in the $Q/I$ and $U/I$ panels to the right
for the Ca\,{\sc i} 4227\,\AA\ line. The recording was made on March
9, 2002, at disk position $\mu =0.2$ near the heliographic S pole with
ZIMPOL-2 at the NSO McMath-Pierce facility at Kitt Peak. }\label{fig:rosette} 
\end{figure*}

The coherent scattering process can also be affected by an external
magnetic field, which causes modifications of the scattering
polarization. It is the set of magnetically induced alterations of the
scattering polarization that we refer to as the {\it Hanle effect},
discovered by Wilhelm Hanle in G\"ottingen in 1923
\citep{stenflo-hanle24}. An intuitive understanding of the Hanle
effect is provided by a classical description in terms of the electric
dipole oscillations of the scattering atom, as illustrated in the left
portion of Fig.~\ref{fig:rosette}. The
unpolarized incident radiation induces dipole oscillations in the
transverse plane. Seen by an observer receiving light scattered at
$90^\circ$, these oscillations are linear and perpendicular to the 
scattering plane. In the absence of a magnetic field we would then get
100\,\%\ linear polarization along this direction (which is assumed to be
the vertical direction in Fig.~\ref{fig:rosette}). If we introduce an
external magnetic field along the scattering direction, the damped
oscillator will precess around the field and describe a rosette
pattern. We have a competition between the radiative damping rate and
the rate of Larmor precession, which is proportional to the magnetic
field strength. For the three rosette patterns illustrated in
Fig.~\ref{fig:rosette} the field strength increases from left to
right. If the precession rate is much smaller than the damping rate,
the emission process gets completed before much precession has had
time to take place. Then the scattering polarization will not deviate
much from the nonmagnetic case. If on the other hand the precession
rate is much larger than the damping rate, the orientations of the
dipole oscillations get randomized, resulting in unpolarized
radiation. In the intermediate case, when the two rates are
comparable, the net effect on the scattered radiation is rotation of
the plane of linear polarization combined with partial
depolarization. The effect on the Stokes $Q$ and $U$ parameters is
found through Fourier transformation of the rosette patterns. 

The right side of Fig.~\ref{fig:rosette} shows an example of what the
observational signatures of the Hanle effect look like in the Sun's
spectrum. The recording was made with the ZIMPOL polarimeter in a
weakly magnetic region near the solar limb with the spectrograph slit
oriented perpendicular to the radius vector (parallel to the nearest
limb), and with positive Stokes $Q$ defined as linear polarization
with the electric vector oriented along the slit. The spectral window
is centered around the Ca\,{\sc i} 4227\,\AA\ line,
which happens to be the line that exhibits the largest scattering
polarization in the whole visible solar spectrum. The core of
  this line is formed in the low chromosphere well above the
  temperature minimum. In the absence of 
magnetic fields, scattering polarization would only show up as positive
$Q$ with no variations along the slit, while $U$ and $V$ would be
zero. All the signatures that we see in the fractional circular polarization
($V/I$) are due to the longitudinal Zeeman effect, while the symmetric
doublet patterns seen in the surrounding blend lines in the fractional
linear polarization ($Q/I$ and $U/I$) are due to the transverse Zeeman
effect. In the core of the strong 4227\,\AA\ line, however, we notice
a qualitatively completely different type of signature that has
nothing to do with the transverse Zeeman effect but is due to the
Hanle effect and varies along the slit because of spatially structured
magnetic fields. The variations of $Q/I$ along the slit are caused by
spatially varying Hanle depolarization, while the signatures in $U/I$
are due to spatially varying Hanle rotation of the polarization plane. 

Note that there are also two bright bands in $Q/I$ on either side of
the 4227 line core, which are due to nonmagnetic scattering
polarization in the wings of this chromospheric line. These bands show
no variations along the slit and are absent in $U/I$. The reason for
this behavior is that the effectiveness of the Hanle effect is limited
to the Doppler cores of
spectral lines but is absent in the wings, where the scattering
polarization behaves like in the nonmagnetic case. 

In quantum mechanics the Hanle effect can be understood in terms of
the coherent superposition of the partially split magnetic $m$
substates. In a scattering process that is undisturbed by collisions, the
excited, intermediate state represents a coherent superposition (with
mutual phase relations) of the $m$ states (a ``Schr\"odinger cat
state''). It is the quantum interferences between the $m$ states that
give rise to the polarization phenomena that represent the Hanle
effect. A good overview of the physics and diverse applications of the
Hanle effect is provided in \citet{stenflo-mor91}. 

The polarized scattering experiments in 1923-1924 that led to the
discovery of the Hanle effect \citep{stenflo-hanle24} played a key
role in the early conceptual development of quantum mechanics, since
they demonstrated experimentally the principle of linear superposition of atomic
states with quantum interference and the partial decoherence caused by
external magnetic fields that break the symmetry. The theoretical
edifice of quantum mechanics 
was built on these concepts.  

The Second Solar Spectrum is the playground for the Hanle effect,
since it represents the spectrum that is exclusively due to coherent
scattering, and it is the magnetic-field modifications of this
spectrum that constitutes the Hanle effect. The exploitation of the
rich potential of the Hanle effect requires a good understanding of
the Second Solar Spectrum and the underlying physical mechanisms. A
high-precision Atlas of the Second Solar Spectrum from the deep UV to
the near infrared as recorded with the ZIMPOL polarimeter has been produced by
\citet{stenflo-gandorf00,stenflo-gandorf02,stenflo-gandorf05}. For an
overview of the various types of physical effects revealed by this
spectrum, like $J$ state interference, scattering at molecules and
rare earth elements, continuum polarization, hyperfine structure and
isotope effects, optical pumping, and still unexplained enigmatic physics, see
\citet{stenflo-s04}. 

The Hanle and Zeeman effects are highly complementary and diagnose 
different aspects of solar magnetism, both because of their 
different field sensitivities and their different symmetry
properties. In the case of the Zeeman effect the field-strength
sensitivity depends on the ratio between the Zeeman splitting and the
line width, which is mainly determined by the Doppler width. In the
case of the Hanle effect the sensitivity depends on the ratio between
the Zeeman splitting and the damping width. Since the radiative
damping width is smaller than the Doppler width by typically a factor
of 25, the Hanle effect is sensitive to correspondingly weaker
fields. This is of particular importance for the diagnostics of
chromospheric magnetic fields, which are weak while the chromospheric
spectral lines are generally broad.

\subsection{Hanle effect signatures of turbulent fields}\label{sec:hanleturb}
The Zeeman effect is blind to a turbulent field that has mixed
polarities on a scale much smaller than the spatial resolution of the
telescope, because the polarization contributions of the opposite
polarities cancel each other. The Hanle rotation of the plane of
polarization has a similar type of
cancelation, but the Hanle depolarization effect does not, because it
responds to the field with only one ``sign''
(depolarization). 

This difference between the Hanle and Zeeman effects has its origin in
the different symmetry properties of the Hanle effect in its
dependence on the field orientation. Maximum Hanle depolarization is
obtained when the magnetic field is oriented along the line of
sight. The depolarization becomes complete in the strong-field limit,
regardless of the field polarity. If the field is vertical and we
assume that the illumination of the scattering atomic system is
axially symmetric around the field vector, then the Hanle effect
vanishes and the scattering polarization behaves like in the
nonmagnetic case, regardless of the magnitude of the field strength. For a quantitative
interpretation of the measured Hanle depolarization in terms of field
strengths we therefore need some information or assumption about the
field orientation. 

This need can be satisfied if we interpret the observations  not in terms
of individual magnetic elements, but in terms of {\it ensembles} of
unresolved elements within each spatial resolution element. The
assumption that such ensembles of unresolved structures really exist
on the Sun is validated by the behavior of the scattering polarization
as observed in photospheric spectral lines: there is an absence of
significant Hanle rotation effects (implying the existence of subresolution
cancelation effects), at the same time as one finds very substantial
Hanle depolarization that does not seem to fluctuate much on resolved
spatial scales. This points to the real existence of a statistically
nearly invariant ocean of unresolved structures that are much smaller
than the resolved scales, so that the ensemble averages over each
resolution element do not vary significantly. 

Let us note that this behavior only applies to photospheric but not to
chromospheric spectral lines. For the chromospheric lines the Hanle
rotation and depolarization effects are of similar magnitude, as we
expect when the fields are resolved. Furthermore, these effects are found to
exhibit large fluctuations on the resolved spatial scales. Therefore
our discussion of the turbulent Hanle effect does not apply to the
chromosphere, it is exclusively a property of the photosphere. 

Observations of the Hanle depolarization effect
reveal that regions that look entirely empty as voids in
high-resolution magnetograms cannot be voids at all but must be
seething with an ocean of 
turbulent fields that carry a highly significant amount of magnetic
flux and energy. As the smallest scales of the resolved domain are only slightly
larger than the transition scale of about 100\,km between the optically
thick and thin regimes, we can conclude that the great majority of the
flux elements that contribute to the observed Hanle depolarization are
optically thin. This is the microturbulent regime, in which we
can do direct ensemble averaging over the Hanle scattering phase matrix
before solving the radiative transfer problem \citep{stenflo-s82}. 

For the conversion of Hanle depolarization to field strength we need
to characterize our field ensemble with a single free parameter that
is uniquely determined by the single observational constraint (the
observed depolarization). The simplest model is to assume
that the ensemble is characterized by a single turbulent field
strength $B_t$, that the angular distribution of field vectors is
axially symmetric around the vertical direction, and then examine how
the assumption for the distribution of inclination angles affects the
result. 

\begin{figure*}
\centering
\includegraphics[width=0.5\textwidth,angle=90.]{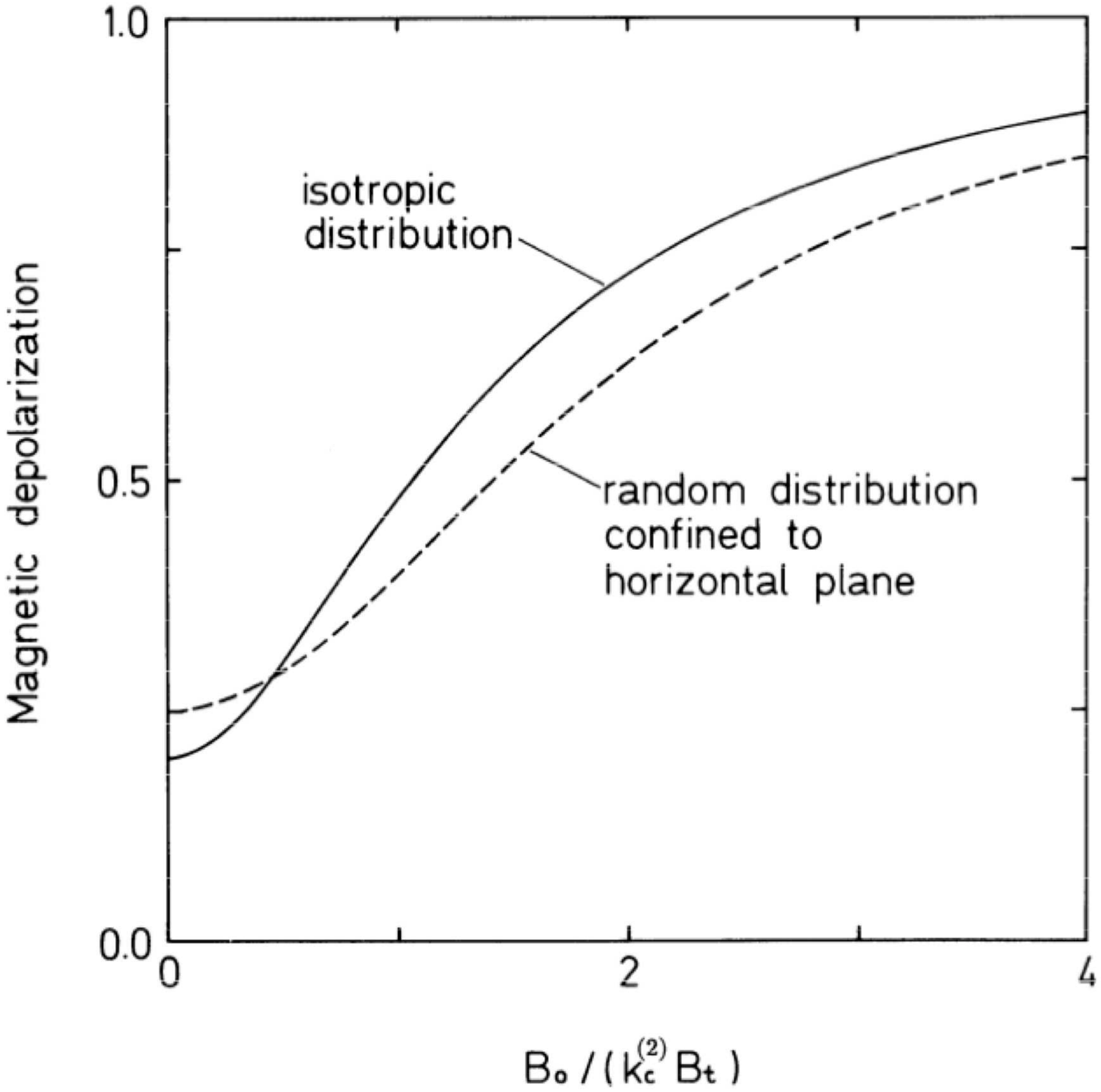}\caption{Hanle
  depolarization factor due to a turbulent magnetic field of strength
  $B_t$. The parameter $B_0$ depends on the atomic transition and
  represents the field strength for which the Larmor precession rate,
  multiplied by the Land\'e factor of the excited state, equals the
  radiative damping rate. $k_c^{(2)}$ is a collisional depolarization
  factor. The Hanle depolarization depends on the assumed angular
  distribution of the field vectors, as illustrated for two special
  cases. From
\citep{stenflo-s82}.}\label{fig:hanledepol} 
\end{figure*}

Two special types of distributions are of particular interest:
an isotropic distribution, and a flat distribution that is
confined to the horizontal plane. Figure \ref{fig:hanledepol}
illustrates how the resulting Hanle depolarization depends on field
strength $B_t$ for these two cases.  The field
strength sensitivity is greatest where the Larmor precession rate is
comparable in magnitude to the inverse life time of the excited
state.  The difference between the two curves in
Fig.~\ref{fig:hanledepol} is not particularly large, which indicates
that the interpretation is not very sensitive to the assumptions for
the angular distribution. In general there is no good physical
justification for the choice of the flat distribution for the
optically thin scales in the photosphere. As these scales are much
smaller than the atmospheric scale height, we expect the distribution
to be close to isotropic. 

Realistic ensembles of magnetic flux elements do not have a single
field strength but are characterized by a probability density function
PDF that describes the field strength distribution. Various such PDFs
have been applied to the interpretation of the observed Hanle
depolarization \citep{stenflo-trujetal04,stenflo-s12aa1}, and all
physically reasonable choices lead to average field strengths that are
substantially larger than what is obtained with the single-valued
assumption. The observational constraints may be enhanced by the
application of the {\it differential Hanle effect} 
\citep{stenflo-setal98}, using combinations of spectral lines with
different sensitivities to the Hanle effect and combinations of solar regions
where the observed depolarizations in the chosen set of lines are found to be
different.

\subsection{Forward scattering Hanle effect}\label{sec:forward}
While the scattering polarization with the Hanle effect is normally
best observed close to the solar limb due to favorable scattering
geometry (large-angle scattering that resembles the $90^\circ$
scattering case), horizontal magnetic fields can generate scattering
polarization everywhere on the solar disk, even at disk center
\citep{stenflo-truj01,stenflo-trujetal02}. This 
so-called forward-scattering Hanle effect has considerable potential
as a diagnostic tool for horizontal magnetic fields in the solar
chromosphere. 

The principle can be understood by considering for simplicity the
idealized case of extreme limb darkening, i.e., when the incident radiation
field is assumed to be almost entirely in the vertical direction. Then
the transverse plane of the incident radiation is the horizontal
plane. Unpolarized incident radiation induces dipole oscillations
in the scattering atomic system, which are axially symmetric, i.e.,
they have a circular distribution when viewed from above, as is done
when the center of the solar disk is observed. In the absence of
magnetic fields the emitted radiation
in the forward direction is therefore unpolarized. 

If we now introduce a horizontal magnetic field, the exponentially
damped dipole oscillations along the field are unaffected, but the components
perpendicular to the field are subject to Larmor precession around
the field vector. The
result will be a rosette pattern, which when viewed from above has an
elliptical shape with the long axis along the field. The 
emitted radiation will then be linearly polarized along the field
direction, and the polarization amplitude will depend on the ratio
between the Larmor precession rate and the damping rate as for the
large-angle Hanle effect. 

The large- and small-angle Hanle effect therefore act in opposite
directions. In the case of large-angle scattering the maximum
polarization amplitude corresponds to the nonmagnetic case, and the
magnetic field causes depolarization and rotation of the plane of
polarization. In the case of small-angle scattering the nonmagnetic
case is nearly unpolarized, and magnetic fields induce linear
polarization along the field direction. 

\begin{figure*}
  \vspace{-2mm}
\centering
\includegraphics[width=0.75\textwidth,angle=90.]{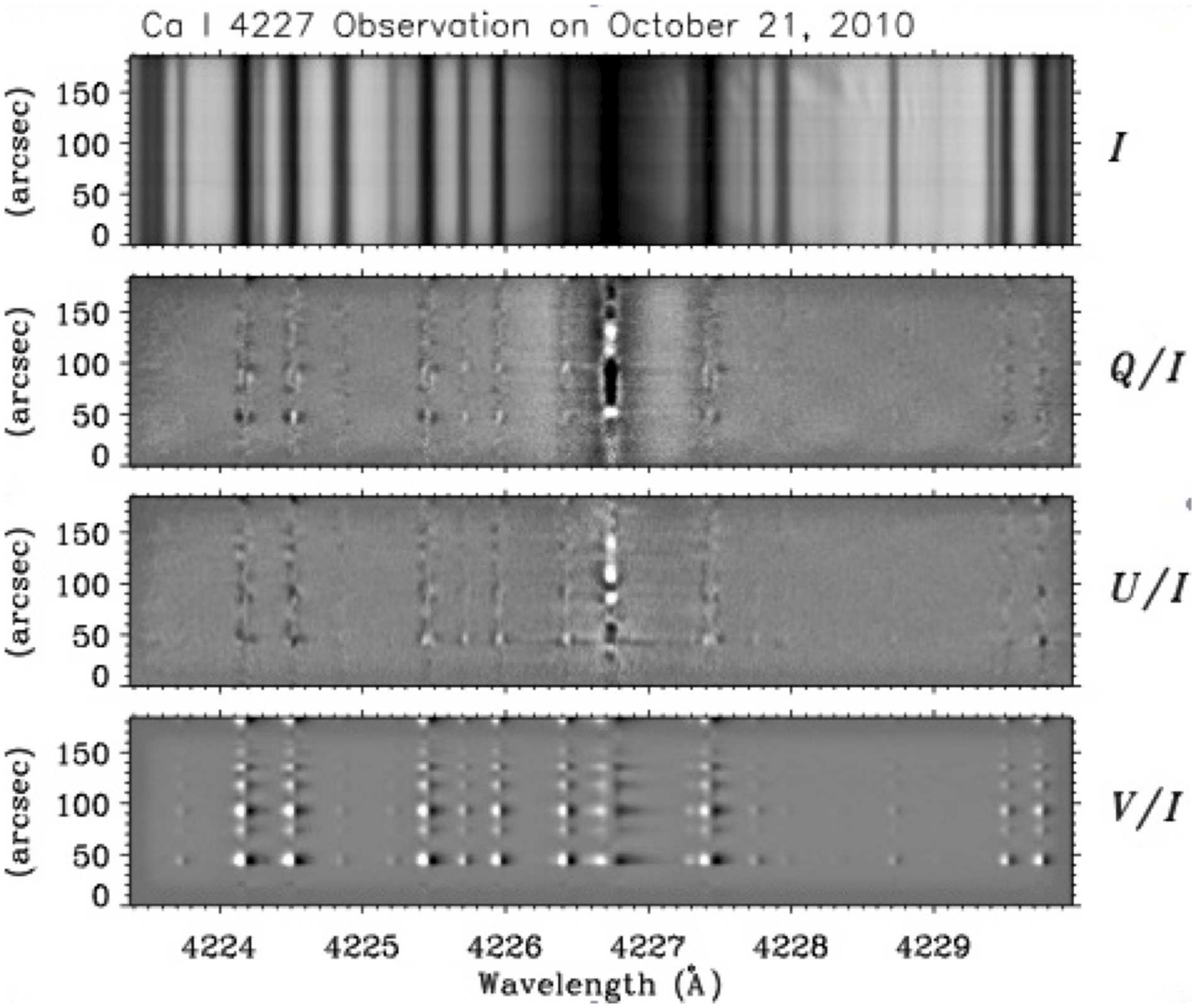}
  \vspace{-8mm}
\caption{Signatures of the forward-scattering Hanle effect in the
  chromospheric core of the Ca\,{\sc i} 4227\,\AA\ line in Stokes
  $Q/I$ and $U/I$. The
  recording was made near disk center, at $\mu\approx 0.94$, with the
  ZIMPOL-3 polarimeter at IRSOL (Istituto Ricerche Solari
  Locarno). In combination with $V/I$ these signatures can be used to
  diagnose vector magnetic fields in the
  Sun's chromosphere. From \citet{stenflo-bianda11fwd}. }\label{fig:fwdhanle} 
\end{figure*}

Figure \ref{fig:fwdhanle}, from \citet{stenflo-bianda11fwd}, illustrates an example of the
forward-scattering Hanle effect in the Ca\,{\sc i} 4227\,\AA\
line. The Hanle signatures are confined to the Doppler core of this chromospheric
line and are very structured along the slit, in
different ways for $Q/I$ and $U/I$. They reveal the presence of spatially resolved
horizontal magnetic fields in the chromosphere that vary in
both magnitude and orientation along the slit. The
signatures in the circular polarization $V/I$ are caused by the
longitudinal Zeeman effect and represent the line-of-sight component of
the field. The $Q/I$, $U/I$, and $V/I$ signals provide three
observational constraints, from which the chromospheric vector
magnetic field can be derived. The radiative-transfer theory for the
interpretation of the forward-scattering Hanle effect has been
developed by \citet{stenflo-anushaetal11} and applied to interpret the
data in Fig.~\ref{fig:fwdhanle}. 

The forward-scattering Hanle effect is of particular importance for
the diagnostics of horizontal magnetic fields in the solar
chromosphere because of the absence of good alternatives. Since the
chromospheric spectral lines are broad and the chromospheric fields
are weak, the transverse Zeeman effect is generally too small to be
used. The Hanle effect on the other hand is not diminished by the line
width, and it is sensitive to much weaker fields than the Zeeman
effect.

\section{The Sun's global dynamo}\label{sec:dynamo}
Magnetic fields govern almost all variability in the universe on
intermediate time scales. The generation of macroscopic magnetic
fields is generally understood in terms of dynamo processes
\citep{stenflo-larmor1919,stenflo-elsasser1946,stenflo-elsasser1956},
through induction effects when the highly conductive medium
dynamically interacts with a seed field. Dynamo processes are believed
to be responsible for the build-up and maintenance of planetary,
stellar, interstellar, and intergalactic magnetic fields. The Sun with
its 11-yr activity cycle represents an oscillatory dynamo with a
period of 22\,yr (the Hale cycle), since the solar magnetic field
reverses sign every 11 years. The Sun is a prototypical dynamo that
because of its proximity serves as a unique laboratory, where the
dynamo processes can be explored in detail. 

The dynamo origin of the global properties of the magnetic-field
pattern and its cyclic variation lies in the rotation of the Sun,
which through the Coriolis force statistically breaks the left-right
symmetry of the convective motions, making them cyclonic with a net
{\it helicity} \citep{stenflo-parker55,stenflo-steenkrause69}. It is
the interaction of the magnetic field with the convective or turbulent
motions in a highly conductive and rotating medium that causes the
global magnetic field of not only the Sun, stars, and planets, but
also of galaxies. 

This type of dynamo, in which rotation plays a vital, symmetry-breaking
role, is generally referred to as the {\it global dynamo}, to
distinguish it from the {\it local dynamo}, which does not need
rotation, does not contribute to any stellar activity, and is
statistically time invariant. It will be 
dealt with in the next section.  

An early phenomenological description of how the global solar dynamo operates
has been provided by \citet{stenflo-babcock61} and
\citet{stenflo-leighton69}. The frozen-in field lines of an initially
poloidal field get wound up by differential rotation. This 
leads to the generation and amplification of a toroidal
field. Buoyancy forces cause sections of subsurface 
toroidal flux ropes to be brought to the Sun's
surface. The field breaks through the surface in the form of bipolar
magnetic regions, which may contain sunspots where the flux
concentration is sufficiently large to inhibit convective energy
transport from the Sun's interior. 

When Coriolis forces act on the buoyantly rising flux ropes, they
acquire a systematic tilt with respect to the E-W direction. The tilt 
leads to an N-S bipolar moment that serves as the seed for the new
poloidal field with reversed polarity. The magnetic field from all the
emerged bipolar fields of all sizes spread by turbulent diffusion to
replace the old global poloidal field with a new one having opposite orientation. 

The dynamo of the real Sun is much more stochastic and fluctuating than the
Babcock-Leighton scenario with its coherent subsurface flux ropes
suggests. The rotation is differential both in latitude and depth,
meridional circulations play a role, and the main location of the
dynamo action is still under debate (tachocline dynamo at the bottom
of the convection zone, or distributed dynamo throughout the
convection zone). For more details, we refer to various reviews 
\citep[e.g.][]{stenflo-brandsub05,stenflo-brandenburg05,stenflo-charb10}.

\begin{figure*}
\centering
\includegraphics[width=0.75\textwidth,angle=90.]{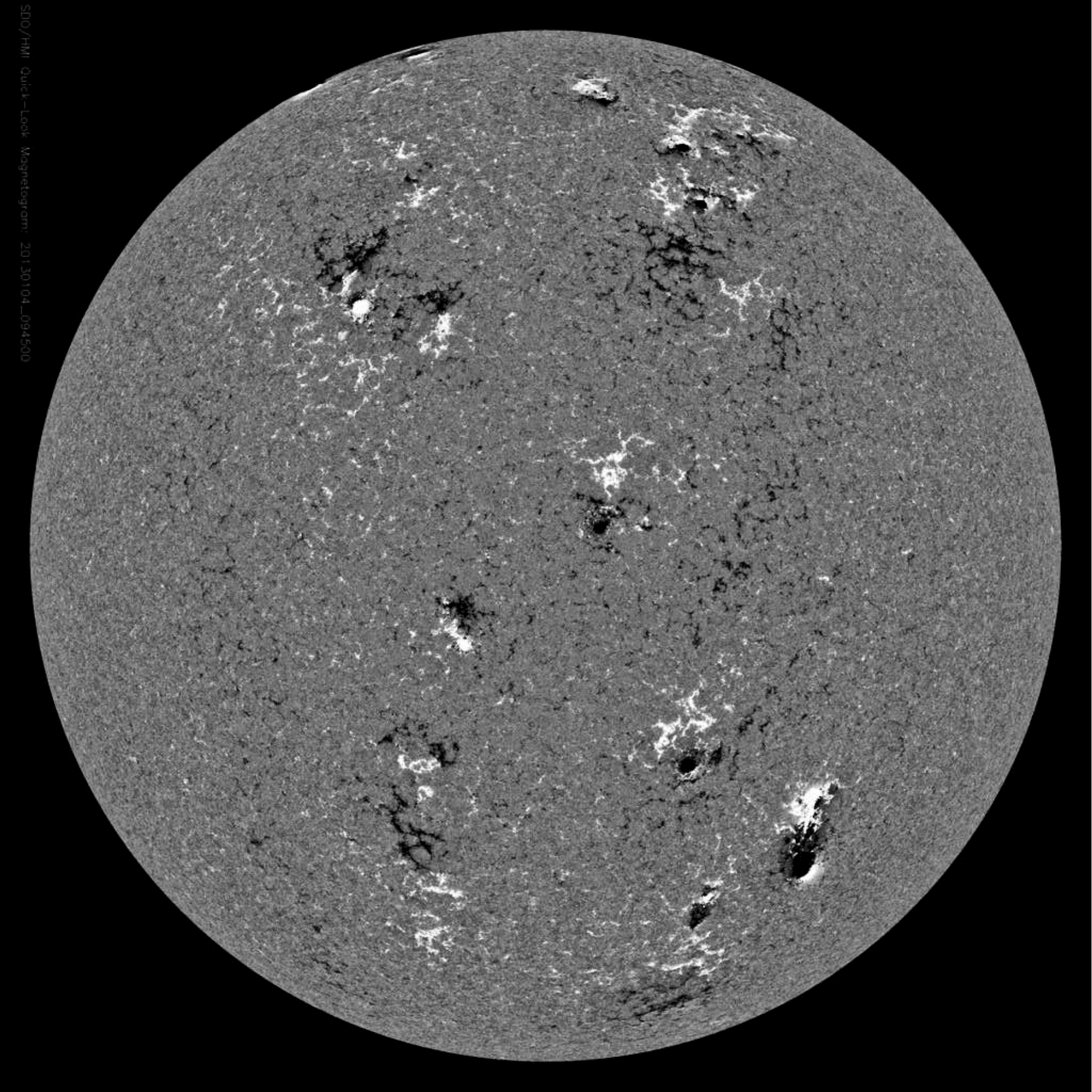}
\caption{Full-disk
magnetogram recorded by the HMI instrument on the SDO spacecraft
\citep{stenflo-scherrer12,stenflo-schouetal12} on 
January 4, 2013. The map, representing the line-of-sight component of
the magnetic flux density, is based on recordings of the circular
polarization in the Fe\,{\sc i} 6173\,\AA\ line with a spatial
resolution of 1\,arcsec. Notice that there is not only one latitude
zone of activity in each hemisphere but a coexistence of several
distinctly different latitude zones. Courtesy of NASA/SDO and the AIA, EVE, and HMI science teams.}\label{fig:fulldisk} 
\end{figure*}

\subsection{Hale's polarity law}\label{sec:halelaw}
The full-disk magnetogram in Fig.~\ref{fig:fulldisk} illustrates one
of the fundamental properties of the magnetic pattern produced by the global
solar dynamo: the anti-symmetric E-W polarity orientation of the
bipolar magnetic regions. In the figure, heliographic N is up, E to
the left. Brighter regions represent positive polarity (field directed
towards us), darker regions negative polarity. We notice that in the
northern hemisphere the E-W polarity orientation is $+-$ (bright to
the left of dark), while in the southern hemisphere it is the opposite
($-+$). 

This anti-symmetry with respect to the solar equator can be understood
in terms of the systematic winding up of the poloidal field by the
differential rotation. In the Babcock-Leighton scenario the toroidal field is
first amplified at higher latitudes, where the first sunspots of the
new cycle appear, after which the activity zone gradually migrates to
lower latitudes during the course of the cycle. This latitude
migration is generally represented as 
isocontours in latitude-time space (``butterfly diagrams''). 

According to this scenario we would at any given time expect to have
one distinct activity belt of 
bipolar magnetic regions in each hemisphere, which migrates with the
phase of the cycle.  Note however that this is not at all the case for
the pattern illustrated in Fig.~\ref{fig:fulldisk}, which reveals the
coexistence of several toroidal flux belts in each hemisphere. The
high- and low-latitude bipolar regions in the 
northern hemisphere have the same polarity orientation, implying that
they belong to the same activity cycle. 

The onset of a new 11-yr activity cycle is marked by the appearance of
high-latitude bipolar regions with the reversed polarity orientation,
followed by a new migration towards the equator. Two cycles may overlap, with
a coexistence of low- and high-latitude bipolar regions with opposite
polarity orientations. Note, however, that the pattern seen in
Fig.~\ref{fig:fulldisk} does not represent a case of overlapping
cycles. 

The property of the magnetic pattern that the E-W polarity orientation
of the bipolar magnetic regions is anti-symmetric with respect to the
equator and reverses sign with each new 11-yr cycle is called {\it
  Hale's polarity law} \citep{stenflo-haleetal19}. The full magnetic
cycle is therefore $2\times 11$\,yr, the 22-yr {\it Hale cycle}. 

\begin{figure*}
\centering
\includegraphics[width=0.85\textwidth]{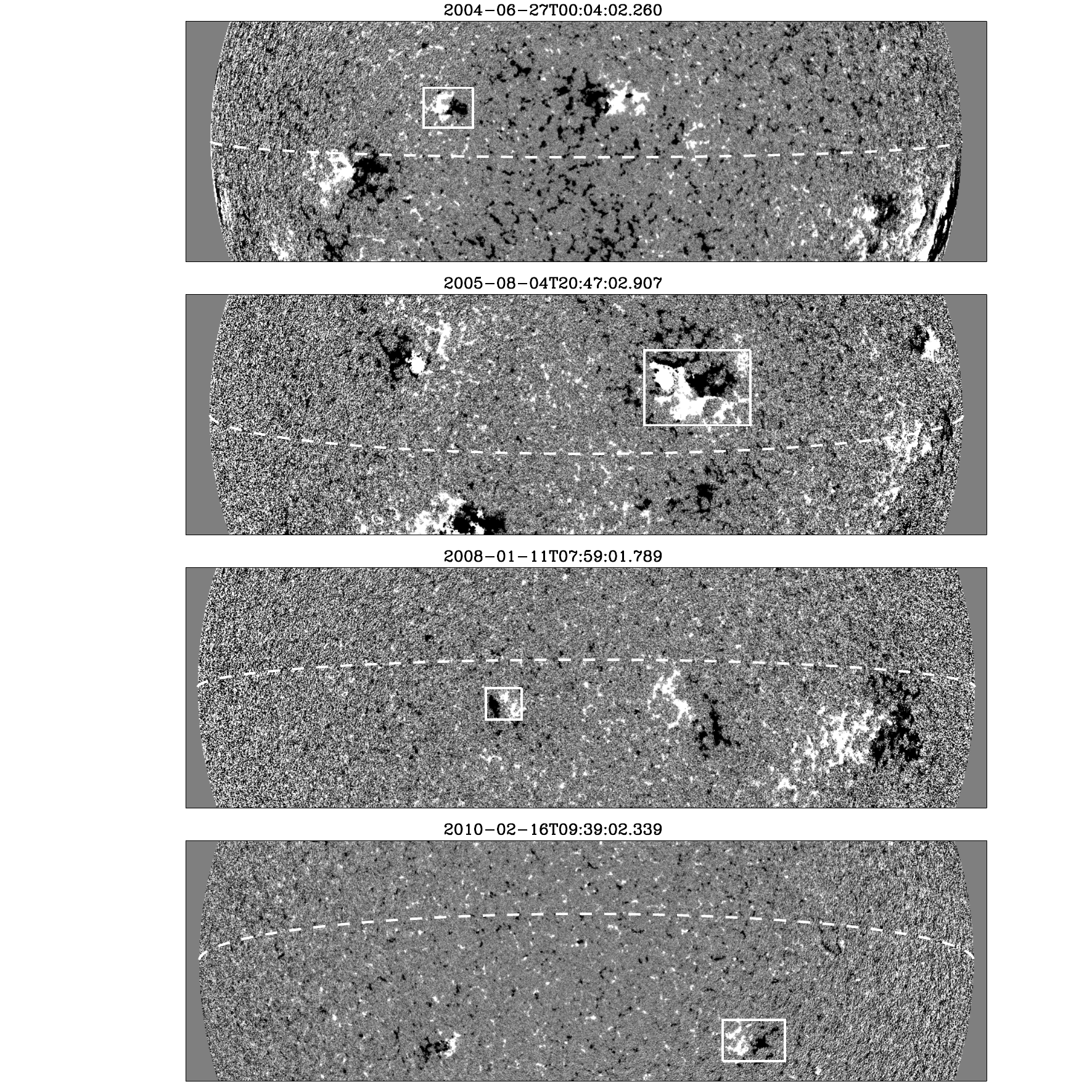}\caption{Examples
  of violations of Hale's polarity law by medium or large-size bipolar
  magnetic regions from different phases of the
  solar cycle between 2004 and 2010, from \citet{stenflo-sk12}. The
  violating regions with reversed orientations are marked by the rectangular boxes, the solar
  equator by the dashed lines. In all cases there are bipolar regions
  with the correct orientation in the same latitude zone of the same
  magnetogram, which is evidence that they cannot originate from a
  common subsurface toroidal flux system.}\label{fig:violate} 
\end{figure*}

In the Babcock-Leighton scenario with coherent subsurface flux ropes
wound up by the Sun's differential rotation, one would expect strict
obedience to Hale's polarity law. There are however numerous
violations of this law. In Fig.~\ref{fig:violate}, from
\citet{stenflo-sk12}, four such violations from different phases of
the solar cycle are illustrated. They represent cases where medium or
large-size bipolar regions with opposite E-W polarity orientations
occur side by side in the same magnetogram. They clearly cannot be
part of the same subsurface flux rope, but reveal the coexistence of
toroidal subsurface flux systems with opposite orientations at the
same latitudes, something that cannot be accounted for in the
Babcock-Leighton scenario. 

The time sequences of
magnetograms have been inspected to verify that the violating regions
do not exhibit any significant rotation but represent stable polarity
orientations. For medium to large-size bipolar regions
\citet{stenflo-sk12} find that typically
4\,\%\ (possibly less for the largest regions) violate Hale's
law. This is consistent with 
the conclusions of \citet{stenflo-richardson48}, 
\citet{stenflo-wangsheeley89}, \citet{stenflo-khlystova2009}, and
\citet{stenflo-sokoloff2010}. As we go to smaller bipolar regions,
however, the fraction of violating regions is found to increase
dramatically, which indicates that the polarity orientations get
randomized in the small-scale limit. 

The violations of Hale's law are not only in the form of reversed E-W
orientation, but there is also a similar frequency of regions with N-S
polarity orientation \citep{stenflo-sk12}, which should not occur for
an origin in terms of coherent toroidal flux ropes. This shows that the simplistic
phenomenological model of
coherent flux ropes needs to be replaced by a dynamo scenario in which
fluctuations on all scales play a major role.

\subsection{Joy's law}\label{sec:joy}
Another fundamental law discovered through the pioneering Zeeman-effect work of
Hale and collaborators \citep{stenflo-haleetal19} is {\it Joy's law}
that governs the tilt angles 
of bipolar magnetic regions. Although the main orientation is in the
E-W direction, there is a systematic tilt with respect to this
direction, such that the preceding (westward) part of the region is
systematically closer to the equator than the following part. It is
the average tilt angle and its latitude variation that is governed by
Joy's law. 

While Hale's polarity law expresses how a toroidal magnetic field is
generated from a poloidal field, Joy's law embodies the generation of
a poloidal magnetic field from a toroidal one, because the tilt implies
an emerged N-S bipolar moment that 
serves as a seed for the new large-scale poloidal field. Since the
following polarity  of the bipolar regions is opposite to the polarity
of the polar field and spreads by turbulent
diffusion more towards the pole than the preceding polarity, it will
eventually cause polar-field reversal and the creation of a
global poloidal field with reversed orientation every 11 years. 

\begin{figure*}
\centering
\includegraphics[width=0.7\textwidth]{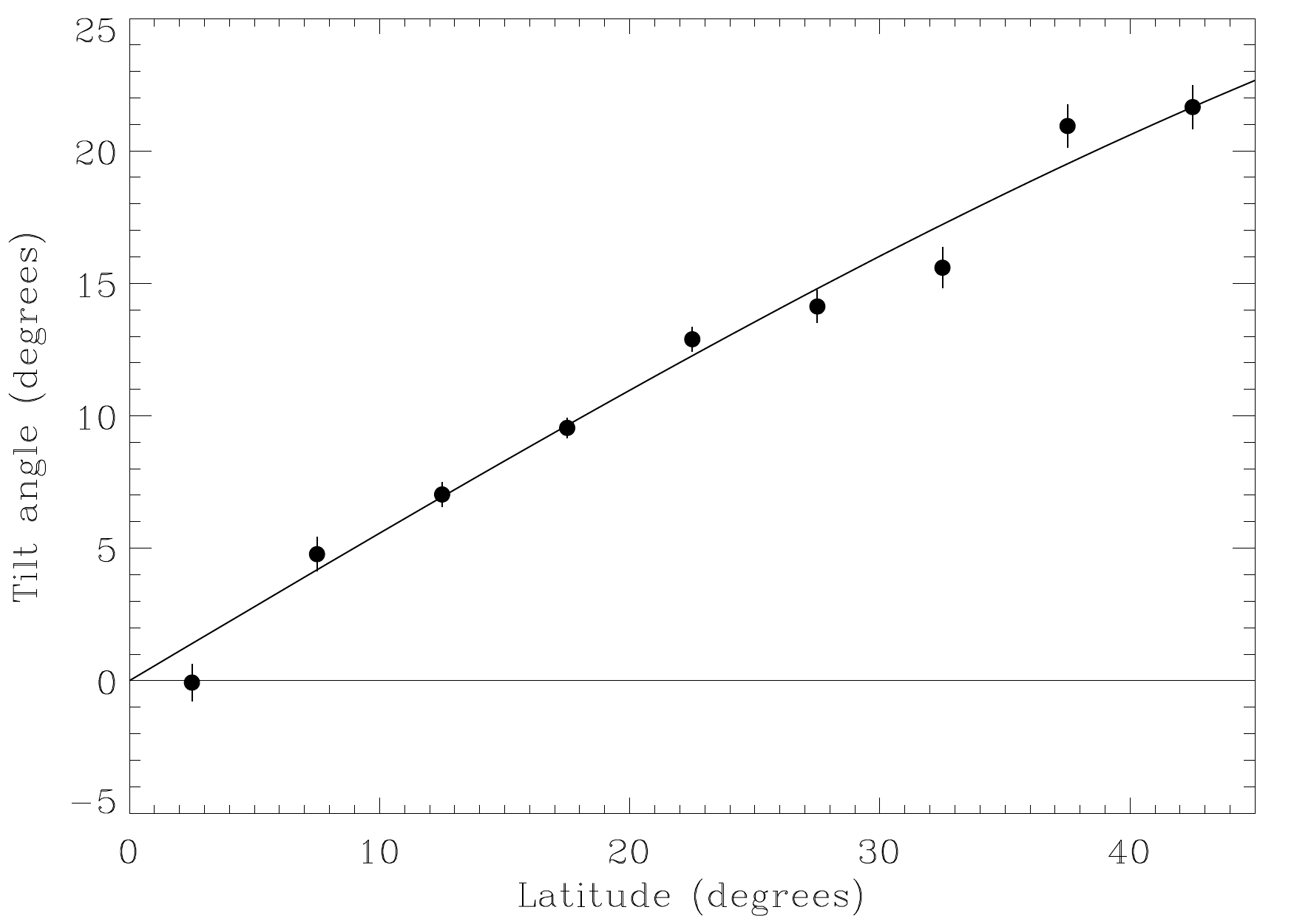}\caption{Latitude
variation of the average tilt angles of bipolar magnetic regions, expressing
Joy's law. Positive tilt means that the preceding polarity is
equatorwards of the following polarity. The solid line is the
analytical fit function $32.1^\circ $ sin(latitude). From \citet{stenflo-sk12}.}\label{fig:joy} 
\end{figure*}

The tilt angles of the bipolar regions have a large intrinsic
scatter due to the fluctuating nature of the solar dynamo, so an
accurate determination of Joy's law has only recently become possible 
with massive statistics. Through analysis of the 73,838
magnetograms of the entire SOHO/MDI data set with a computer algorithm that
automatically identifies bipolar magnetic regions of various sizes,
the precise latitude variation of the average tilt angle could be
determined. The result is illustrated in Fig.~\ref{fig:joy},  
from \citet{stenflo-sk12}. The empirically determined tilt angles can
be closely represented by the simple analytical function $32.1^\circ $
sin(latitude). The main source of the systematic tilt angles is
believed to be the Coriolis force, which also varies with the sine of the
latitude. 

The rather large tilt angles implied by
Fig.~\ref{fig:joy} are in good general agreement with the results of
a number of other studies 
\citep{stenflo-wangsheeley89,stenflo-liulrich12,stenflo-tlatov13}, but disagree 
with the much flatter latitude dependence that was found from Mount Wilson and
Kodaikanal data by \citet{stenflo-espuig11} and derived from numerical
simulations by \citet{stenflo-schbau06}. \citet{stenflo-liulrich12}
  searched for temporal variations over the period 1974-2012, which
  spans nearly four solar cycles, and found the tilt angle for each
latitude zone to be time invariant. 

There is no indication that Joy's law as expressed by
Fig.~\ref{fig:joy} has any significant dependence on region size. 
  According to the study of \citet{stenflo-sk12}, bipolar regions
differing as much as four orders of magnitude in their flux content
seem to follow the same law. Although the {\it average} tilt (for a
given latitude zone) seems to be the same, the {\it scatter} of the
tilt values increases dramatically as we go to smaller regions. 
  It should however be mentioned that \citet{stenflo-tlatov13}, while
  agreeing with the results of \citet{stenflo-sk12} for the larger
  bipolar regions, claim that small, ephemeral-type bipolar regions
  have similar magnitudes for the tilt angles but reversed
  orientations with respect to the large regions. This intriguing
  claim is difficult to understand and needs to be verified.

\subsection{Scales that feed the oscillatory
  dynamo}\label{sec:dynamoscales}
Bipolar magnetic regions occur on all scales and have a size spectrum
that follows a power law
\citep{stenflo-harveyphd93,stenflo-harveyzwaan93,stenflo-schrharv94,stenflo-parnelletal09},
from the largest regions that harbor major sunspots to 
ephemeral active regions \citep{stenflo-harveymartin73,stenflo-harveyetal75,stenflo-martinharvey79} without sunspots, and
the still smaller internetwork fields. The global contribution to the
overall flux emergence rate dramatically increases as we go down in
scale size \citep{stenflo-zirin87}, suggesting that the global flux
balance could be 
dominated by the smallest scales. Let us therefore next try
to determine which scales contribute the most to the
regeneration of the poloidal field from the toroidal one. 

\citet{stenflo-harveyphd93} found from analysis of Kitt Peak full-disk
magnetograms that the emergence rate $R$ of bipolar
magnetic regions has a
smooth and continuous scale dependence. Within the
error bars the histogram 
distribution ${\rm d}R/{\rm d}A$ as a function of 
region area $A$ may be described in terms of the power law
\begin{equation}\label{eq:drda}
{\rm d}R/{\rm d}A \sim A^{-p}
\end{equation}
with a power law index of $p\approx 3$. 

Using a feature recognition algorithm \citet{stenflo-parnelletal09}
analyzed magnetograms recorded by SOHO/MDI and by 
Hinode/SOT to determine the size distribution of solar magnetic
structures. They found that the distribution follows a power law over
5 orders of magnitude in flux $\Phi$. If 
$N$ is the number of magnetic features at a given time, 
\begin{equation}\label{eq:dndphi}
{\rm d}N/{\rm d}\Phi \sim \Phi^{-\alpha}
\end{equation}
with a power law index of $\alpha =1.85\pm 0.14$. 

Let us now examine how these two relations are related to each other. Each structure has
an average life time $\tau$ before it dissolves, most likely through
fragmentation, cascading to higher wave numbers in the magnetic energy
spectrum. Various studies of the life time $\tau$ of solar magnetic
structures indicate that $\tau\sim A\sim d^{\,2}$, where $d$ is the
size of the structure
\citep[cf.][]{stenflo-s76iau,stenflo-harveyphd93}. Such a size
dependence of the life time is expected if the evolution of the
structure obeys a diffusion equation (fragmentation through the
interchange instability followed by turbulent
diffusion). Since the emergence rate $R$ is related to the number of
features through $N=R\,\tau$, it follows that 
\begin{equation}\label{eq:dnda}
{\rm d}N/{\rm d}A \sim A^{-p+1}\,.
\end{equation}
The magnetic flux of a magnetic element can be written as $\Phi
=BA=B\,d^{\,2}$, where $B$ is the field strength. Most of the flux that is
visible in magnetograms is in collapsed kG form and the dependence of
$B$ on size $d$ is generally weak. If we disregard this dependence,
then $\Phi \sim A$, and we retrieve Eq.~(\ref{eq:dndphi}) from
Eq.~(\ref{eq:dnda}) if $p=\alpha +1$. With $\alpha =1.85$ we then get
$p=2.85$, which agrees within the uncertainties with the value of 3
estimated from the data of \citet{stenflo-harveyphd93}. 

Of importance for the regeneration of the poloidal field is
not only the amount of emerged flux $F$, but also the separation $S$ between
the opposite polarities. The key parameter is the bipolar moment,
defined in \citet{stenflo-sk12} as $M=\textstyle{1\over 2}F_{\rm tot}
S$, where $F_{\rm tot}$ is the sum of the unsigned positive and
negative magnetic flux of the bipolar region. The N-S component of the
bipolar moment is then obtained through multiplication of $M$ with
the sine of the tilt angle (which appears to be independent of region
size). 

Since $S$ scales with the size $d$ of the structures, $M\sim
B\,d^{\,3}$. Let us by $R_{\rm NS-moment}=M\,N/\tau$ denote the
emergence rate of the N-S bipolar moment. If we
disregard the weak $d$-dependence of $B$ and the average tilt angle, and make use
of Eq.~(\ref{eq:dnda}) and $\tau\sim d^{\,2}$, we get  
\begin{equation}\label{eq:nsmom}
R_{\rm NS-moment}\sim d^{-2p+5}\,.
\end{equation}
With our previous value $p=3$ from \citet{stenflo-harveyphd93} we findß
$R_{\rm NS-moment}\sim 1/d$, which implies that it is the smallest
scales that contribute the most to the regeneration of the poloidal
field. The rate diverges in the small-scale limit unless there is a
cut-off mechanism. Such a mechanism is provided by the randomization
of the tilt angles. 

In the analysis of \citet{stenflo-sk12} the histograms of tilt angles
for each latitude zone were fit with an analytical model that
represented a decomposition with two components, one oriented component
with a Gaussian tilt angle distribution, and one flat distribution
representing random tilt angles. The average tilt angle for the
Gaussian component is what has been plotted in Fig.~\ref{fig:joy} to
represent Joy's law. The random component does not contribute to or
affect this average, but it determines the fraction of the emerged flux
that contributes to the regeneration of the poloidal field. 

\begin{figure*}
\centering
\includegraphics[width=0.7\textwidth]{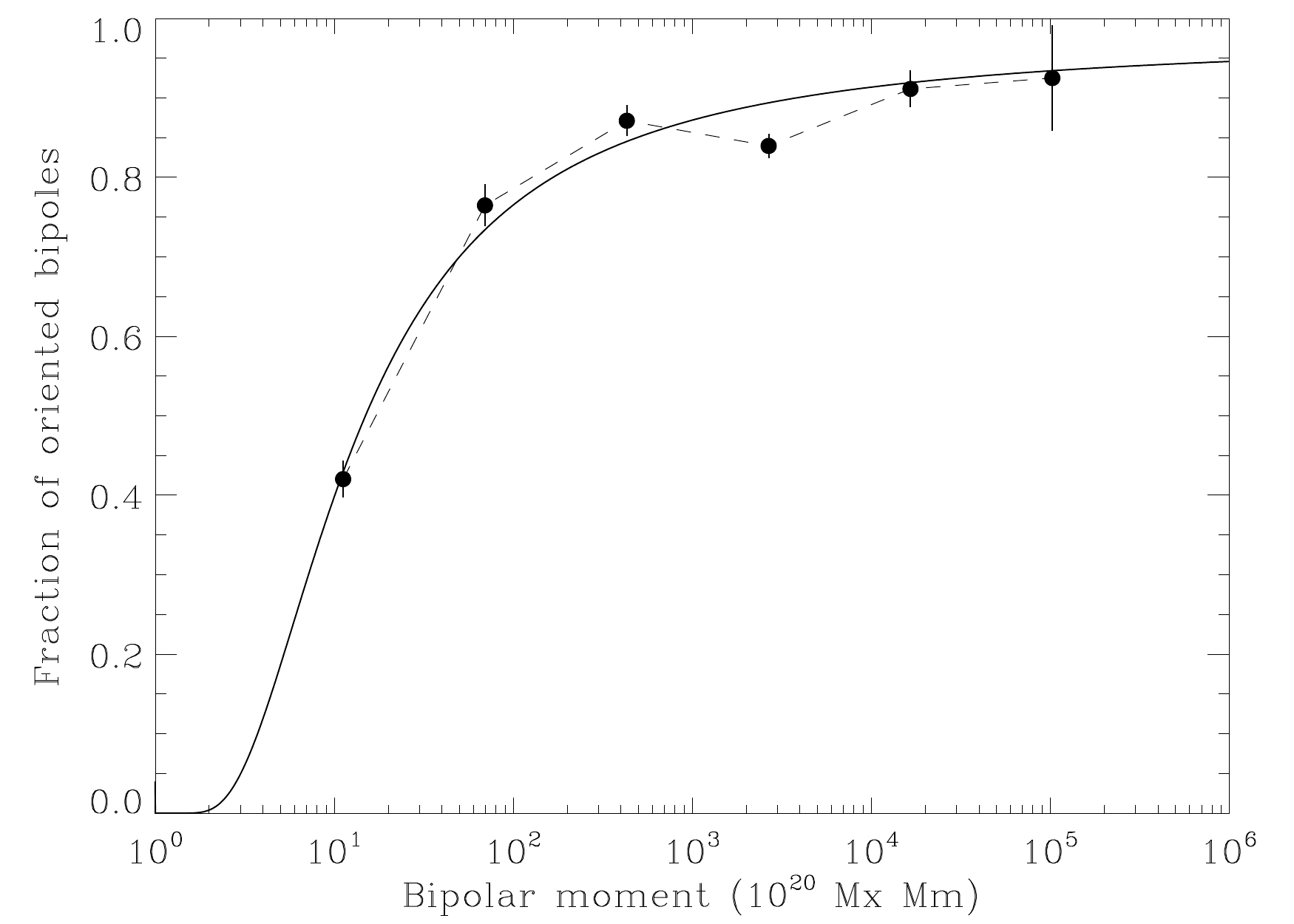}
\caption{Relative
  fraction of bipolar magnetic regions of various sizes that may
  contribute to the
  global N-S bipolar moment, which regenerates the
  Sun's poloidal magnetic field. The cut-off range of 10-100$\times
  10^{20}$\,Mx Mm corresponds to an approximate size range of 2-8\,Mm. The
  data points (filled circles and dashed line) have been extracted
  from the analysis material of \citet{stenflo-sk12}, while the solid
  line is an analytical fit curve.}\label{fig:nsmoment} 
\end{figure*}

If we denote the relative occurrence of the oriented and random
components (integrated over all tilt angles) by $f_{\rm oriented}$ and
$f_{\rm random}$, then it is the fraction $r=f_{\rm
  oriented}/(f_{\rm oriented}+f_{\rm random})$ that contributes to the
poloidal field regeneration. Based on the data from the analysis of
\citet{stenflo-sk12} we have in Fig.~\ref{fig:nsmoment} plotted this
fraction as a function of bipolar moment. We find that it
has a rather sharp cut-off near a bipolar moment of $M\approx$
10-100$\times 10^{20}$\,Mx Mm, below which there is no significant
contribution to the N-S bipolar moment. 

Since we may write $M=mB\,d^{\,3}$, where $m$ is a dimensionless
number of order 2-10, depending on how the polarity separation scales
with size $d$ of each of the  two main polarity structures, we can
estimate the scale size $d$ to which a given $M$ corresponds. With our
generous allowed range for parameter $m$ and assuming
$B=1$\,kG, we find that the cut-off of Fig.~\ref{fig:nsmoment}
corresponds to scales of order $d=2$-8\,Mm. 

According to Eq.~(\ref{eq:nsmom}) it is the smallest scales that
contribute the most to the regeneration of the poloidal field through the
emergence of bipolar regions, as long as we are above the cut-off. From our
previous estimates it then follows that bipolar magnetic
regions with sizes of order 10\,Mm appear to be the main
contributors. 

This result might seem to contradict claims that the
  largest bipolar regions provide sufficient flux to explain
  the surface field evolution in
  terms of the Babcock-Leighton flux transport model of the solar
  cycle \citep[e.g.][]{stenflo-sheeley85,stenflo-jiangetal13}. However, since
  these models are idealized and contain adjustable parameters, and
  since the large bipolar 
  regions and the accumulated
  contributions from the small bipoles affect
  the surface pattern in qualitatively similar ways, the flux
  transport models cannot be used as 
  evidence against a dominant role of the smaller bipoles. By
  adjusting the models to compensate for the neglected small-scale
  contributions, the impression may arise that these contributions do
  not play a significant role.

\section{Local dynamo}\label{sec:local}
The global dynamo is responsible for all of solar activity, with
sunspots, flares, CMEs, etc. The cyclic variation has its origin in the left-right
symmetry breaking of the turbulent motions induced by the Sun's
rotation via the Coriolis force. The word ``global'' does not imply
``large-scale''. The dynamo action takes place on all
scales. The generated magnetic fields at the larger scales cascade down the
magnetic energy spectrum until they reach the magnetic diffusion limit
around a scale of order 25\,m, where the magnetic Reynolds number becomes
unity and the field lines decouple from the plasma, because the
frozen-in condition is not valid for smaller scales
\citep[cf.][]{stenflo-s12aa1}. The global dynamo is therefore a source of 
structuring throughout the entire scale spectrum.  It is not
meaningful to separate the global dynamo into a small-scale and a
large-scale part, all scales are connected. 

Based on theoretical considerations and numerical simulations it has
however been suggested that there exists a second, qualitatively
different dynamo that also
operates on the Sun and is responsible for the statistically
time-invariant small scale structuring of the magnetic field
\citep{stenflo-petrovay93,stenflo-cattaneo99,stenflo-vs07}.  

The circumstance that the magnetic field is highly structured on small
scales, far below the resolution limit of current telescopes, is
sometimes incorrectly taken as evidence for a local dynamo. 
Small-scale structuring is also produced by the global dynamo via the
turbulent cascade. What distinguishes the two dynamos from each other
is that the local dynamo is decoupled from solar activity and
contributes to a background quiet-sun magnetic pattern that is
statistically constant over the Sun and does not vary with the
cycle. We call such magnetic flux {\it basal} flux. 

\begin{figure*}
\centering
\includegraphics[width=0.75\textwidth]{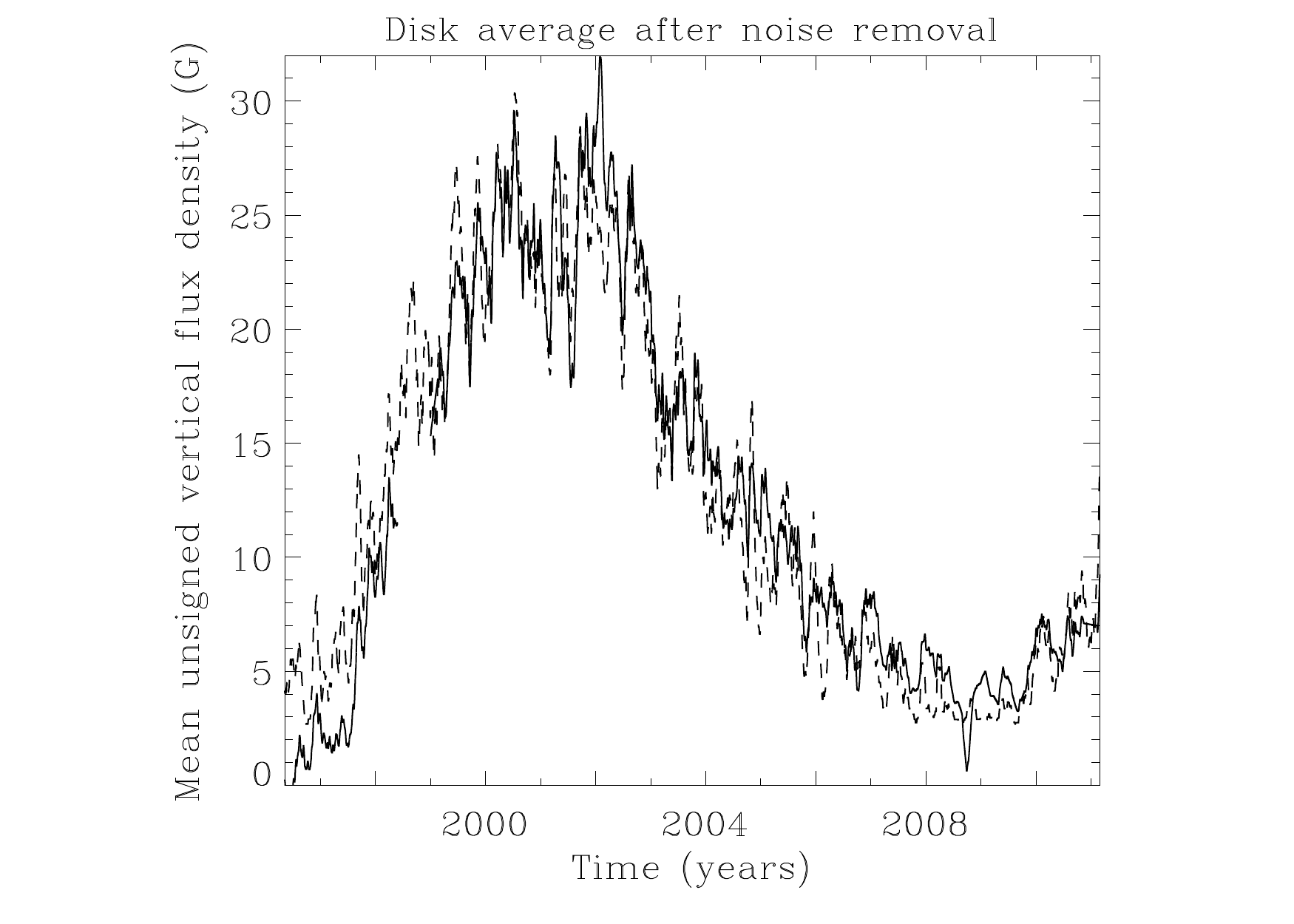}\caption{Illustration
  of the nearly perfect correlation between the disk average of the
  unsigned vertical magnetic flux density derived from SOHO/MDI
  full-disk magnetograms (solid line) and a second-order
  polynomial fit of the relative sunspot number $R_z$ (dashed line). It shows that the
  average basal flux density of the Sun, in
  the absence of sunspots, is 2.7\,G for the 4\,arcsec scale of the MDI
  recordings. From \citet{stenflo-s12aa2}.}\label{fig:rzcorr} 
\end{figure*}

Analysis of the full SOHO/MDI data set of full-disk magnetograms shows
that there is an almost perfect correlation between the disk average
of the unsigned vertical magnetic flux density and the sunspot number
$R_z$ \citep{stenflo-s12aa2}, as illustrated in
Fig.~\ref{fig:rzcorr}. In the absence of sunspots the MDI basal flux
density is found to be 2.7\,G, which represents an {\it upper limit}
to the possible contributions from a local dynamo at this spatial
scale (4\,arcsec). Much of this apparently basal flux may well
originate from the
global dynamo, since the background flux pattern that has been
generated by the breakup and turbulent diffusion of bipolar magnetic
regions will not suddenly disappear when solar activity is switched
off. Much of the background pattern may be expected to survive
throughout the cycle minimum until the new cycle begins to supply new
flux from the Sun's interior. 

The average unsigned vertical flux density $B_{\rm ave}$ depends on scale size $d$
according to the {\it cancelation function} 
\begin{equation}\label{eq:canc}
B_{\rm ave} \,\sim\, d^{-\kappa}
\end{equation}
as determined from Hinode SOT/SP data for a scale range from the
Hinode resolution scale (230\,km) up to and beyond the MDI scale (4\,arcsec)
\citep{stenflo-pietarila09,stenflo-s11aa}. Consistent
results are obtained with a {\it cancelation exponent} $\kappa
=0.13$. Analysis of
magnetograms obtained with SDO/HMI during the deep cycle
minimum 2010-2011 gives a basal vertical flux density at disk center
of 3.0\,G for the HMI spatial scale (1\,arcsec), which is consistent
with the MDI value of 2.7\,G and the scaling law of
Eq.~(\ref{eq:canc}) with $\kappa =0.13$. Extrapolated with this law to
the Hinode spatial scale, the basal flux density is 3.5\,G \citep{stenflo-s12aa2}. 

\begin{figure*}
\vspace{-1mm}
\centering
\includegraphics[width=0.8\textwidth]{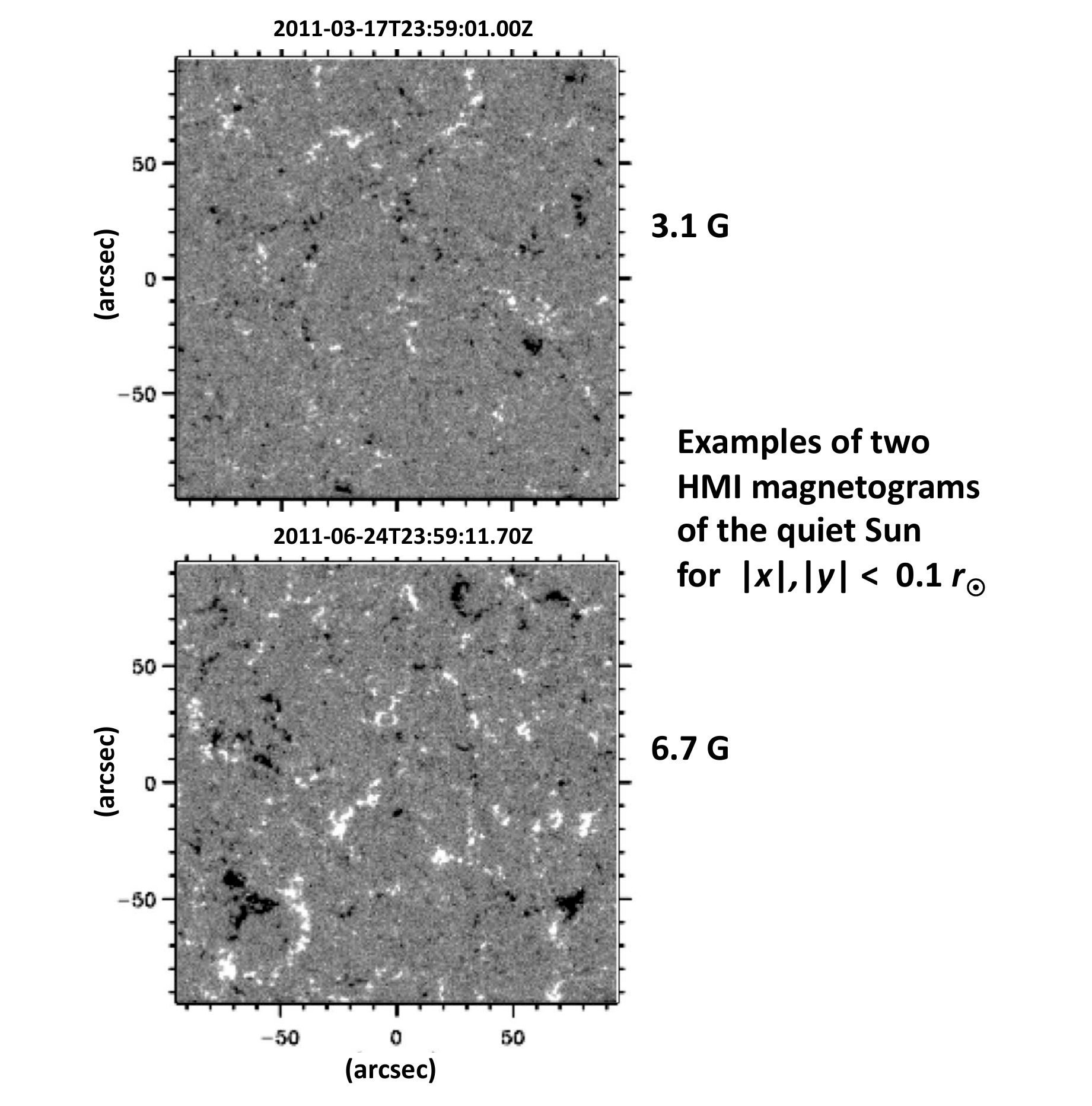}
\vspace{-1mm}
\caption{Comparison
of two SDO/HMI magnetograms of the quiet Sun at disk center ($x$ and
$y$ are the coordinates relative to disk center), recorded
at times 2011.2086 (upper panel) and 2011.4796 (lower panel). The
grey-scale cuts are the same for both panels ($\pm 50$\,G). The top
panel represents a region with the basal flux density level (3\,G) for the 1\,arcsec
scale of HMI, while the bottom panel shows a region with 2.2 times
more flux. From \citet{stenflo-s12aa2}.}\label{fig:hmiquiet} 
\end{figure*}

The average flux density at the quiet-sun disk center varies greatly
with time, as illustrated in Fig.~\ref{fig:hmiquiet}. Such variations
have nothing to do with a local dynamo, which must be
statistically constant. Instead they indicate that much, possibly
all, of the flux that we see on the quiet Sun is supplied by the global
dynamo. The magnetic pattern in the upper panel of
Fig.~\ref{fig:hmiquiet}, which is representative of the basal flux
density, is qualitatively similar to the pattern in the bottom panel,
which represents 2.2 times more flux. Both patterns exhibit network-like
structuring on the supergranular scale, the difference lies only in the
relative amounts of available flux. 

Since the upper limit to the possible contribution from a local
dynamo to magnetic structuring at the Hinode scale is as low as 3.5\,G,
while the average unsigned flux densities that have been reported from Hinode
SOT/SP observations of the quiet-sun disk center are typically three
times larger \citep{stenflo-litesetal08,stenflo-s10aa}, it is clear
that most, if not all, of the magnetic structuring revealed by Hinode on the quiet
Sun has its origin in the global dynamo, not in a local dynamo. 

It should be mentioned that this conclusion is still
  controversial, in particular because of claims that the ubiquitous and
  statistically time invariant horizontal
  magnetic fields inferred from linear-polarization measurements with
  Hinode give evidence for a local dynamo at the Hinode spatial scale
  \citep[e.g.][]{stenflo-buehler13}. However, as indicated by
  Fig.~\ref{fig:pdfnoise}, the Hinode linear polarization recorded on
  the quiet Sun is dominated by noise, and the instrumental noise is time
invariant. As shown in \citet{stenflo-s13aa1} and described in
Sect.~\ref{sec:angular} below, the photospheric magnetic fields at the
center of the quiet solar disk have a preferentially vertical, not
horizontal, angular distribution,
contrary to repeated claims from analysis of Hinode data. This
conclusion is both resolution and model independent, as will be
explained in Sect.~\ref{sec:angular}. 

Although there may not be a significant role of 
a local dynamo at any of the resolved spatial scales, it is
likely that the local dynamo plays a dominating role at scales
below about 10\,km. The reason is that the vast amounts
of ``hidden'' magnetic flux that have been revealed by the observed
Hanle effect depolarization (cf. Sects.~\ref{sec:hanleturb} and
\ref{sec:hanresults}) cannot be explained only in terms of the 
turbulent cascade of flux generated by the global dynamo, an
additional source of flux is needed. 

\begin{figure*}
\vspace{2mm}
\centering
\includegraphics[width=0.7\textwidth]{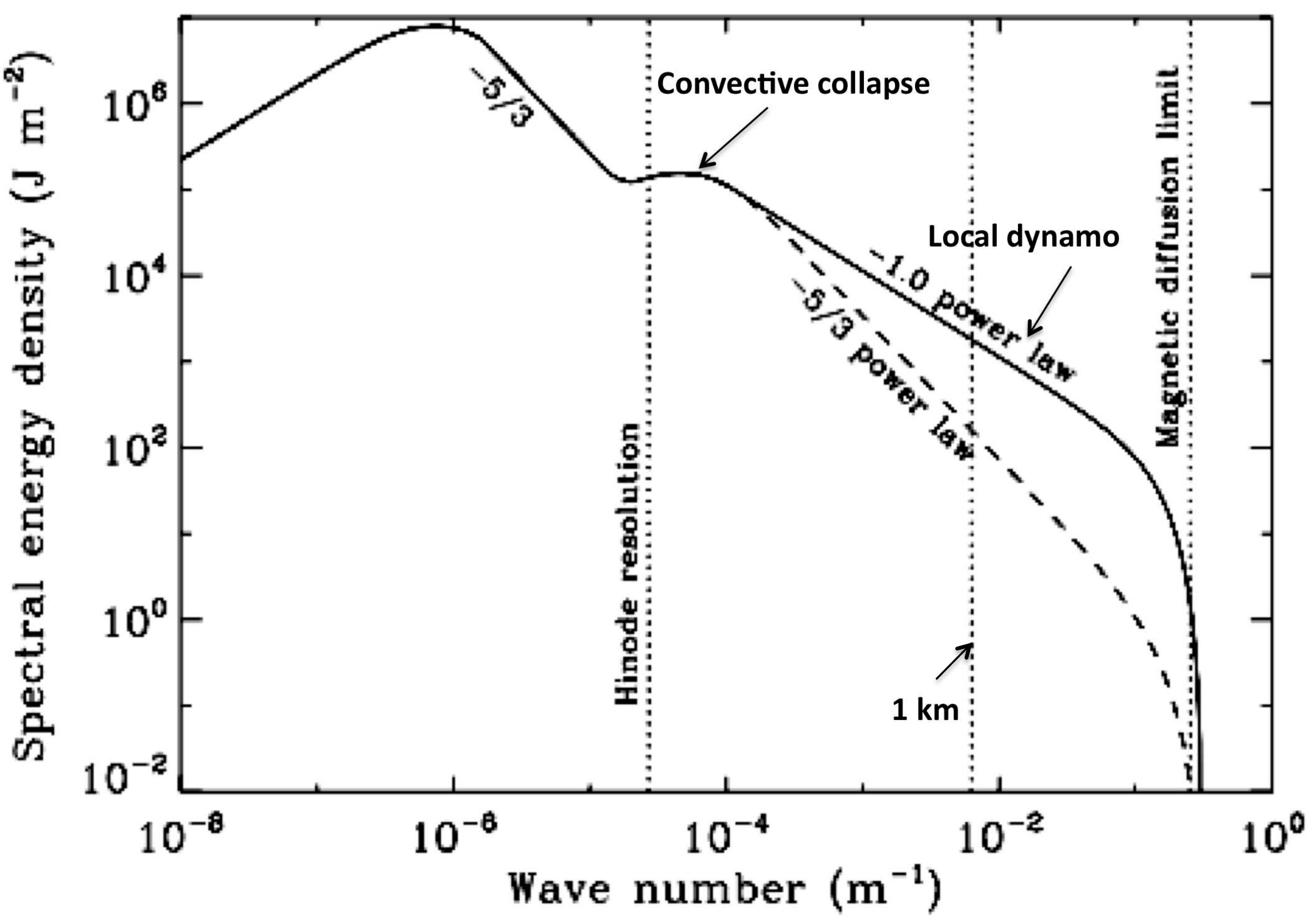}
\caption{Magnetic
  energy spectrum of the solar photosphere, spanning seven orders of
  magnitude in wave number, adapted from \citet{stenflo-s12aa1}. A
$-5/3$ Kolmogorov power law in the spatially unresolved domain would
not give rise to sufficient magnetic 
structuring at small scales to account for the observed Hanle
depolarization. If the spectrum is raised to follow a $-1.0$ power law
instead, it becomes consistent with the Hanle constraint. As
so much power at small scales is inconsistent with the scaling
behavior at the resolved scales, which are governed by the global dynamo,
it is conjectured that a local dynamo is operating as an additional
source of magnetic flux at scales below
about 10\,km. }\label{fig:enmaglocdyn} 
\end{figure*}

Figure \ref{fig:enmaglocdyn} gives an overview of the magnetic
energy spectrum of the quiet Sun. The part to the left of the
vertical dotted line (the Hinode resolution limit at 230\,km) has been
directly determined from the observations, since it represents the
spatially resolved domain. The bump in the scale range of about
10-100\,km has been inferred from Hinode Stokes $V$ line ratio data
and represents the population of kG type fields that have been
spontaneously generated by the convective collapse instability. As we
go to higher wave numbers the structures get much smaller than the 
atmospheric scale height and therefore experience an increasingly isotropic
environment. According to \citet{stenflo-kolmogorov41} theory one
would then expect a $-5/3$ power law for the turbulent cascade in this
inertial range, which ends at the magnetic diffusion limit (here
estimated to be at the 25\,m scale), where the magnetic Reynolds
number becomes unity and the magnetic field lines cease to be
frozen-in. 

The observed Hanle depolarization (cf. Sects.~\ref{sec:hanleturb} and
\ref{sec:hanresults})
implies the existence of a vast hidden ocean of
tangled or turbulent fields \citep{stenflo-s82} with an average unsigned flux density of at
least 60\,G \citep{stenflo-trujetal04}. It is largely invisible in
magnetograms due to cancelation of the contributions from
the opposite polarities that are mixed on subresolution scales. The
scaling law represented by the cancelation function of
Eq.~(\ref{eq:canc}), which was empirically determined in the resolved regime
that is dominated by the global dynamo, is much too shallow to provide
anything close to the flux density that is needed when one
extrapolates it down to
the magnetic diffusion limit. Similarly the $-5/3$ power law of the
magnetic energy spectrum is much too steep to provide sufficient
magnetic structuring at small scales needed to satisfy the Hanle
constraint. To restore consistency with the Hanle observations we find
it necessary to raise the energy spectrum to the level of a $-1.0$
power law in the small scale inertial range. The cancelation function
that is valid in the resolved domain then cannot be expected to retain 
its validity at the smallest scales, since the flux there is
supplied by a different mechanism, which is insignificant at resolved scales.

\section{Global evolution}\label{sec:global}
The appearance of the corona during a total solar eclipse
indicates that the Sun is a magnetized sphere that possesses a global,
dipole-like magnetic field. Soon after his discovery of magnetic
fields in sunspots \citep{stenflo-hale08}, George Ellery Hale and his
team began to apply the Zeeman effect to determine the 
dipole-like structure of the general magnetic field. In a series of
papers \citep[e.g.][]{stenflo-haleetal18} they reported successful
results. Many of Hale's assistants did independent visual analysis of
the many photographic plates of the Zeeman effect. About half of them
obtained null results, but Hale's conclusions were largely based on one
of his assistants, Van Maanen, who consistently got well-defined
positive results with small error bars. 

These results later turned out to be entirely inconsistent with modern
photoelectric measurements (made possible with the
\citet{stenflo-babcock53} magnetograph), which revealed a general
magnetic field weaker by an order of magnitude and with a polarity
that according to the Hale cycle should
have been opposite to the one found by Hale and Van Maanen at the time of
their measurements. For this reason a batch of about 400 photographic
plates analysed by Van Maanen in 1914 were located and
remeasured with a modern digitized microphotometer
\citep{stenflo-s70hale}. While Van Maanen, using the identical plate
material, found a strong dipole-like field with a well-defined latitude
variation that was anti-symmetric around the equator, the modern reanalysis
gave null results with error bars that ruled out his
results, which can only be understood as the result of severe subjective bias. 

Van Maanen also measured the proper motions of the stars on
photographic plates of the spiral nebula M33. The large proper motions
that he found could only be understood if M33 were a nearby
object. They were therefore seen as strong evidence against the extragalactic
nature of spiral nebulae. We now know
this result to be wrong, another example of subjective bias
\citep[cf.][]{stenflo-s70hale}.  

Full-disk solar magnetograms have been recorded on a daily basis,
starting at the Mt Wilson Observatory in August 1959, with higher
spatial resolution at the National Solar Observatory/Kitt Peak since
December 1976, with the SOHO/MDI magnetograph in space May 1996 -
April 2011, and presently continued by SDO/HMI with 1\,arcsec
resolution and 45\,s cadence. A convenient overview of the
evolutionary information contained in the sequence of full-disk
magnetograms is provided by synoptic maps, showing the distribution of
the line-of-sight component of the magnetic field as a function of
latitude and longitude over the whole surface of the Sun. The
different longitudes are covered by sampling the longitude strip around the
central meridian in each magnetogram. During the course of one full
solar rotation, all longitudes get covered. 

While the solar plasma rotates differentially, with the period of
rotation varying with latitude (and depth), but coordinate systems
rotate rigidly, the definition of the solar longitude system is chosen
by convention. The standard choice is the Carrington system, which has
a period of 27.2753\,days. This is the time window covered by a
synoptic map. Evolutionary effects over this time scale are mixed with
longitude variations in the synoptic maps, since we cannot view the
whole range of longitudes simultaneously. 

Because the observations only give us the line-of-sight component of the
magnetic flux density, the polar fields have poor visibility since
they are always located near the solar limb. A physically more
relevant representation is in terms of the vertical flux density, but
since it is not directly measured, it has to be obtained through  
projection. This is done by assuming that on average the field is
vertical at the atmospheric 
level where it is measured. The vertical flux
density is then obtained from the line-of-sight flux density simply through
division by $\mu$, the cosine of the heliocentric angle. 

The vertical assumption gets its justification from the observation 
(through application of the Stokes $V$ line-ratio technique,
cf. Sect.~\ref{sec:kg}) that more than 90\,\%\ of the
line-of-sight magnetic flux in magnetograms with moderate or low
spatial resolution has its origin in strong, kG type intermittent flux
bundles \citep{stenflo-hs72,stenflo-fs72,stenflo-s73}, usually
referred to as flux tubes. These flux tubes are anchored in the
subphotospheric layers and get forced towards an upright orientation
in the photosphere because they are highly buoyant. 

\begin{figure*}
\centering
\includegraphics[width=0.4\textwidth,angle=-90.]{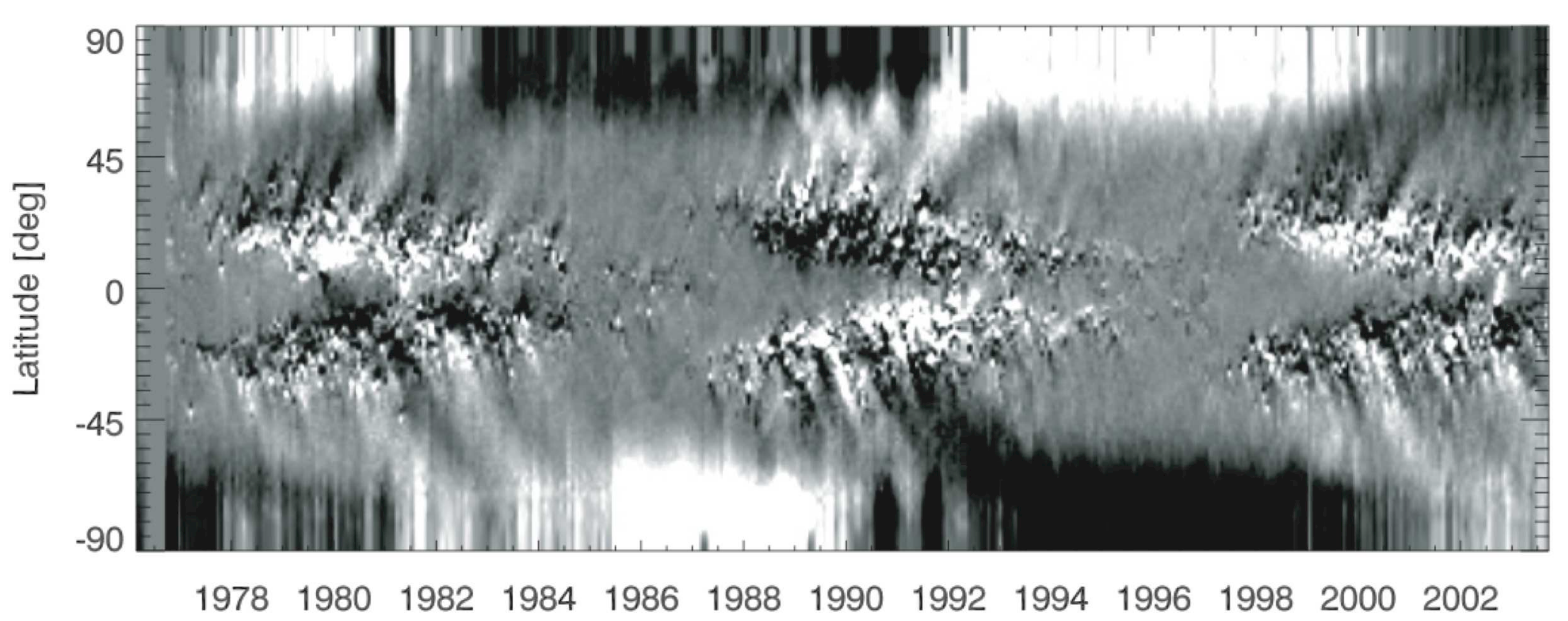}\caption{``Butterfly
diagram'' of the axisymmetric component of the Sun's radial magnetic
field, based on synoptic magnetic maps recorded at NSO/Kitt
Peak. Bright regions indicate fields directed out of the Sun, dark
regions fields directed towards the Sun. From \citet{stenflo-schbau06}.}\label{fig:butterfly} 
\end{figure*}

The solar magnetic cycle represents a global oscillation between a
toroidal and a poloidal field. The dominating features of solar
magnetograms are the bipolar regions, which represent sections of the
subsurface toroidal field that have buoyantly risen to protrude through
the solar surface, where we can observe them. The poloidal field is
less conspicuous and much in the form of a large-scale background
pattern. To make the global evolution of the poloidal field more
visible one can extract its axisymmetric component by averaging each
synoptic magnetic map over all longitudes to obtain the latitude
variation of this average as a function of time. Such a representation
is illustrated in Fig.~\ref{fig:butterfly}, which represents the
vertical component of the axisymmetric field (obtained from the
line-of-sight field via the vertical assumption). A diagram that
represents the sunspot number in latitude-time space is called a
butterfly diagram, but this name is often used for the magnetic-field
representation as well. 

The low-latitude belts that migrate towards the equator in
Fig.~\ref{fig:butterfly} coincide with the activity belts with
sunspots. The polarity pattern is anti-symmetric with respect to the
equator and reverses each 11\,yr, as required by Hale's polarity law,
although the phase is latitude dependent. Each low-latitude belt is 
dominated by one polarity, the preceding polarity of the bipolar
magnetic regions. Discrete pulses of the opposite (following)
polarity shoot out towards the polar regions from the poleward sides
of the belts. They supply the polar regions with the new flux that
replaces and reverses the old polar field. During the period of
minimum solar activity the polar fields reach their maximum amplitude
and the global pattern resembles that of a dipole.

\subsection{Harmonic decomposition}\label{sec:harmonic}
The natural mathematical description of a pattern on a spherical surface is in
terms of spherical harmonics. They are used in the formulation of the
dynamo equations as an eigenvalue problem
\citep[e.g.][]{stenflo-steenkrause69}. To bring the observational data
into a form that is suited for interpretation in terms of dynamo theory,
the observed pattern of vertical magnetic fields on the Sun can be
decomposed in its spherical harmonics. The global 
evolution of the field can then be explored in terms of time series
analysis of the various harmonic modes \citep{stenflo-altschuler74,stenflo-sv86global,stenflo-sg88global,knaack_harm2005,stenflo-derosa12}. 

While such time series analysis reveals a number of intermittent,
quasi-periodic variations, in particular biennial oscillations in the
period range 1.2-2.5\,yr \citep[cf.][]{knaack_harm2005}, the truly
resonant 22\,yr oscillation dominates the odd, axisymmetric modes
(which represent patterns that are anti-symmetric with respect to the
equator). In contrast, this resonance is nearly absent in the even
(symmetric) modes. 

Let $B(x,\varphi,t)$ be the vertical flux density as a function of
$x=\cos\theta$ (where $\theta$ is the colatitude), longitude
$\varphi$, and time $t$. Then the axisymmetric field is 
  \begin{equation} \label{eq:axisym}
    \bar B (x,t)={1\over 2\pi}\,\int_{-\pi}^\pi B(x,\varphi,t)\,{\rm d}
    \varphi\,.
  \end{equation}
It can be expanded in terms of the Legendre polynomials $P_\ell(x)$: 
 \begin{equation} \label{eq:expansion}
    \bar B (x,t)=\sum_{\ell=0}^\infty\,b_\ell(t) P_\ell(x)\,.
  \end{equation}
Using the orthonormality relations for the Legendre polynomials we can 
invert Eq.~(\ref{eq:expansion}) to obtain the time series $b_\ell(t)$ of the 
expansion coefficients for each value of spherical harmonic degree $\ell$: 
  \begin{equation}\label{eq:bcoeff}
    b_\ell(t)={1\over 2} \,(2\ell+1)\,\int_{-1}^{+1}\bar B (x,t) 
    P_\ell(x)\,{\rm d}x\,.
 \end{equation}

\begin{figure*}
\centering
\includegraphics[width=0.75\textwidth]{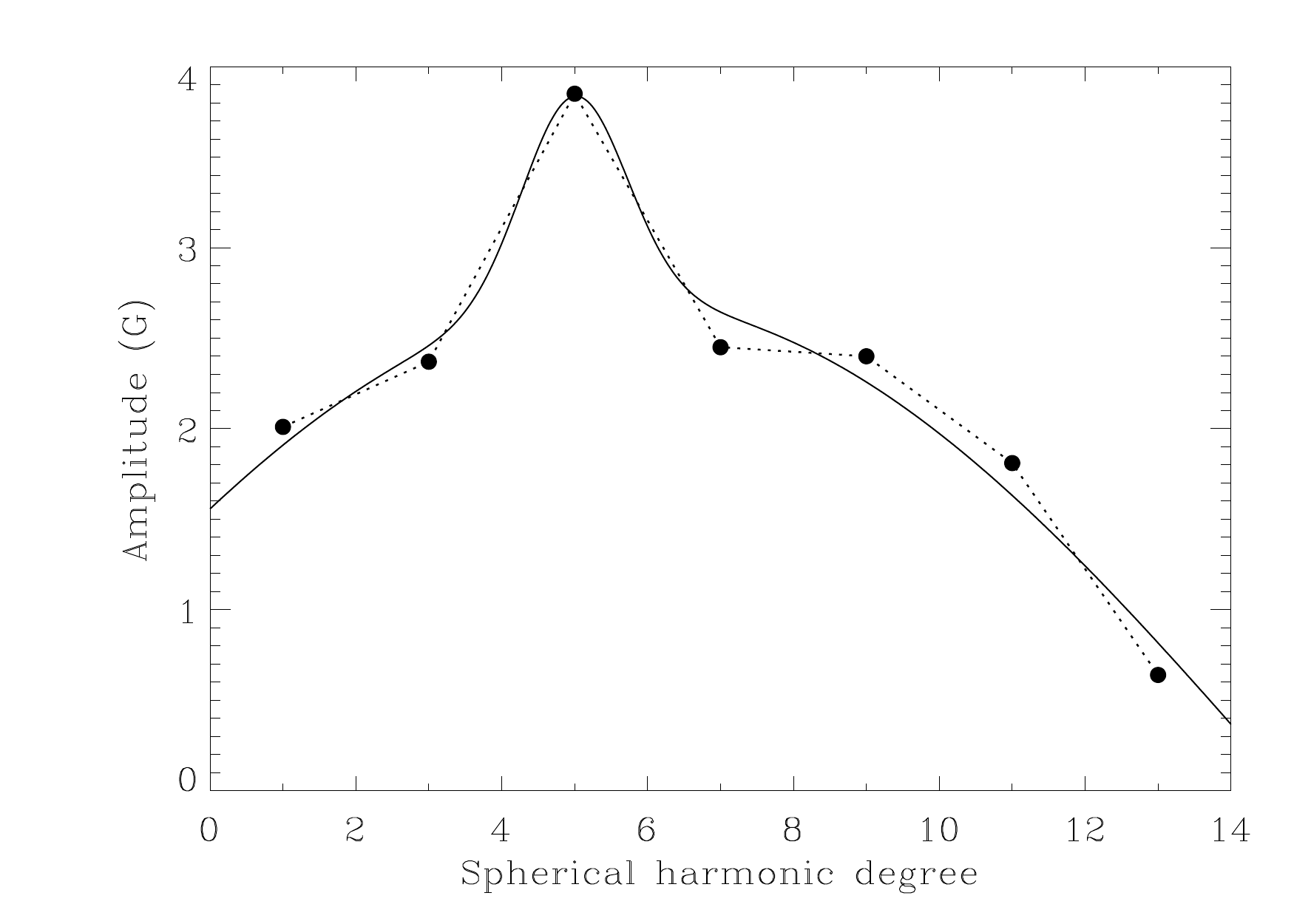}\caption{Amplitude
  of the 22\,yr periodic variations of the axisymmetric modes of the
  Sun's vertical magnetic-field 
  pattern as a function of spherical harmonic degree $\ell$, as
  determined from harmonic decomposition of a 33\,yr data base
  compiled from Mt
  Wilson and Kitt Peak synoptic magnetic-field maps for the period
  1959.6-1992.8 (filled circles and dotted line). The solid line
  represents an analytical fit function to get a smoother
  representation. Note that the dominating contributor to the
  axisymmetric pattern is the mode
  with $\ell =5$. Adapted
  from \citet{stenflo-s94cycle}.}\label{fig:harmonic} 
\end{figure*}

Since the power spectra of $b_\ell(t)$ show the power to be
concentrated around the 22\,yr resonance for the odd modes
\citep[cf.][]{stenflo-sv86global,stenflo-sg88global}, it is a good
approximation to use a sinusoidal representation of the odd modes as 
  \begin{equation}\label{eq:btime}
    b_\ell(t)=u_\ell +\,a_\ell\,\cos[\,\omega(t-t_\ell)]\,,
 \end{equation}
where $\omega =2\pi/22$\,yr$^{-1}$ and $-\omega t_\ell$ is the phase
angle. $u_\ell$ represents a time-invariant, fossil field, while
$a_\ell$ is the amplitude (constrained to be positive) of the mode
with spherical harmonic degree $\ell$. 

\begin{figure*}
\vspace{1cm}
\centering
\includegraphics[width=0.75\textwidth]{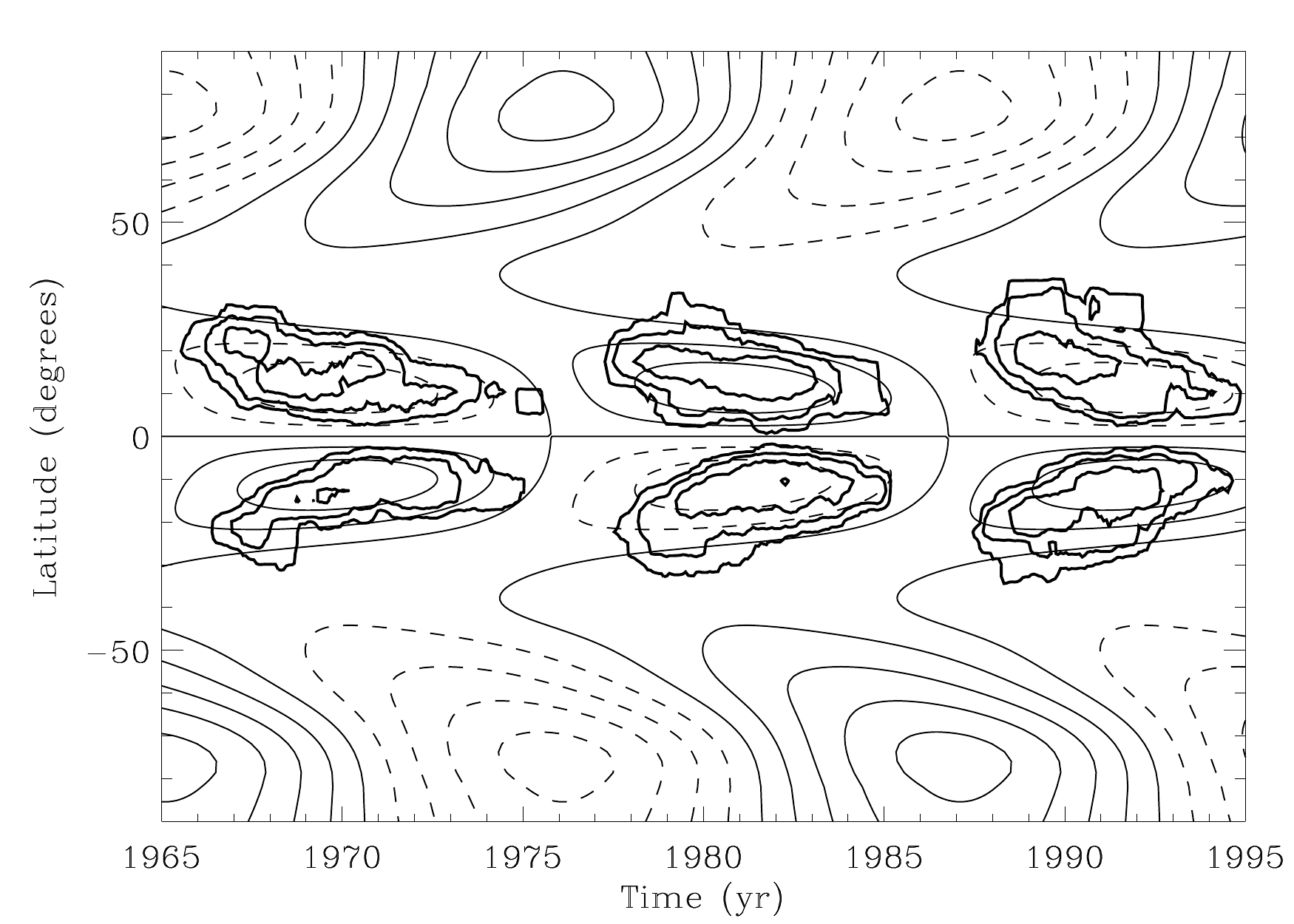}\caption{Superposition
of the sunspot butterfly diagram (thick solid contours) on the
time-latitude distribution of the axisymmetric vertical magnetic field
(thin contours), based on the spherical harmonic decomposition with
the amplitudes given by the filled circles in Fig.~\ref{fig:harmonic}
together with their corresponding phases.}\label{fig:bflyiso} 
\end{figure*}

Using  a 33\,yr long data base (1959.6-1992.8) of Mt Wilson and Kitt
Peak synoptic maps, the sinusoidal model of
Eq.~(\ref{eq:btime}) has been applied with an iterative least squares
technique to determine the three free parameters $u_\ell$, $a_\ell$,
and $t_\ell$ for each $\ell$ \citep{stenflo-s94cycle}. 

The result for the amplitudes $a_\ell$ is shown in
Fig.~\ref{fig:harmonic}. It has a pronounced maximum at $\ell
=5$. This harmonic degree dominates the behavior of the
axisymmetric modes, not the dipole component. The ``fossil'' field
component $u_\ell$ does not 
differ significantly from zero. The upper limit (1-$\sigma$ error)
of the fossil dipole component $u_1$ is approximately 0.2\,G. 

Using the amplitudes and phases for the seven odd modes (with $\ell
=1$-13) determined with the
model of Eq.~(\ref{eq:btime}), we can reconstruct the butterfly
diagram. The result is shown in Fig.~\ref{fig:bflyiso}, where we for
comparison have overplotted the isocontours for the sunspot
number. The diagram is consistent with the full butterfly diagram of
Fig.~\ref{fig:butterfly}, although it does not show details on the
smaller time scales, since it has been constructed exclusively from
the 22\,yr contributions. In particular we do not see the herringbone
pattern of discrete flux pulses that steeply migrate from the activity
belts towards the poles, since they vary on a quasi-periodic biennial
time scale.

\section{Magnetic intermittency}\label{sec:intermitt}
\begin{figure*}
\centering
\includegraphics[width=0.43\textwidth,angle=-90.]{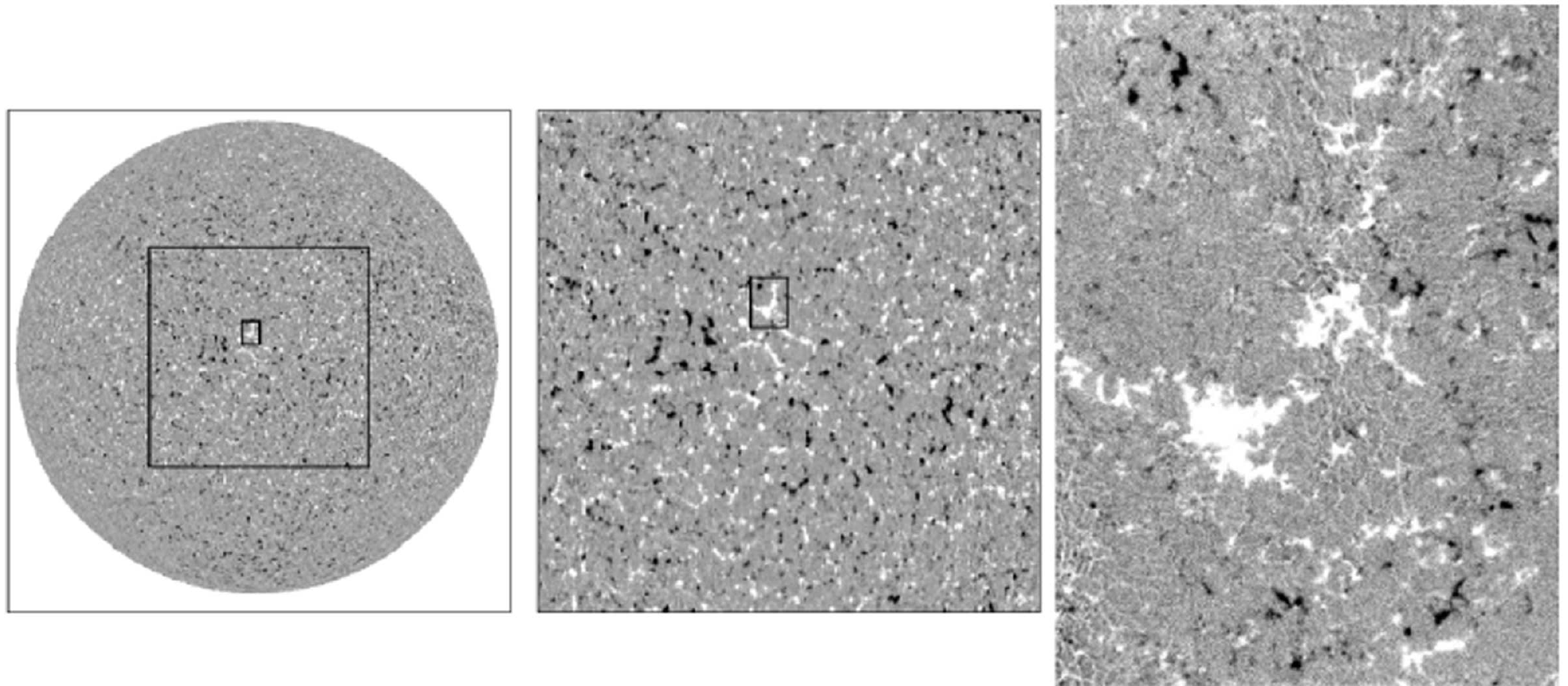}\caption{Illustration
of the fractal-like pattern of quiet-sun magnetic fields, recorded on
February 9, 1996, when there were no spots present on the Sun's
disk. The left panel shows the full-disk Kitt Peak magnetogram, the
middle panel the magnified central portion of it. The right panel is a
high-resolution recording at the Swedish La Palma Observatory
(courtesy G\"oran Scharmer) of the small rectangular area marked near
the middle of the other panels. The pattern is characterized by flux
concentrations separated by voids with little net flux. Indirect
evidence from Hanle depolarization measurements reveals that they are
not voids at all but are seething with an ``ocean'' of turbulent
fields with polarities mixed on scales beyond the resolution limit. From \citet{stenflo-evershed10}.}\label{fig:fractal} 
\end{figure*}

When we zoom in on ever smaller scales on the quiet Sun, a fractal-like
pattern with a high degree of self-similarity is revealed, as
illustrated in Fig.~\ref{fig:fractal}. As we saw from
the magnetic energy spectrum in Fig.~\ref{fig:enmaglocdyn}, the
magnetic structuring is expected to continue down to scales that are about four orders
of magnitude smaller than the currently resolved scales. 

Since all observations of quiet-sun magnetic fields represent
averages over spatially unresolved structures, we make a distinction
in our terminology between {\it field strength} and {\it flux
  density}, although both are measured in the same units. In the limit
of infinite spatial resolution the two concepts are
identical. Each resolution element averages over
unresolved regions with different field strengths. The average field
strength is the integrated flux divided by the area over which we
integrate. We call this ``flux density''. Since the circular polarization
is almost linearly related to the line-of-sight component of the field
strength, the observed quantity in Stokes $V$ recordings is the
line-of-sight component of the flux density. This is not the case for
the linear polarization, since its relation to field strength is
highly nonlinear (nearly quadratic). Therefore the quantity
that is measured is related to the average energy
density and not to flux density. 

\subsection{Network and internetwork}\label{sec:network}
\begin{figure*}
\centering
\includegraphics[width=0.75\textwidth]{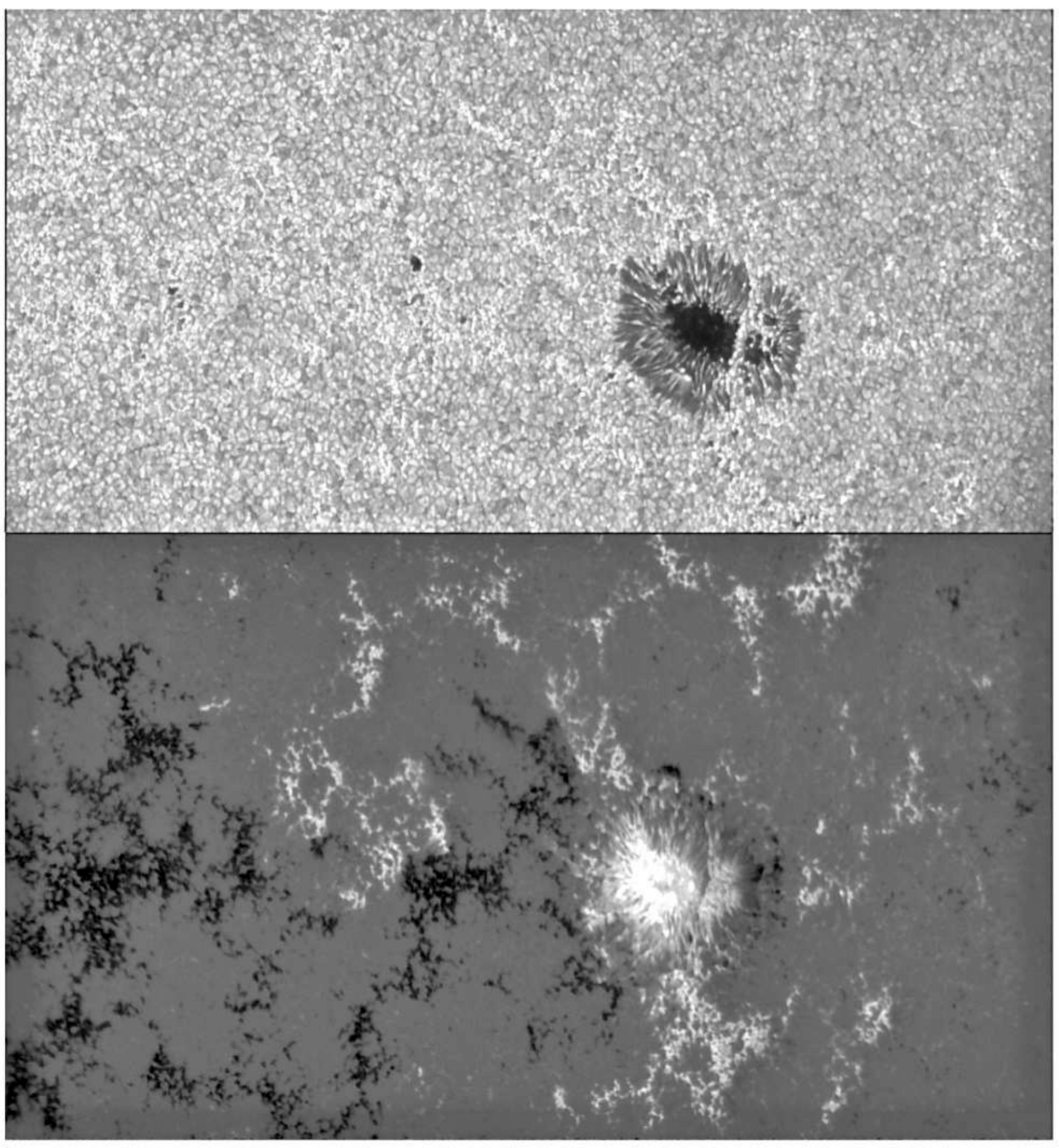}\caption{Recording
with Hinode/SOT on September 27, 2012. The top panel is a G-band
filtergram of active region AR 11575, the bottom panel a line-of-sight
magnetogram recorded with the NFI (Narrow-band Filter Imager) on SOT. Hinode is a Japanese mission developed and launched by ISAS/JAXA, with NAOJ as domestic partner and NASA and STFC (UK) as international partners. It is operated by these agencies in co-operation with ESA and NSC (Norway).}\label{fig:network} 
\end{figure*}

The frozen-in magnetic field lines are carried by the convective
motions to the cell boundaries. The pattern of flux concentrations
therefore develops a cell structure. On the solar surface the two main
types of velocity cells are represented by the granulation (cell size
$\sim 1$\,Mm) and the supergranulation (cell size $\sim 30$\,Mm). The flux
concentrations at the supergranular cell boundaries are conspicuous in
magnetograms in the surroundings of active regions, where there is an 
abundance of magnetic flux, as illustrated in 
Fig.~\ref{fig:network}. In quiet regions only fragments of the cell
boundaries are filled with flux, which makes the cell structure 
appear less obvious (cf. Figs.~\ref{fig:hmiquiet} and \ref{fig:fractal}). 

The term ``network'' that is often used for the flux concentrations at the supergranular
cell boundaries was originally introduced to describe the brightness pattern
in monochromatic images of the Sun recorded in various photospheric or
chromospheric spectral lines. The brightness enhancements with implied
heating of the upper atmosphere are however correlated both with the
concentrations of magnetic flux and with downdrafts, since the largely
horizontal flows in 
the interior of the supergranulation cells become vertical at the 
cell boundaries. 

By definition, what is not network is called internetwork,
supposed to represent the interior of supergranular cells. It should
however be stressed that ``network'' and ``internetwork'' are
statistical concepts, since supergranulation cells undergo statistical
fluctuations, they form and dissolve over time scales of typically
24\,hr. There is no way to tell whether a given magnetic element is
part of the network or the internetwork. For this reason questions
like whether large flux concentrations can also be found in the
internetwork are not meaningful and cannot be answered. We always have
a continuous distribution of flux densities of all magnitudes, there
is no well-defined dichotomy between network and internetwork magnetic flux. 

\begin{figure*}
\centering
\includegraphics[width=0.75\textwidth]{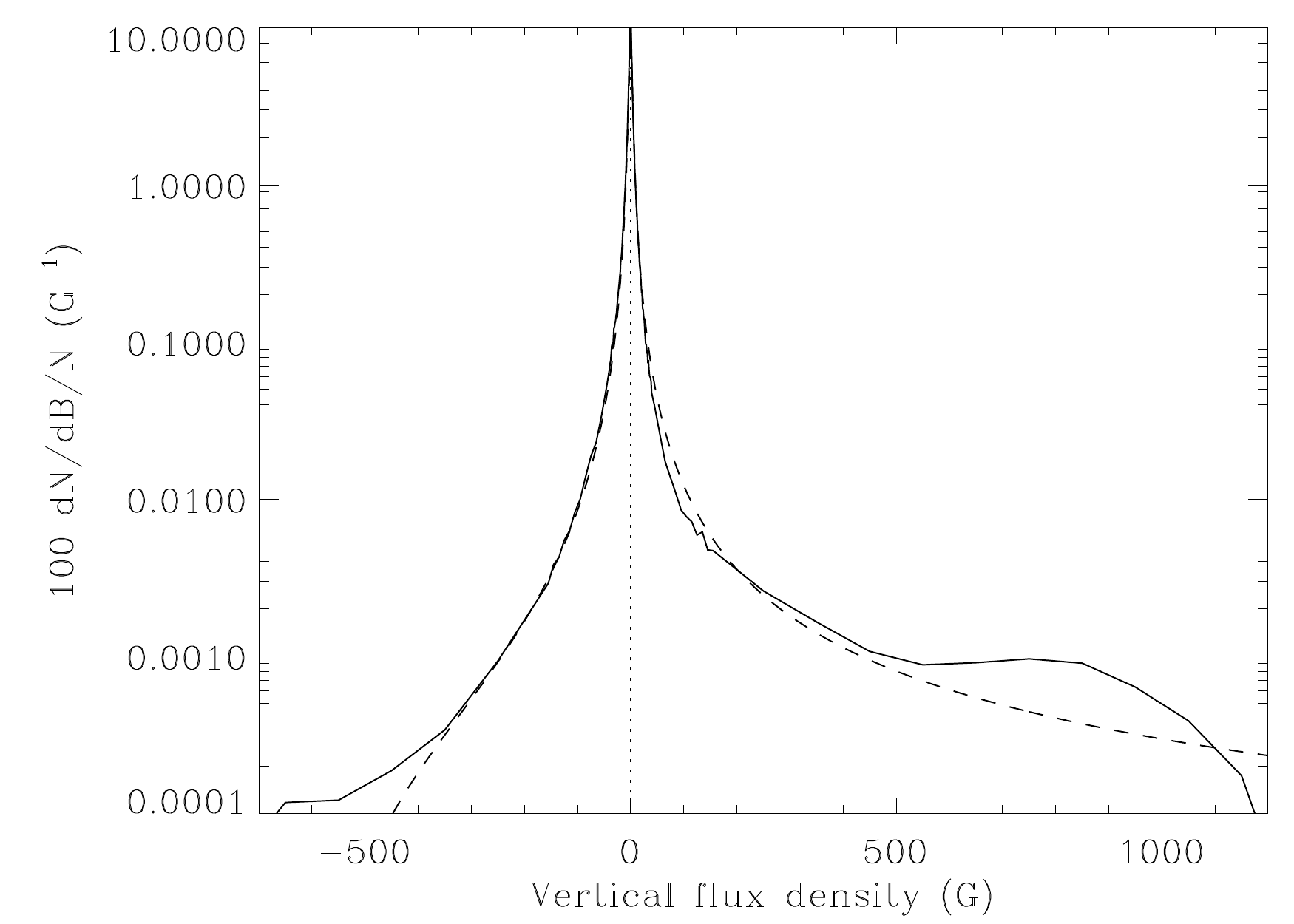}\caption{Probability
density function (PDF) for the vertical magnetic flux density at the
disk center of the quiet Sun, as recorded by the SOT/SP instrument on
Hinode on February 27, 2007 (solid line), from \citet{stenflo-s10aa}. The Gaussian noise
broadening has been removed. The dashed curve is an analytical
representation with a core in the form of a stretched exponential and
quadratically declining, extended wings. This PDF shape is typical for flux
densities at resolved spatial scales. In the limit of infinite
resolution we expect the field-strength PDF to have a 100-200\,G wide
core region in order to be consistent with the observational
constraints from the Hanle effect \citep{stenflo-s12aa1}.}\label{fig:bvpdf} 
\end{figure*}

The absence of such a dichotomy in the observed flux densities becomes
clear if we have a look at the histograms or probability density
functions (PDF) of the measured flux densities, like in
Fig.~\ref{fig:bvpdf} derived from Hinode SOT/SP observations of the
quiet-sun disk center. All such PDFs, after being corrected for noise
broadening, are characterized by an extremely narrow and peaked core
centered at zero flux density, surrounded by slowly declining wings,
which in observations with high spatial resolution (like with Hinode)
may extend to kG flux densities (implying that such fields are nearly
resolved). Since the quiet Sun exhibits large-scale patterns that are
dominated by one of the two 
polarities, the PDF extracted for a given quiet region is
generally asymmetric between plus and minus, like the one in
Fig.~\ref{fig:bvpdf}. The bump that we see in the solid curve for
large positive flux densities is not a statistically significant
general property of the PDF shapes. 

The dashed curve in Fig.~\ref{fig:bvpdf} is an analytical
representation that is characterized by a core region in the form of a
stretched exponential and quadratically declining ``damping''
wings. It is typical for the shape of quiet-sun PDFs at resolved
scales. The apparent scale invariance of this PDF shape must however end 
at very small scales (below about 10\,km) to be compatible with the
observed Hanle depolarization (cf. Sects.~\ref{sec:hanleturb} and
\ref{sec:hanresults}), which
reveals the existence of a hidden ocean of turbulent fields. It can be
shown that the PDF damping wings cannot contribute significantly to
the observed depolarization, which must instead originate in the PDF
core region. A sufficiently large effect can only be obtained
with a wide 
core region that has a half width of order 100-200\,G
\citep{stenflo-s12aa1}. The PDF for the field
strengths is therefore expected to be very different from the scale
dependent PDF for the flux densities. 

The visual appearance of the magnetic pattern in Hinode magnetograms
like the one in Fig.~\ref{fig:network}, which gives the impression
that all the flux is in the network, with large empty voids in
between, is simply a consequence of the choice of grey-scale cuts in
the representation. These cuts are always set high enough to suppress
the influence of the noisy PDF core region. All
that is made visible then comes from the PDF damping
wings, beyond  typically 50\,G. These parts of the PDF represent a tiny fraction of all
the pixels in the image. The damping wings can be understood as
signatures of intermittency, because the combination of high flux density and
low occurrence probability implies spatial separation. The PDF core region on the
other hand represents the apparent voids in the magnetograms.

\subsection{Extreme intermittency: kG-type flux on the quiet
  Sun}\label{sec:kg}
The measured flux densities only represent {\it lower limits} to the
actual field strengths. Assume for instance that instead of being
resolved, the flux element that is the source of the measured
polarization occupies only 1\,\%\ of the resolution element. Then the
actual or intrinsic field strength is 100 times larger than the
apparent, average field strength or flux density. 

To allow us to distinguish between these two scenarios we need another
type of observable, which is provided by the magnetic line ratio. It
is the ratio between the Stokes $V$ signals measured 
simultaneously in the line wings of the pair of Fe\,{\sc i} lines at 5250.22 and
5247.06\,\AA. This particular line pair is the only known one for
which the magnetic-field effects decouple from the thermodynamic
effects so that intrinsic field strengths can be measured in a nearly
model independent way, independent of any particular assumption for
the thermal structure of the atmosphere inside the flux concentrations. 
When the line-ratio technique was introduced four decades ago, it
immediately led to the discovery that more than 90\,\%\ of the
magnetic flux visible in magnetograms with moderate (a few arcsec)
resolution has its origin in strong kG type flux bundles
\citep{stenflo-s73}, usually 
referred to as flux tubes, which occupy a small fraction (typically
1\,\%\ of the photospheric volume)
\citep[cf. also][]{stenflo-hs72,stenflo-fs72}. 

\begin{figure*}
\vspace{-1.7cm}
\centering
\includegraphics[width=0.95\textwidth]{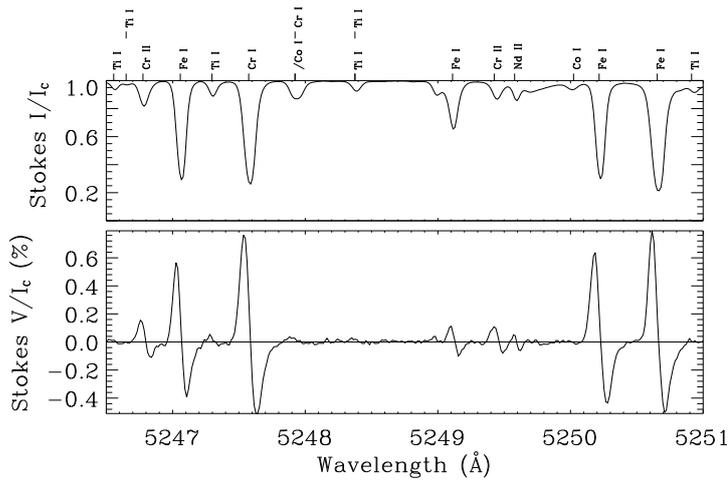}
\vspace{-6.5cm}
\caption{Sample
spectrum of the 5247-5250\,\AA\ region from the FTS Stokes $V$ atlas
\citep{stenflo-setal84} 
of a weak plage at disk center. }\label{fig:fts5250} 
\end{figure*}

The idea behind the line-ratio technique is to make use of the
deviation from linearity in the relation between Stokes $V$ and field
strength when the Zeeman 
splitting becomes comparable to the line width. The {\it differential
  nonlinearity} (difference in amount of Zeeman saturation) between two lines
that are identical in all respects except for their Land\'e factors
is a direct function of intrinsic field strength. This
differential effect expresses itself in the ratio between Stokes $V$ of the two
lines. Since the magnetic filling factor divides out in the ratio, the
effect is independent of the amount of measured flux and only depends
on intrinsic field strength. $V$ alone gives the flux, so in combination with
the $V$ ratio we get the filling factor. 

\begin{figure*}
\centering
\includegraphics[width=0.75\textwidth]{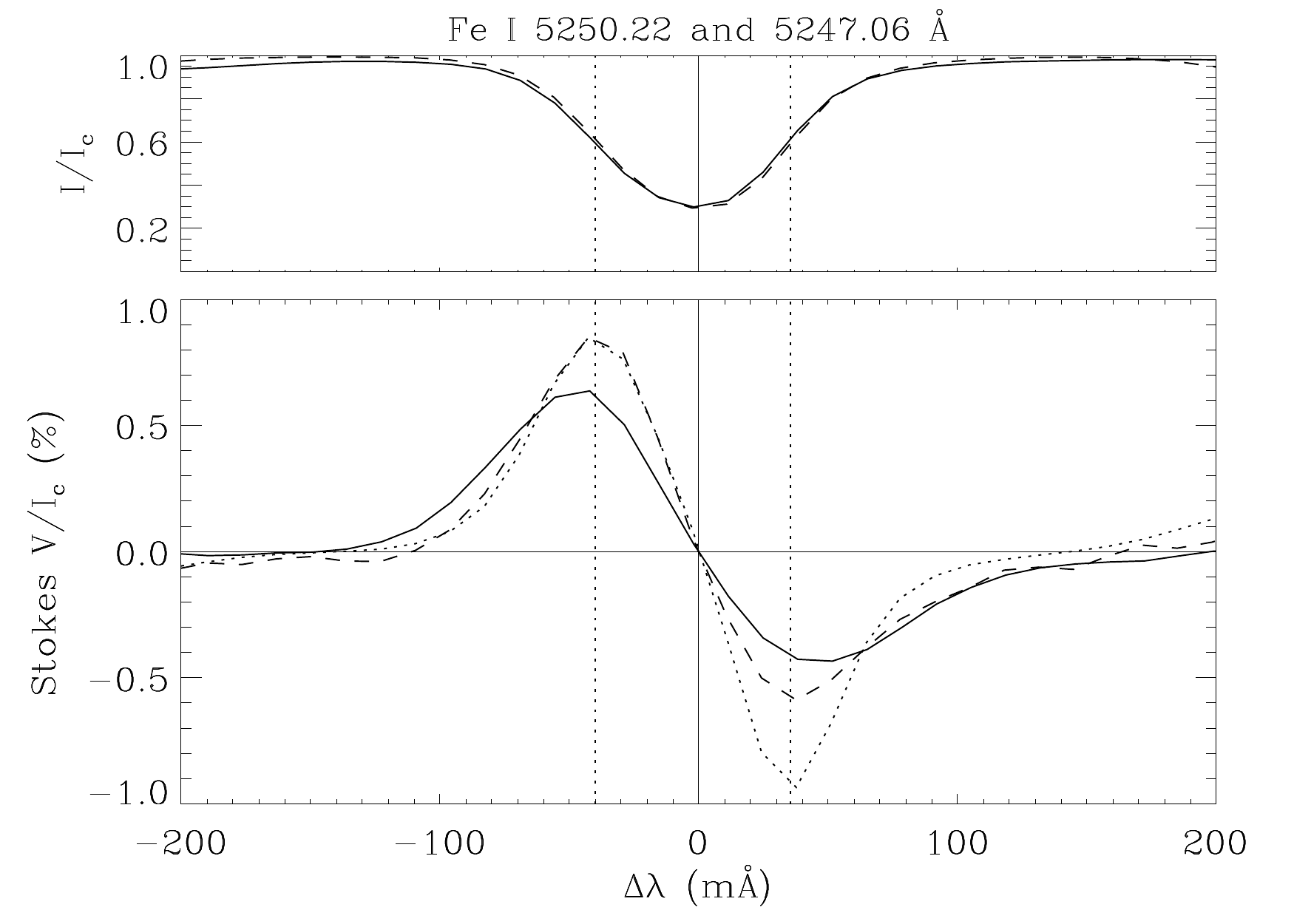}\caption{Superposition
of the $I$ and $V$ line profiles in Fig.~\ref{fig:fts5250} of the 5250.22\,\AA\ line (solid) and
5247.06\,\AA\ line (dashed) on a common wavelength scale relative to
the respective line center. The Stokes $V$ profile of the 5247\,\AA\
line has been scaled up by a factor of 3/2 to compensate for the
Land\'e factor difference with respect to the 5250.22\,\AA\ line. The
dotted curve represents $-\partial I/\partial\lambda$ for the
5247\,\AA\ line, normalized to the blue-lobe Stokes $V$ amplitude. The
locations of the two inflection points of the $I$ profile are indicated
by the vertical dotted lines.}\label{fig:vsuperpose} 
\end{figure*}

Figure \ref{fig:fts5250} shows the spectrally fully resolved $I$ and
$V$ profiles for a small section of the FTS atlas of a weak plage at
disk center that includes the 5250-5247\,\AA\ line pair, from the data
set used in 
\citet{stenflo-setal84}. While the 5250.22 and 5247.06\,\AA\ lines
both belong to iron multiplet no.~1, have the same line strength and
excitation potential, and are therefore formed in the same way in
whatever atmosphere that we may have, they differ in their Land\'e
factors, which are 3.0 and 2.0, respectively, both among the largest
in the visible spectrum but significantly different from each other. 

To illustrate how the differential nonlinearities affect the $V$
profiles, we superpose in Fig.~\ref{fig:vsuperpose} the $I$ and $V$
profiles of the 5250.22\,\AA\ line (solid curves) on top of the 5247.06\,\AA\ line
(dashed curves), whose $V$ profile has been scaled with the factor 3/2 to compensate
for the difference in Land\'e factor between the two lines. A common
wavelength scale relative to the respective 
line centers has been used. Since for intrinsically weak fields $V\sim \partial
I/\partial\lambda$ according to Eq.~(\ref{eq:vweak}), we plot for
comparison as the dotted curve $-\partial I/\partial\lambda$,
normalized to the blue-lobe $V$ amplitude of the 5247.06\,\AA\ line. 

The upper panel verifies that the two Stokes $I$ profiles are indeed
identical. The differential Zeeman saturation expresses itself in two
ways. The $V$ amplitude of the line with the
larger Land\'e factor gets suppressed (saturated) with respect to the
other line, and the profile gets broadened and more extended. We notice
that the blue lobe of the 5247.06\,\AA\ line has a shape that is
nearly identical to that of $-\partial I/\partial\lambda$, indicating
negligible Zeeman saturation for this line. The red $V$ lobes are both
suppressed, in the same proportions, relative to the $-\partial
I/\partial\lambda$ profile. While the {\it areas} of the blue and red
$\partial I/\partial\lambda$ lobes are always exactly balanced (since
integration must lead to the same continuum intensity on both sides of
the line), they are significantly unbalanced for the $V$ profiles. Such
a Stokes $V$ area asymmetry has its origin in correlations between the gradients {\it along
  the line of sight} of the magnetic field and the Doppler shifts
\citep{stenflo-illing75,stenflo-auer78,stenflo-setal84,stenflo-steiner00}. 

\begin{figure*}
\centering
\includegraphics[width=0.75\textwidth]{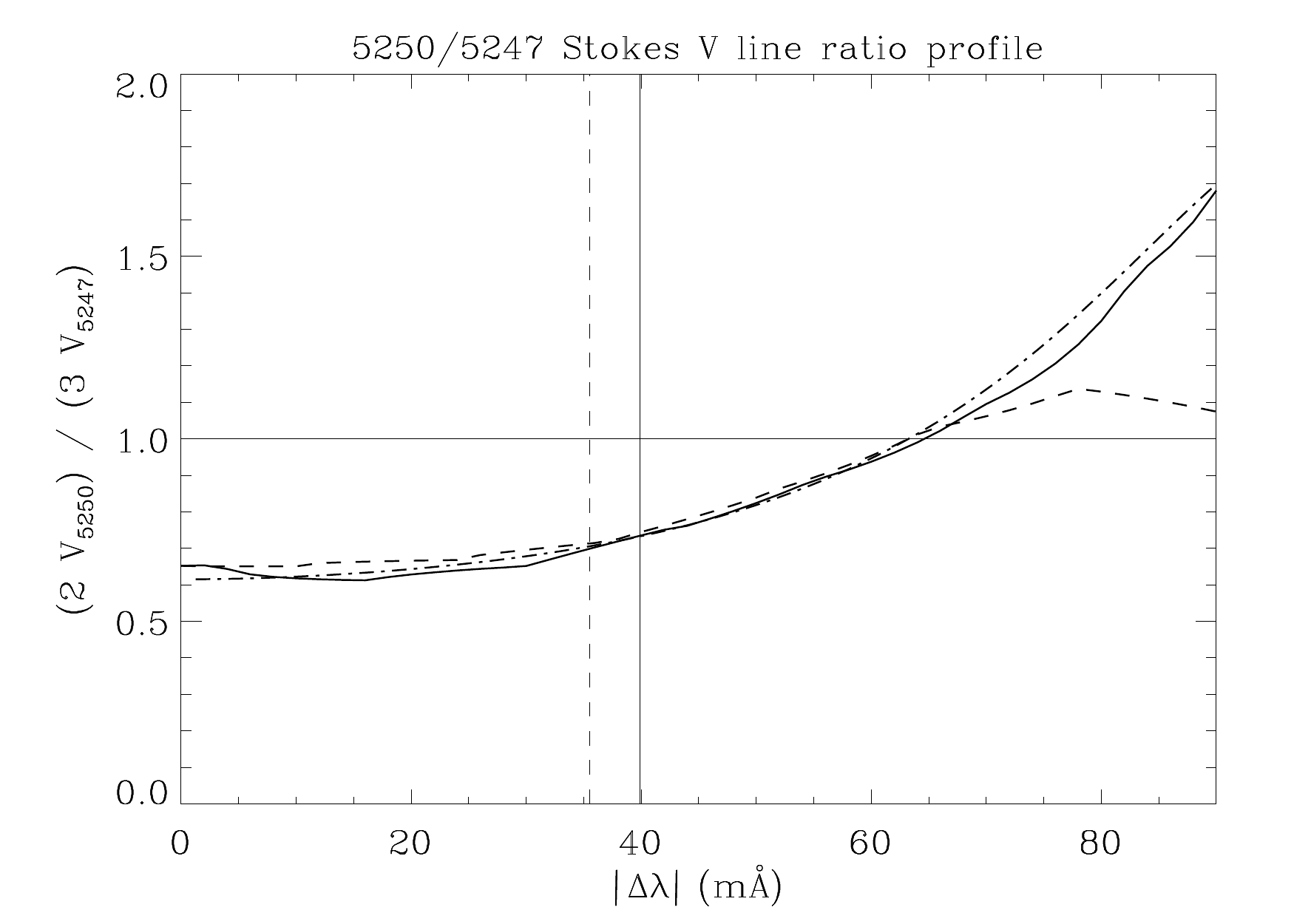}\caption{Ratio
between the solid and dashed curves in Fig.~\ref{fig:vsuperpose} as a
function of absolute wavelength distance from the line center. Solid
line: blue lobe. Dashed line: red lobe. Dash-dotted line: Theoretical
line-ratio profile, based on a field-strength PDF from
\citet{stenflo-s12aa1} and an angular distribution from
\citet{stenflo-s10aa}, as described in the text. The vertical lines
mark the locations of the Stokes $I$ inflection points for the blue
(solid) and red (dashed) line wings.}\label{fig:vratioprof}
\end{figure*}

We can now form the ratio between the Stokes $V$ 5250 and 5247\,\AA\ profiles in
Fig.~\ref{fig:vsuperpose}. This ratio profile is plotted in Fig.~\ref{fig:vratioprof} vs. distance
$\vert\Delta\lambda\vert$ from line center, separately for the blue
lobe (solid curve) and the red lobe (dashed curve). We see that the
blue and red lobe ratio profiles are in perfect agreement all the way
from line center to far beyond the inflection points of the $I$
profile. They diverge only in the distant wings, where $V$ gets small and
the red lobe is severely affected by the line-of-sight
correlations between the magnetic and velocity field gradients. 

If the fields were intrinsically weak ($\la 500$\,G), the $V$ ratio
profile would not differ significantly from the horizontal level of
unity. The more it differs from unity, the larger is the intrinsic
field strength. The conversion of the line ratio to field strength is
insensitive to the model atmosphere, as long as the chosen model
(e.g. a simple Milne-Eddington atmosphere) is able to
reproduce the half width of the intensity profile (since the magnitude
of the line-ratio effect depends directly on the ratio between 
intrinsic Zeeman splitting and line width). Spatially unresolved
line weakenings caused by a different thermodynamic structure of the
magnetic elements have no effect as long as the line width does not
change appreciably. 

The observed line ratio reveals field strengths 
of order 1\,kG almost everywhere on the quiet Sun as observed with
moderate spatial resolution. In contrast the measured $V$ amplitudes
correspond to flux densities (average field strengths) that are usually only
a few G. These two findings can only be reconciled if the flux is
extremely intermittent, with a mixture of very strong (kG) and weak
fields within the resolution element. The simplest description of this
situation is in terms of a {\it 2-component model}, introduced as an
interpretational tool in the initial applications of the line-ratio
method \citep{stenflo-s71,stenflo-s73}. One component is magnetic with field
strength $B$ and occupies a fraction $f$ (the {\it filling factor}) of
the line-forming {\it volume} of the atmosphere (the angular
resolution element times the depth of line formation along the line of
sight). We stress here that the filling-factor concept relates to the
volume and not to the surface area, because later applications of
filling factors to horizontal fields and the transverse Zeeman effect
have often been confused and incorrectly interpreted, due to a lack of
understanding of the meaning of filling factors when the fields
are highly inclined with respect to the solar surface. All confusion
is eliminated if one understands that a filling factor
is a volume occupation fraction. 

In the standard application of the 2-component model the second
component, with filling factor $1-f$, is assumed to be
``nonmagnetic'', because its contribution to the measured
polarimetric signal is disregarded. It is obvious that this is an
assumption made exclusively for mathematical and conceptual
convenience. The
extremely high electrical conductivity of the solar plasma effectively
prohibits the existence of strictly nonmagnetic regions, any seed
field will be tangled up and amplified by the turbulent motions. For
instance, we now know from Hanle depolarization measurements that the
``nonmagnetic'' component is seething with an ocean of hidden,
turbulent magnetic flux that may even dominate the
energy balance of the solar atmosphere
\citep{stenflo-s82,stenflo-trujetal04}. The 2-component model has
however served as a very useful conceptual tool for the construction
of a series of empirical flux tube models at increasing levels of
sophistication \citep{stenflo-sol93}. Typically these models consist of
a magnetic element with axial symmetry that expands with height
according to the laws of magnetohydrostatics, surrounded by
nonmagnetic surroundings with inflows to a downdraft region at the
outside of the flux tube. 

Although the interpretation of the observed line-ratio profile in
Fig.~\ref{fig:vratioprof} in terms of a single-valued field strength
(as implied by the 2-component model) is able to give a satisfactory
fit to the observations, a much more realistic representation of the
Sun is in terms of continuous distribution functions. For the derivation of the theoretical
dash-dot fit curve in Fig.~\ref{fig:vratioprof} we have therefore used distribution
functions for both the field strengths \citep[from][]{stenflo-s12aa1}
and inclination angles according to their empirically determined
statistical dependence on field strength
\citep[from][]{stenflo-s10aa}. Note that for the field-strength PDF
one cannot use the distribution obtained for the flux densities (like the one
in Fig.~\ref{fig:bvpdf}), but one needs a PDF that represents the case
of infinite spatial resolution. The field-strength PDF
inferred in \citet{stenflo-s12aa1} has a broad core region needed to 
satisfy the Hanle constraint, and quadratically declining damping
wings that extend out to 2\,kG, where it is assumed that the PDF ends
(because stronger fields cannot be in pressure balance with the
ambient gas pressure). The calculations show that the PDF core region
is nearly irrelevant to the line ratio, although it dominates the
occurrence probability. The line ratio requires the existence of
strong fields and gets its main contributions from the extended wings
beyond about 500\,G, although the integrated filling factor for these
strong fields is small. We note that the theoretical fit in the far
line wings improves when we account for the full anomalous Zeeman splitting
pattern of the 5247.06\,\AA\ line rather than treating it as a normal
Zeeman triplet with an effective Land\'e factor.

\subsection{The 6302/6301 line ratio and its renormalization}\label{sec:redratio}
Although it is only for the Fe\,{\sc i} 5250.22 - 5247.06\,\AA\ line
pair that the magnetic-field effects (differential Zeeman saturation)
can be cleanly separated from the thermodynamic and line-formation
effects, useful applications with other line combinations are not
ruled out. The Hinode SOT/SP instrument, which in
recent years has delivered Stokes vector data that are by far the best
in terms of high and stable angular resolution, is limited by design to the use
of the Fe\,{\sc i} 6302.5 - 6301.5\,\AA\ line pair. The $V$
ratio of this pair is severely contaminated by thermodynamics and line-formation
effects, since the two lines have different line strengths and are
therefore formed at different heights in the atmosphere. 

This contamination seriously complicates the interpretation and
conversion of the line-ratio values into intrinsic field
strengths. \citet{stenflo-khocoll07} tested the 6302/6301 line ratio
through numerical simulations and found it to systematically give (if
interpreted in a straightforward way) 
much too strong and generally unphysical values for the field
strengths, in contrast to the 5250/5247 ratio, which behaved as
expected. However, in scatter plots of $V$ for the 6302.5\,\AA\ line vs. $V$
for the 6301.5\,\AA\ line based on Hinode quiet-sun disk center data
one finds a very well-defined linear 
regression relation with a slope that is clearly much smaller than 
expected if the fields were intrinsically weak. The small slope is a
tell-tale signature of intrinsically strong fields.  The same scatter
plot however also reveals a small population of points with a
different slope, consistent with the weak-field slope that one expects
from scatter plots of $\partial I/\partial\lambda$ for the two
lines. Assuming that the secondary population represents truly weak,
``uncollapsed'' fields, it is possible to recalibrate the 6302/6301
line ratio scale to enable a determination of the intrinsic field
strengths for the main flux population \citep{stenflo-s10aa}. 

Most field strengths determined this way are of order 1\,kG and
consistent with previous determinations based on the 5250/5247 line
ratio. An {\it upper limit} to the size of the corresponding
``collapsed'' flux elements is obtained by assuming that there is one flux tube
per resolution element. It is given by ${\sqrt{ f\,A}}$, where $f$ is
the filling factor and $A$ the area of the resolution element. If
there is more than one flux tube per resolution element, then the size
of each flux tube is naturally smaller. In this way it is possible to
infer from the Hinode line-ratio data a histogram for the flux tube
sizes. It shows that the main kG-type flux tube population on the quiet Sun
has sizes in the range 10-100\,km \citep{stenflo-s11aa}. 

\begin{figure*}
\centering
\includegraphics[width=0.75\textwidth]{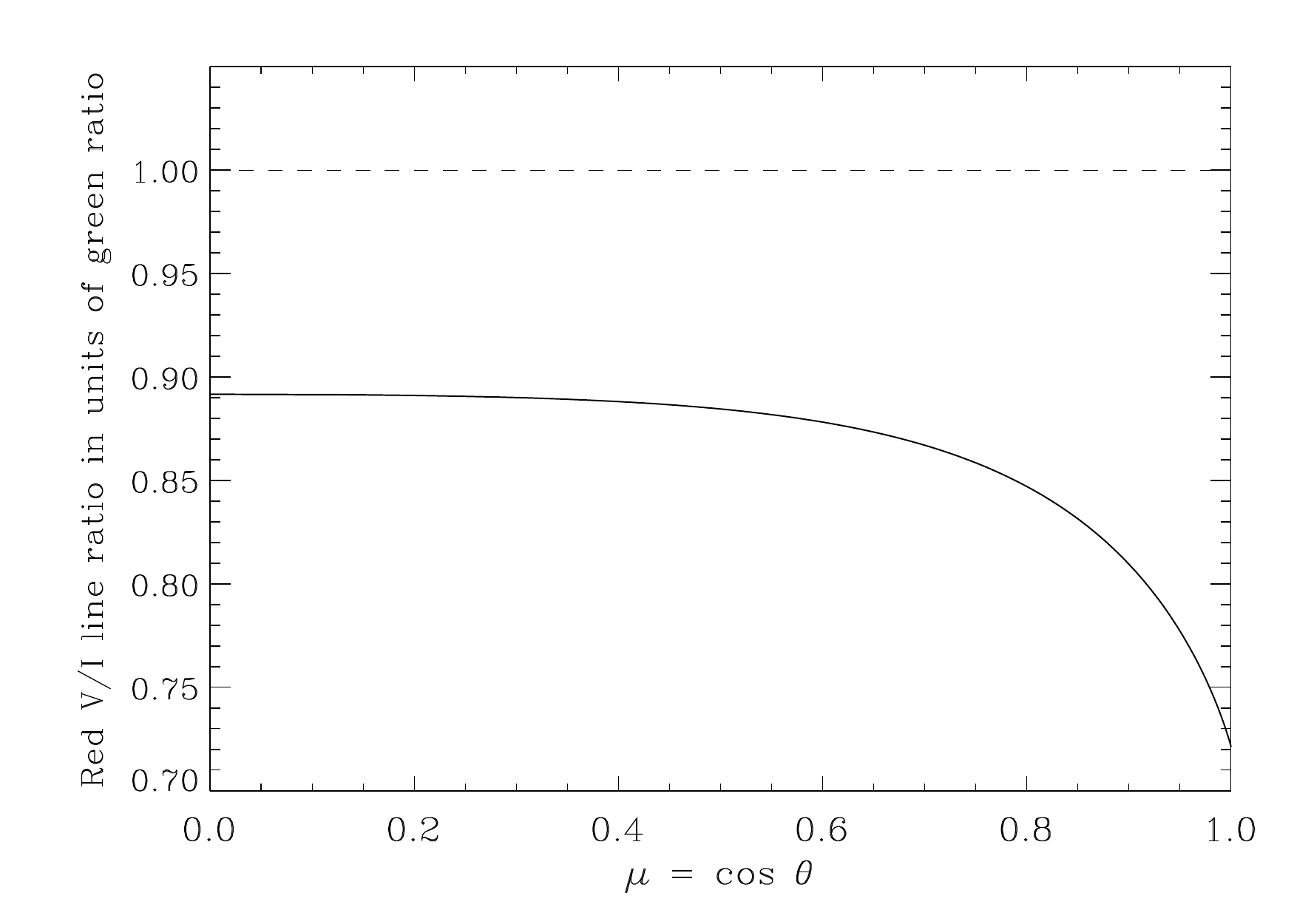}\caption{Ratio
  between the ``red'' 6302/6301 Stokes $V/I$ line ratio and 
the corresponding ``green'' 5250/5247 line ratio as a function of
center-to-limb distance. The curve allows the
``contaminated'' red line ratio to be translated to the green line ratio. From
\citet{stenflo-setal13aa}.}\label{fig:redovergreen}  
\end{figure*}

Recently it became possible to calibrate the contaminated 6302/6301
line ratio in terms of the uncontaminated 5250/5247 line ratio
\citep{stenflo-setal13aa}, using a special setup with the
ZIMPOL polarimeter at IRSOL in Locarno, which allowed both line pairs
to be recorded simultaneously on the same CCD sensor. The resulting
``calibration curve'' is shown in Fig.~\ref{fig:redovergreen} as a
function of center-to-limb distance. It means that the measured
6302/6301 line ratio values have to be ``renormalized'' through
division by the values represented by the solid line, before one can
convert them into intrinsic field strengths, otherwise fictitiously
high and unphysical field strengths would be found. At disk center the
division factor is 0.72, which agrees with the renormalization factor
that was used in \citet{stenflo-s10aa} and based on the assumption that
the secondary flux population in the scatter plot represented truly
weak fields. The independent calibration represented by
Fig.~\ref{fig:redovergreen} provides a validation of the
renormalization used.

\subsection{Convective collapse}\label{sec:collapse}
The kG fields that were revealed by the line-ratio technique four
decades ago \citep{stenflo-s73} presented an enigma, since the dynamic
forces of the convective flows that concentrate the flux at the cell
boundaries are not strong enough to amplify the fields beyond
the hectogauss range. The only force that is sufficiently strong to
confine kG fields is 
the gas pressure. In a static situation the magnetic pressure is 
balanced by the difference in gas pressure between the exterior
and interior of the magnetic flux element. If the flux region is
entirely evacuated, the magnetic pressure equals the ambient gas
pressure, which determines the maximum field strength that can be
confined. For the photosphere it is of order 2\,kG but depends on the
height level that we refer to. 

The enigma was resolved by the mechanism of convective collapse, which
was proposed in the late 1970s
\citep{stenflo-parker78,stenflo-spruit79,stenflo-sprzw79,stenflo-unnoando79}. It
was developed and can best be described in terms of a 2-component
scenario, with an initially weak magnetic flux element embedded in a
nonmagnetic environment. It can be shown that this configuration is
unstable in the superadiabatic region at the top of the convection
zone (immediately below the photosphere). If the flux region is
optically thick in the horizontal direction, an initially
small downdraft will generate an adiabatic gradient inside the
flux region, when the time scale for thermal exchange with the
surroundings is larger than the convective time scale. This will
create a temperature and pressure deficit with respect to the
surroundings because of the external superadiabatic stratification. The
deficit in gas pressure leads to compression, which amplifies the
downdraft with the adiabatic cooling, causing evacuation of the
flux region and spontaneous collapse. As a consequence the field gets
concentrated until an equilibrium is reached and the external gas
pressure is unable to amplify the field more. 

\begin{figure*}
\centering
\includegraphics[width=0.8\textwidth]{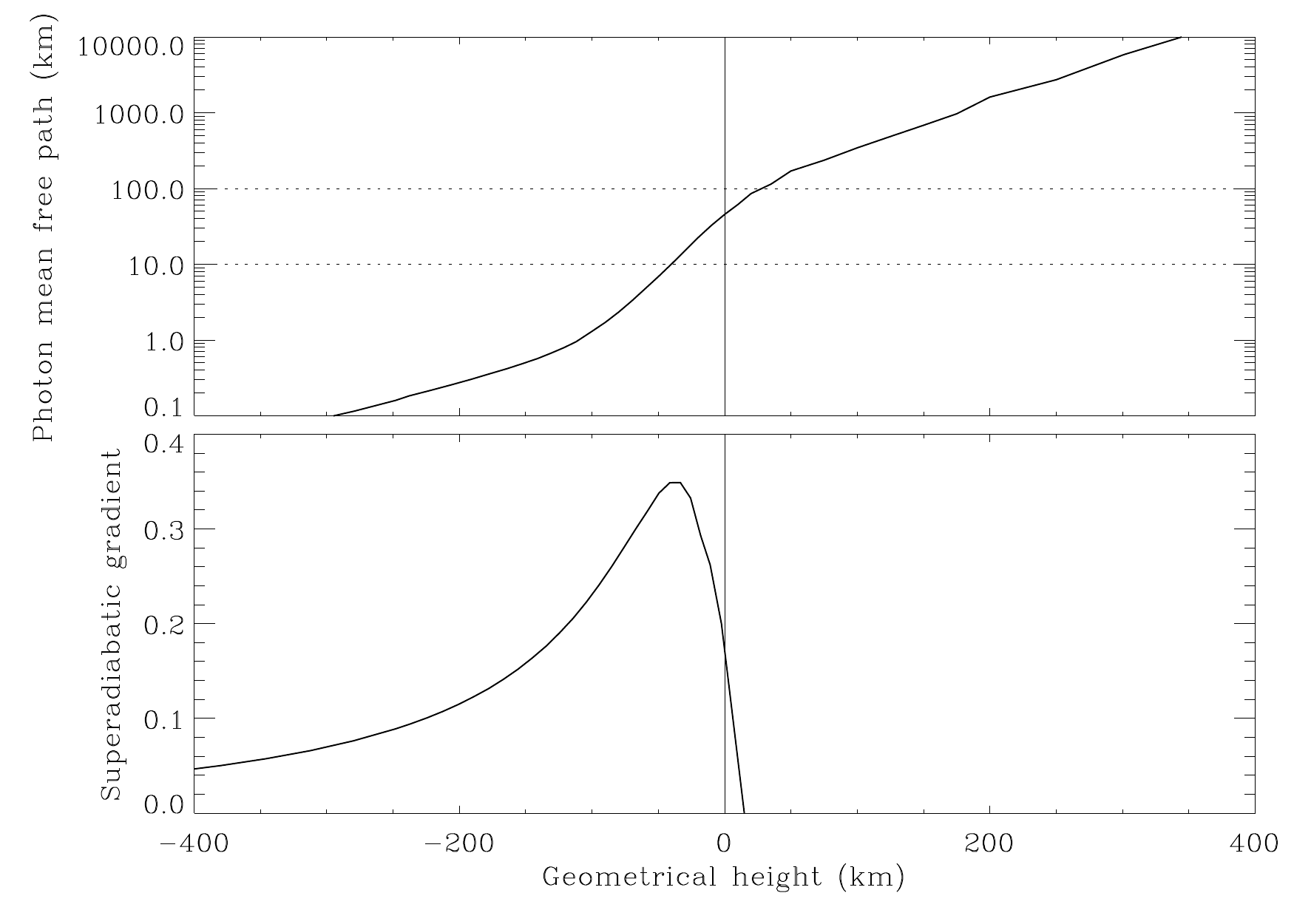}\caption{The two
  main parameters that govern flux tube collapse, the photon mean free
  path and the superadiabatic temperature gradient, vs. height in the
  Sun's surface layers. The conditions favor a flux tube population in
  the 10-100\,km size range, which is indicated by the horizontal
  dotted lines in the top panel. From \citep{stenflo-s11aa}}\label{fig:collapse} 
\end{figure*}

The convective collapse instability is most efficient where the
superadiabatic gradient is the largest, and the collapsing flux
regions need to be optically thick, i.e., their cross sections need to be
larger than the photon mean free path in the lateral direction across
the flux tube. In Fig.~\ref{fig:collapse} we show for an atmospheric model of
\citet{stenflo-maltby86} and a convection zone model of
\citet{stenflo-spruit77} how the photon mean free path (in
the continuum at 5000\,\AA) and superadiabatic gradient (difference between
the magnitudes of the actual logarithmic temperature gradient and the
corresponding adiabatic temperature gradient) vary with height. The
zero point of the geometrical height scale, marked by the solid
vertical line in the figure, is 
where the optical depth in the continuum at 5000\,\AA\ is unity and
represents the bottom of the photosphere. Below this level we enter
the subsurface layers. 

The superadiabatic gradient has a sharp maximum but drops to zero as
we enter the photosphere, marking the end of the convection zone. In
the near surface layers, where the convective collapse instability may
occur, the photon mean free path is of order 10-100\,km, as indicated
by the two dotted horizontal lines in the upper panel. Below this size
range the instability will lose efficiency. We would therefore expect
kG flux tubes to be larger than $d\approx 10$\,km. After formation, flux
tubes will have a limited life time proportional to $d^{\,2}$ due
to the fluting instability (cf. Sect~\ref{sec:dynamoscales}) and will
fragment to smaller scales, but kG-type fields may not be maintained
below about 10\,km. Such theoretical considerations are consistent
with the finding from line-ratio analysis of Hinode data that the bulk
of the collapsed flux population has sizes in the range 10-100\,km
\citep{stenflo-s11aa}. 

Although most of the collapsed flux tubes are expected to have sizes
just beyond the resolution capabilities of current telescopes, the
distribution has a tail that extends well into the resolved
domain. Thus it has been possible to directly resolve such flux tubes
on the quiet Sun \citep[cf.][]{stenflo-keller92,stenflo-lagg10}. The
convective collapse process has received support both from observations
  \citep{stenflo-nagata08,stenflo-fischer09} and from 3-D
  MHD simulations \citep{stenflo-danilovic10}.

It should be noted that because the theoretical scenario that
describes the convective collapse mechanism uses the idealization of 
a magnetic flux region embedded in nonmagnetic surroundings, the
spontaneous instability leads to a dichotomy, where the flux ends up
being in either collapsed or uncollapsed form. This scenario has
its counterpart in the 2-component interpretational model used for the
line-ratio observations. In the real Sun, however, there is no sharp
distinction between magnetic and nonmagnetic, the conditions are
better described in terms of continuous distribution functions
connecting the weak and strong fields. In such an environment the
occurrence of the collapse instability is not an either-or question,
and the amplitude and end result of the instability can be expected to
have a continuous range of values.

\section{Small-scale properties of internetwork magnetic fields}\label{sec:weak}
The magnetic intermittency described in the previous section implies
that much of the photospheric magnetic flux exists in highly
concentrated form. As these flux elements have small filling factors, the
bulk of the photosphere must be filled with much weaker fields, which
get tangled by the turbulent motions and as a consequence become
nearly invisible in
magnetograms because of cancelation effects, unless the angular resolution
is sufficiently high to separate the opposite magnetic polarities. In
the present section we will focus the attention on the properties of
these weaker, tangled fields, which with commonly used terminology may
be classified as ``internetwork'' fields, in contrast to the collapsed
network-type fields that were dealt with in the previous section. 

We will first address the angular distribution of the internetwork
fields and then deal with their intrinsic field strengths. Since most
of the structuring of these fields takes place on scales far smaller
than the resolved ones, the diagnostic methods used need to be
resolution-independent and be based on statistical ensemble
averages. The determination of the angular distributions is based on
the symmetry properties of the transverse Zeeman effect, the
determination of the field strengths on observations of the Hanle
depolarization effect.

\subsection{Angular distribution of the field
  vectors}\label{sec:angular}
In an isotropic medium the angular distribution of the field vectors
would also be isotropic. The stratification of the outer solar
envelope with its nearly exponential decrease of the gas
pressure with height has however a number of important
consequences for the field orientation: (a) The flux tubes that are anchored in the deeper and
denser layers are being pushed towards a vertical orientation by buoyancy. (b) The
exponential decrease of the external gas pressure leads to an
expansion with height of the flux tubes, whose field lines 
flare out to become increasingly horizontal with a canopy structure in the higher
layers \citep{stenflo-giovanelli80,stenflo-jonesgiov83}. (c) The many 
flux loops of various sizes, which represent the topological structure of the 
turbulent field, are predominantly vertical in the deeper layers where
their footpoints are anchored, but are more horizontal near the loop
tops \citep[cf.][]{stenflo-steiner10}. For these various reasons
we expect the angular distribution 
of the field vectors on the quiet Sun to have a preference for the
vertical orientation in the lower layers, but the preference should
shift in favor of the horizontal orientation as we move up in
height. However, we need observations to verify this expectation and
to tell us at what height the
transition from vertical to horizontal preference takes place and how
it depends on field strength. 

It came as a big surprise when initial analysis of Hinode SOT/SP data
for the quiet-sun disk center was proclaimed to show that the
horizontal fields not only dominate, but that there is as much as five
times more horizontal than vertical magnetic flux
\citep{stenflo-orozcoetal07,stenflo-litesetal08}. Subsequent analyses
of the identical Hinode data set however gave completely different 
results,  quasi-isotropic angular distribution according to
\citet{stenflo-ramos09}, vertical preference but with a flux-density
dependence that approaches the isotropic case in the weak flux-density
limit according to \citet{stenflo-s10aa}. 

The cause of this colossal divergence of the results has to do with
the pitfalls when trying to determine the field inclinations from a
combination of the linear and circular polarization data. In contrast to the
circular polarization, the linear polarization has a highly nonlinear
(nearly quadratic) dependence on the magnetic field
(cf. Sect.~\ref{sec:zeeman}). The interpretation 
of the optical averages over each pixel depends on the model that we use for the
subpixel structuring of the nonlinear quantity over which we
average. In addition the polarimetric noise is larger in the
transverse field in comparison with the longitudinal field by a factor
of approximately 25 (cf. Fig.~\ref{fig:pdfnoise}). The results for the
inclination angles then depend on how we deal with both the noise and the
nonlinearities. 

\begin{figure*}
\centering
\includegraphics[width=0.6\textwidth]{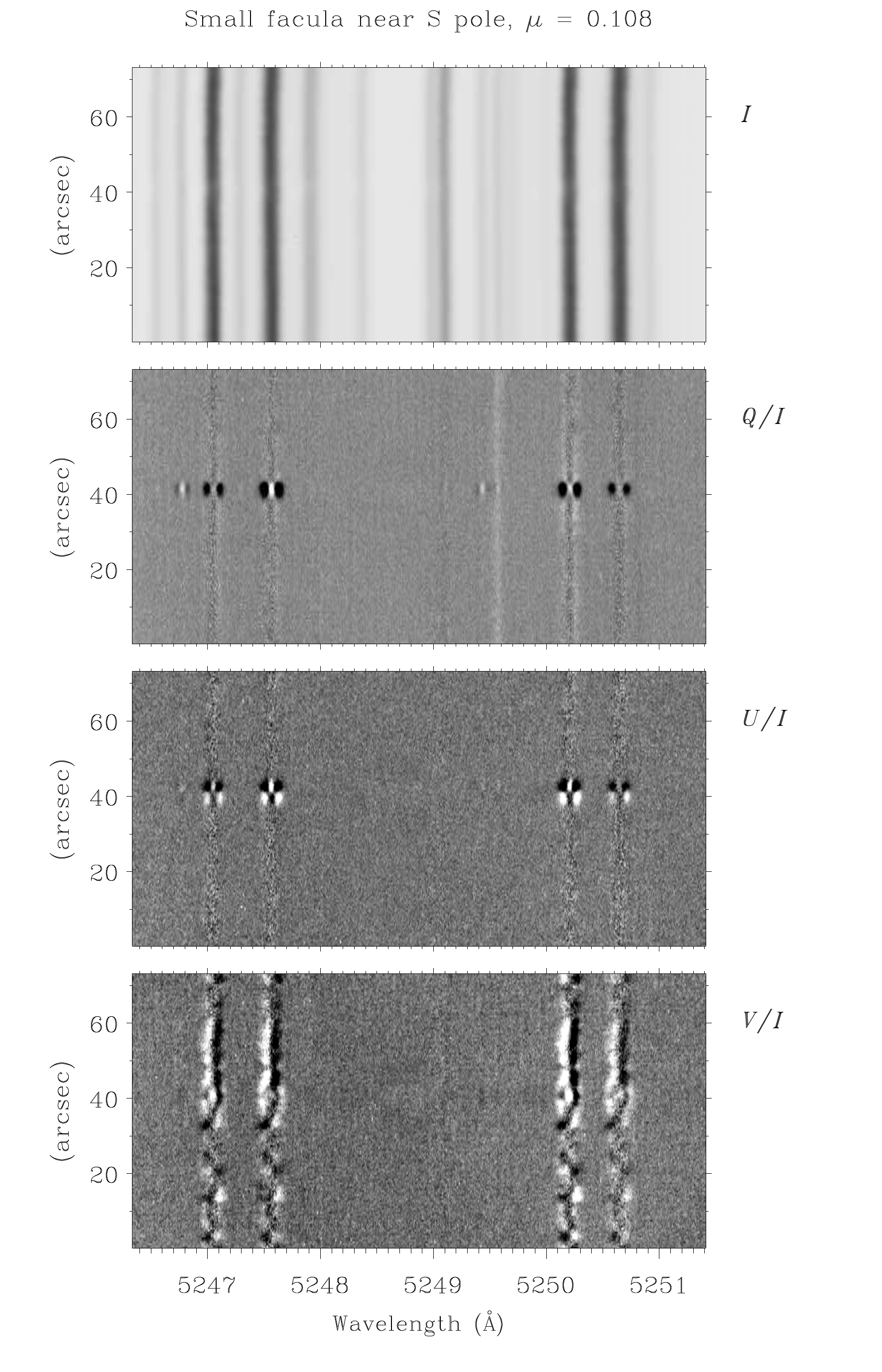}
\caption{Signature of nearly vertical fields for a polar facula at
  $\mu=0.108$ (6\,arcsec inside the limb) near the heliographic S
  pole. As the Stokes coordinate system has been defined such that
  positive Stokes $Q$ represents linear polarization oriented
  perpendicular to the radius vector, the symmetric 3-lobe pattern of
  the transverse Zeeman effect in Stokes $Q$ will have the sign
  pattern $-+-$ if the field is vertical. The recording was made with
  ZIMPOL-2 at the French THEMIS telescope on June 10, 2008. From
\citet{stenflo-s13aa1}.}
\label{fig:faculalimb} \end{figure*}

Recently the angular distribution could be determined with a method
that avoids these pitfalls by not depending on any comparison between
the linear and circular polarization but only on the qualitative
symmetry properties of the Stokes line profiles in linear polarization
due to the transverse Zeeman effect when observing off disk center. To
understand these symmetry properties, let us begin by considering a
vertical magnetic field that is observed relatively close to the solar
limb, and choose the Stokes coordinate system such that the positive
$Q$ direction is defined to be perpendicular to the radius vector on
the solar disk (or, equivalently, parallel to the nearest solar limb). As in
the vertical field case the transverse component is along the radius vector,
the $Q$ polarization of the central $\pi$ component of the line
profile will be
positive for an absorption line. The observed, symmetric Stokes $Q$
line profile will then have the sign pattern $-+-$. This is
illustrated in Fig.~\ref{fig:faculalimb} for a small polar facula
6\,arcsec inside the S polar limb. All spectral lines in the field of
view exhibit the
$Q/I$ sign pattern $-+-$, a tell-tale 
signature of vertical fields. $U/I$, which relates to the $\pm
45^\circ$ directions of the transverse field, is nonzero, indicating
that the facular field is not strictly vertical but has an angular
distribution, but $U/I$ reverses sign across the facula with a balance
between the opposite signs. 

The weak, emission-like feature in $Q/I$ at 5249.58\,\AA\ is due to
nonmagnetic scattering polarization in ionized neodymium and is
unrelated to the present topic. 

For transverse fields perpendicular to the radius vector, which would
represent horizontal fields, the $Q/I$ sign pattern would be the
opposite, $+-+$. We notice in Fig.~\ref{fig:faculalimb} that this sign
pattern, although faint, characterizes everything along the slit outside the polar
facula. 

Let us now instead of fields with a well-defined orientation consider a
statistical ensemble of magnetic elements with an angular
distribution of the field vectors that is axisymmetric around the
vertical direction. Then the distribution of inclination angles
determines whether the overall distribution, in comparison with the
isotropic case, is preferentially vertical (more peaked around the
vertical) or preferentially horizontal
(more pancake like). It can be shown
\citep[cf.][]{stenflo-s13aa1} that whenever there is any vertical
preference, even if it is slight, then the $Q/I$ sign pattern of the
{\it ensemble average} will always be $-+-$, while a horizontal
preference will produce the pattern $+-+$ for the ensemble
average. Therefore we can conclude that the weak background fields
(which may be classified as internetwork fields) in
Fig.~\ref{fig:faculalimb} are preferentially horizontal. 

We have made a number of recordings across different types of flux
concentrations at various center-to-limb distances, from tiny faculae
to small sunspots, and all of them exhibit the $Q/I$ $-+-$ sign
pattern, which tells us that they have a clear vertical preference for
all disk positions, all the way to the limb. The behavior of the weak
background fluxes is different, as we saw in
Fig.~\ref{fig:faculalimb}. We have therefore also made recordings (with the Swiss 
ZIMPOL-2 polarimeter mounted on the French THEMIS telescope on
Tenerife) in a sequence of very quiet regions spanning a range of limb
distances from $\mu =0.08$ to $\mu =0.5$. None of these regions included any
facular or network element. All of them can therefore be classified as
internetwork. The ensemble averages needed to determine whether we
have vertical or horizontal preference are obtained by averaging over the
spatial resolution elements along the spectrograph slit. 
The recordings reveal a systematic variation of the $Q/I$ sign pattern
with $\mu$, with a transition from one sign pattern to the opposite 
at $\mu\approx 0.2$. Closer to the limb there is a horizontal
preference, closer to disk center a vertical preference. 

Since smaller $\mu$ values refer to higher levels of the solar
atmosphere, we can therefore conclude that the internetwork field
favors a vertical orientation in the low to middle photosphere
but has a transition in favor of the horizontal orientation above the
atmospheric level that corresponds to $\mu =0.2$. In contrast, the
network and facular fields preserve a vertical preference over the
entire height range spanned by the various $\mu$ values. 

Note that this result is model independent in the sense that it only
depends on the qualitative symmetry property of the transverse Zeeman
effect represented by the sign pattern of the $Q/I$ profiles. It is
also resolution independent, since it only refers to ensemble
averages, and optical and numerical averagings are equivalent. If we
average the $Q$ signal for a given area of the Sun optically because it is not resolved
(and thus represents one resolution element), or we resolve the same
area with many resolution elements and then do numerical averaging of
$Q$ for each of them, the results will be identical. $Q$ is
proportional to the number of photons from each surface 
element, regardless of whether the element is resolved or not. Therefore our
result for the angular distribution will not change with the
application of future telescopes that have much higher resolving power. 

The idea of using ensemble averages of Stokes $Q$ to determine the
angular distributions for observations off
disk center was introduced long ago \citep{stenflo-s87}, but its
application was limited at that time, since the 
observations used single-pixel detectors (photomultipliers). With the
present availability of high-precision imaging polarimeters we are now
in a position to exploit this technique in great detail. 

Close to the solar limb the properties of the horizontal magnetic
fields in the upper photosphere can also be explored with
circular-polarization magnetograms, since near the limb the line of
sight is nearly horizontal. Thus the circular
polarization as observed with low spatial resolution (3\,arcsec) is
found to exhibit large temporal fluctuations, which reveal
that the horizontal components of the internetwork magnetic fields are
highly dynamic \citep{stenflo-harvey_etal07}.

\subsection{Intrinsic field strengths of the internetwork
  fields}\label{sec:hanresults}
In Sect.~\ref{sec:hanleturb} we described how observations of the Hanle effect
reveal the existence of an ocean of hidden magnetic flux with mixed
polarities on scales much smaller than the spatial resolution of
current telescopes. The scattering polarization in photospheric lines
is observed to be systematically 
reduced and spatially almost invariant, while the orientation of the
plane of polarization does not deviate significantly from the
nonmagnetic case. The only known consistent explanation is
in terms of the Hanle effect. The reduced degree of polarization
requires an abundance of magnetic fields with Zeeman
splittings comparable in magnitude to the combined collisional and radiative
damping widths of the atomic transitions used for the diagnostics. The 
absence of any observed Hanle rotation of the plane of
polarization that one would expect for such fields can only be
explained by cancelation effects of positive and negative rotation
angles when averaging over an ensemble of spatially unresolved flux
elements \citep{stenflo-s82}. Note that this behavior only applies to
photospheric but not to chromospheric lines. The scenario
with an ocean of hidden, tangled magnetic flux is only valid for 
the photospheric levels, not above. 

The conversion of Hanle depolarization to turbulent field strength
$B_t$ was indicated in Fig.~\ref{fig:hanledepol}. Usually the flux
ensemble is for simplicity characterized by a single field strength,
while the angular distribution is assumed to be isotropic, although the
obtained results are not very sensitive to the assumed 
field orientations. The first
applications of the Hanle depolarization effect allowed $B_t$ to be
constrained within the range 10-100\,G
\citep{stenflo-s82,stenflo-s87}. This range is still valid in
spite of rather divergent results since then with detailed
radiative-transfer modeling. 

Most of the modeling has focused on the interpretation of observations
with the Sr\,{\sc i} 4607\,\AA\ line. While initial radiative-transfer
modeling gave values of order 30\,G
\citep{stenflo-faurob93,stenflo-faurobetal95}, the most sophisticated
modeling to date with 3-D atmospheres generated by numerical
simulations gave 60\,G \citep{stenflo-trujetal04}. When the
interpretation is made with plausible PDF field strength
distributions instead of the single-valued assumption, the Hanle
constraints imply an average field strength about twice as large,
suggesting that the total magnetic energy density of the hidden field
may play an important role for the overall energy
balance of the atmosphere \citep{stenflo-trujetal04}. 

These large $B_t$ values are in apparent contradiction with
interpretations based on observations of the {\it differential Hanle
  effect} in pairs of optically thin molecular lines, like the C$_2$
lines around 5141\,\AA\ \citep{stenflo-bf04}. A synoptic program with
this line pair that was started in 2007 with ZIMPOL at IRSOL (Istituto
Ricerche Solari Locarno) has given the value $B_t =7.4\pm 0.8$\,G for
the period 2007-2009, with no evidence for any significant temporal
variations \citep{stenflo-kleint10}. 

\citet{stenflo-trujetal04} have offered a possible explanation for
this apparent contradiction, by demonstrating that the molecular C$_2$
lines are formed exclusively inside the granulation cells, while the
Sr\,{\sc i} 4607\,\AA\ line gets similar contributions from both 
the intergranular lanes and the cell interiors. This implies 
that the $B_t$ value inside the intergranular lanes must be
substantially higher than the 60\,G value, since the cell
interiors with their large filling factors would not contribute much to
the observed Hanle depolarization. 

This interpretation is plausible, since the intermittent, collapsed
kG-type flux concentrations are found to have a strong preference for
the intergranular lanes as well \citep{stenflo-s11aa}, and there is
reason to believe that the tangled fields are supplied with flux
from the decaying flux tube fields. Observational support has 
been provided by \citet{stenflo-snik10}, who analysed Hinode SOT
high-resolution observations with a CN filter and were able to  
statistical separate the cell interiors and the
intergranular lanes. It was found that the scattering polarization was
significantly reduced in the lanes. Detailed radiative transfer
modeling of the Hanle effect in the CN ultraviolet lines leads to values
for $B_t$ that are comparable in magnitude to those found for the Sr\,{\sc i}
4607\,\AA\ line \citep{stenflo-shapiro11}. The difference with respect
to the optically thin 
C$_2$ lines is explained by the much larger optical thickness of the
CN lines, which makes them vary less from cell interior to the
lanes. In spite of these insights we need Hanle effect observations in
various types of spectral lines 
with high spatial resolution to better clarify these issues. 

Another not yet settled issue is whether the hidden flux revealed by
the observed Hanle depolarization varies with the phase of the solar
cycle. The synoptic observations made so far in the C$_2$ lines
suggest that the field is invariant. \citet{stenflo-trujetal04}
have compiled evidence, which indicates that also the depolarization in the Sr\,{\sc i}
4607\,\AA\ line does not vary with the cycle. However, visual inspection of 
recordings of various sections of the Second Solar Spectrum observed since 1995
indicates that the appearance of the spectrum is subject to
significant changes, which imply that there are 
substantial cycle variations of the hidden field \citep{stenflo-s03}. 
We therefore need a synoptic program not only for the C$_2$ lines, but
in particular also for atomic lines like the Sr\,{\sc i} 4607\,\AA\
line, with which such large strengths for the hidden field are obtained. Such a
program is currently being initiated with ZIMPOL at IRSOL. 

In Sect.~\ref{sec:local} it was noted in
Fig.~\ref{fig:enmaglocdyn} in the context of observational
evidence for the operation of a local dynamo that a Kolmogorov-type $-5/3$ power law in
the inertial range that goes down to the magnetic diffusion limit is much too steep to provide
the large amounts of small-scale magnetic flux that are required in
order to satisfy the constraints from the observed Hanle
depolarization. Consistency with the observed Hanle effect can be
restored if the small-scale end of the energy spectrum is raised to
give a slope of $-1.0$ instead. Such a pile-up of small-scale flux
cannot be naturally explained in terms of the turbulent cascade from
larger-scale fluxes generated by the global dynamo but seems to demand
the existence of an additional source of flux, which would be
naturally provided by a local dynamo. 

If the hidden flux revealed by the Sr\,{\sc i}
4607\,\AA\ line observations is due to a local dynamo, then it must be
statistically invariant with respect to the solar cycle. If on the
other hand its source comes from the global dynamo, then
significant cycle variations are unavoidable. Therefore a synoptic
Hanle program for atomic lines should be able to settle this issue.

\section{Concluding remarks}\label{sec:conclude}
New powerful solar facilities, both on ground and in space, are
being implemented or are in construction. A primary objective of most
of these
facilities is the diagnostics of magnetic fields via
spectro-polarimetry with high spatial resolution, to advance our
understanding of the small-scale morphology and evolution of the
magnetic field and its relation to the thermodynamic structuring and
solar activity. It is believed that many of the fundamental 
processes take place on small scales, and that it is therefore crucial
to resolve these scales to understand the underlying physics. The
enormous advances in computing power now allow numerical simulations
of the stratified solar atmosphere with the inclusion of all the
relevant radiative and MHD processes and a spatial resolution in 3-D
that surpasses that of the observations. Progress is often judged by
how well the simulations compare with the observations. 

In the present review we have paid particular attention to the property that the magnetic
structuring continues to scales that are orders of magnitude smaller
than the resolution of current telescopes and also much smaller than
the grid resolution of the numerical simulations. In addition, as the current
telescope resolution is close to the transition around the 100\,km
scale between the optically thick and thin regimes, there is a more
fundamental limitation to the resolution that can be reached by future
telescopes. While there is no principal limitation to the possible angular
resolution in the transversal plane, the {\it resolution along the
  line of sight} is fixed by the optical depth scale and remains of
order 100\,km in the photosphere, no matter what the angular resolution is. The
structuring through tangling of the field lines or through local
dynamo action will ignore this limit and continue far into the
optically thin regime. The observational consequence of this is that
the visibility of structures in magnetograms will decline with smaller scale
size due to cancelation effects or small filling factors along the
line of sight. 

Here we have highlighted three types of techniques that avoid the
cancelation and filling factor problems and lead to {\it
  resolution-independent} fundamental insights into the nature and
properties of the magnetic fields in the spatially unresolved
small-scale domain: (1) Magnetic line-ratio technique to explore the
magnetic intermittency and intrinsic field strengths in the
strong-field tail of the probability density function, without 
dependence on filling factor. (2) Exclusive use of the symmetry properties
of the transverse Zeeman effect in observations away from disk center,
to determine whether the angular distribution of the field vectors favors
the vertical or horizontal orientation in comparison with the
isotropic case. (3) Use of the Hanle effect depolarization
of photospheric spectral lines, to determine the intrinsic 
strength of a field that is tangled with mixed polarities on scales
in the optically thin regime. 

Instead of trying to invert the above polarimetric observables with the
help of idealized interpretational models, more physical insight may
be gained by going in the opposite direction, namely to test the validity
of the MHD simulations of the solar atmosphere by computing synthetic
values for the observables that represent the three mentioned 
techniques used to diagnose the resolution-independent
properties. This has not yet been done. A particularly fundamental test is
to try to reproduce the observed Hanle 
depolarization in lines like the Sr\,{\sc i} 4607\,\AA\ line. If as 
believed much of the Hanle depolarization effect 
occurs on scales beyond the reach of the grid resolutions of
state-of-the-art simulations, then these 
simulations are likely to fail to reproduce this observable. It is a prediction
that can be directly tested with available data and simulation
models. Another fundamental test is to try to 
reproduce the observed sign pattern of the ensemble averages of the
transverse Zeeman effect as a function of center-to-limb
distance. Still another test is the reproduction of the observed
5250/5247 Stokes $V$ line ratio, in particular its small intrinsic
scatter and its calibrated relation to the 6302/6301 line ratio. 

The present review has focused on the photosphere, since almost all
the knowledge gained from Stokes polarimetry about the magnetic-field
properties refer to this 
part of the atmosphere. Future work will increasingly focus on the
chromosphere and the atmosphere above, where Zeeman-effect techniques
are less effective. Here we expect the diagnostic tools provided by
the Hanle effect via the wealth of scattering polarization structures
in the Second Solar Spectrum to come increasingly into the
foreground. The theoretical tools to deal with observations of
scattering polarization in magnetized media are currently under
development (cf. the references in Sect.~\ref{sec:historical} to the
series of Solar Polarization 
Workshops). This work in progress is addressing and trying to solve deep
problems related to the
quantum-mechanical foundation of the 
interaction between radiation and matter in magnetized media with
scattering and collisional processes. These processes need to be
integrated in a polarized radiative transfer 
formulation with partial frequency redistribution, which then needs to
be applied to atmospheres with spatially unresolved magnetic 
structures. The Sun serves as a fascinating quantum  and
plasma physics laboratory for these various phenomena. 

When high-resolution spectro-polarimetric observations from space  of
the chromosphere-corona transition region will become available in the
future, the theoretical foundations developed in the context of the
Second Solar Spectrum may be extended to the vacuum ultraviolet part
of the spectrum and applied to the diagnostics
of magnetic fields in the regions of the Sun's atmosphere, where the
corona is being heated.


\end{document}